\documentclass[journal]{IEEEtran}

\usepackage{cuted}
\usepackage{cite}
\usepackage{amsmath,amssymb,amsfonts}
\usepackage{algorithmic}
\usepackage{graphicx}
\usepackage{textcomp}
\def\BibTeX{{\rm B\kern-.05em{\sc i\kern-.025em b}\kern-.08em
		T\kern-.1667em\lower.7ex\hbox{E}\kern-.125emX}}

\usepackage{lipsum}
\usepackage{color}
\usepackage{bm}
\usepackage{algorithm}

\usepackage{hhline}

\ifCLASSOPTIONcompsoc
\usepackage[caption=false, font=normalsize, labelfont=sf, textfont=sf]{subfig}
\else
\usepackage[caption=false, font=footnotesize]{subfig}
\fi

\usepackage{soul}

\usepackage{babel}
\usepackage{threeparttable}

\begin{document}
%\title{Efficient Wavenumber-Domain Algorithm for Near-Field Imaging with Arrays in a Polyline Setup}
\title{Compressive Sensing Based Sparse MIMO Array Optimization for Wideband Near-Field Imaging}
\author{
	Shuoguang Wang,
	Shiyong Li,~\IEEEmembership{Member,~IEEE,}  
	Ahmad Hoorfar, ~\IEEEmembership{Life Senior Member,~IEEE,}  
	Ke Miao,     \\
    Guoqiang Zhao,
    and  Houjun Sun

\thanks{%Manuscript received 2021. 
	The work was supported by the National Natural Science Foundation of China under Grant 62071043. (\textit{Corresponding authors: Shiyong Li and Ahmad Hoorfar}). }

\thanks{S. Wang, S. Li, K Miao, G. Zhao, and H. Sun are with the Beijing Key Laboratory of Millimeter Wave and Terahertz Technology, Beijing Institute of Technology, Beijing 100081, China. (e-mail: lisy\_98@bit.edu.cn).}

\thanks{S. Li, G. Zhao, and H. Sun are also with the Tangshan Research Institute of  BIT, Tanshan 063007,China. }

\thanks{Ahmad Hoorfar is with the Antenna Research Laboratory, Center for Advanced Communications, Villanova University,Villanova, PA 19085, USA (e-mail: ahoorfar@villanova.edu).}

}

%\markboth{IEEE}
%{Shell \MakeLowercase{\textit{et al.}}: Efficient Wavenumber-Domain Processing for Near-Field Imaging with Polyline Arrays}

\maketitle

\begin{abstract}
In the area of near-field millimeter-wave imaging, the generalized sparse array synthesis (SAS) method is in great demand. The traditional methods usually employ the greedy algorithms, which may have the convergence problem. This paper proposes a convex optimization model for the multiple-input multiple-output (MIMO) array design based on the compressive sensing (CS) approach. We generate a block shaped reference pattern, to be used as an optimizing target. The pattern occupies the entire imaging area of interest in order to involve the effect of each pixel into the optimization model. In MIMO scenarios, we can fix the transmit subarray and synthesize the receive subarray, and vice versa, or doing the synthesis sequentially. The problems associated with focusing, sidelobes suppression, and grating lobes suppression of the synthesized array are examined in details. Numerical and experimental results demonstrate that the synthesized sparse array can offer better image qualities than the sparse arrays with equally spaced or randomly spaced antennas with the same number of antenna elements.

\end{abstract}

% Note that keywords are not normally used for peerreview papers.

\IEEEpeerreviewmaketitle

\section{Introduction}

Microwave/Millimeter-wave imaging has been widely applied in a wide range of applications, such as concealed weapons detection \cite{sheen2010near}, nondestructive testing \cite{zoughi2000microwave}, biomedical imaging \cite{nikolova2011microwave}, and through-the wall radar imaging (TWRI) \cite{amin2017through}, to name a few. Three-dimensional (3-D) imaging can be obtained based on using a 2-D array aperture and wideband signals\cite{zhang2012three,zhang2019three}. 

The state-of-the-art 3-D imaging systems are capable of achieving real-time high-resolution imaging, while facing the trade-off between the imaging accuracy and system cost \cite{gonzalez2013sparse}. The multiple-input multiple-output (MIMO) arrays offer a solution to reduce the system cost with smaller number of antenna elements compared to monostatic arrays \cite{li2021efficient}. To reap the benefits of sparseness, the number of antennas of MIMO can be further reduced by using the sparse array synthesis (SAS) method, without causing high-level grating lobes and sidelobes \cite{anderson1991ultra}.

 Significant efforts have been made in MIMO SAS for near-field imaging. Through selecting various transmit and receive antenna intervals, Lockwood et al. proposed a 2-D SAS framework in \cite{lookwood1996optimizing}. Several different MIMO topologies were investigated by Ahmed et al. \cite{ahmed2009near} for ultra-wideband (UWB) near-field imaging. Yang et al.  designed two different 2-D MIMO arrays using the principle of effective aperture and element projection\cite{yang2009experimental}. Zhuge and Yarovoy designed a MIMO array using the principle of separable aperture functions\cite{zhuge2010near}. They also proposed a 2-D curvilinear array topology with lower element shadowing than the periodic arrays\cite{zhuge2012study}.  These topologies are designed as specific array structures, but lacking generalization properties. 
 
 Some scholars achieve SAS for near-field imaging based on the greedy algorithms. Borja et al. \cite{gonzalez2013sparse} used the simulated annealing (SA) algorithm to minimize the sidelobe levels for optimizing the receive positions in the MIMO array. Yang et al. \cite{yang2008uwb} utilized the particle swarm optimization (PSO) method to determine the antenna positions by designing a fitness function. Tan et al. \cite{tan2016sparse} adopted the  principle of uniformity and lower element shadowing with the greedy algorithm to generate the planar array topologies. However, it is only suitable for the 4M-transmit-4N-receive (M and N are positive integers) UWB MIMO planar array topology. An et al. \cite{an2021task} used the multi-objective covariance matrix evolution strategy (CMA-ES) algorithm to optimally design a MIMO array topology for TWRI. The aforementioned stochastic optimization techniques, however, require a significant amount of computation time to reach global solution, and might face the problem of local convergence if not executed over many random trials with a large number of iterations per trial. Consequently, a generalized MIMO SAS-based method is highly desired for near-field imaging. However, two challenges must first be overcome:

\begin{itemize}
	\item Model challenge: unlike the SAS for beamforming, the combinatorial optimization model of SAS for wideband near-field imaging cannot be presented directly. A representative model is required. 
	
	\item PSF challenge: the point spread function (PSF) is a standard indicator for imaging, since it can show the sidelobe levels, main beamwidth, and the grating lobe positions\cite{yang2009experimental}. In MIMO SAS, it is not easy to ensure that all the PSFs at different positions fulfill the requirements.
\end{itemize}

In regards to the first challenge, we propose a convex optimization based model that is deduced from the electromagnetic wave propagation model. We include a sampling matrix into the optimization model to be used for selecting antennas. Therefore, the SAS problem is transformed into reconstruction of the sampling matrix, which then falls into a compressive sensing (CS) framework\cite{donoho2006compressed}.   As for the reference pattern, we set it as the imaging result of the continuous scatterer targets filling over the imaging region of interest, which is proved to cover the PSF challenge. In addition, the proposed SAS method has the flexibility of being implemented in optimization of the transmit subarray and the receive subarray separately.   

The remainder of this paper is organized as follows. In Section II, the construction of the CS-based sparse MIMO array synthesis model is detailed for wideband near-field imaging, with the corresponding reference pattern synthesizing strategy.  Numerical results are shown in Section III to analyze the performance of the MIMO topology synthesized by the proposed method. Section IV further experimentally verifies the efficacy of the proposed method. Finally, Section V summarises the results and concludes the paper.

\section{Multistatic Sparse Array Synthesizing Methodology for Near-Field Imaging}

\subsection{CS Based Sparse Array Synthesis}

\begin{figure}[!t]
	\centering	
	\includegraphics[width=2.6in]{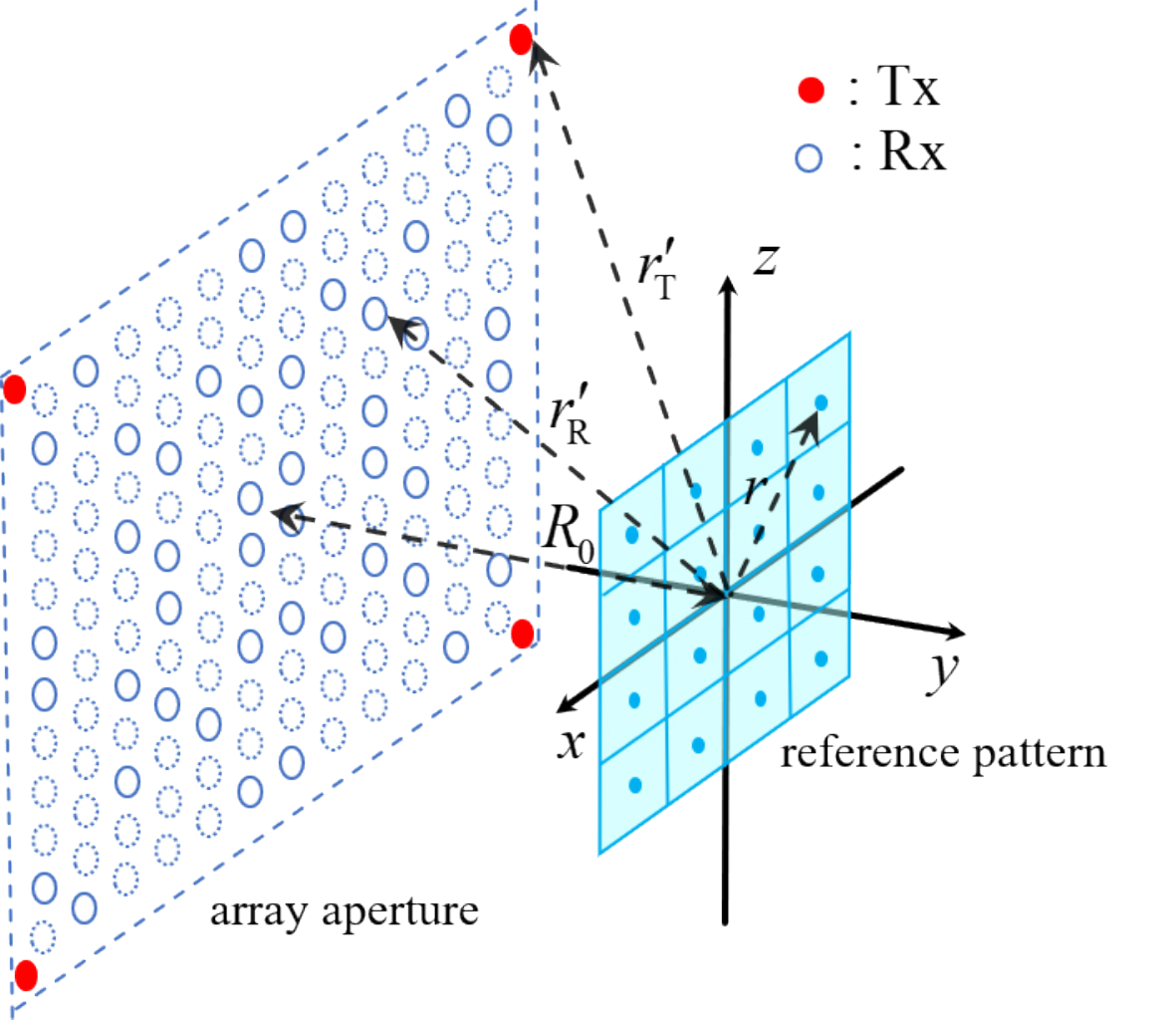}
	\hfill
	\\	
	\caption{MIMO sparse planar array for  wideband near-field imaging, where the reference pattern is set to cover all the imaging area at a fixed distance.}
	\label{fig1_MIMO_array_geometry}
\end{figure}
%%
%The width of the main lobe (the cross-range resolution) of a wideband system is determined by the ratio between the wavelength at the center operating frequency and the size of the aperture\cite{zhuge2010short}.

We first consider a wideband multistatic planar sparse array, of which the $N_{\rm T}$ transmit antennas are fixed (on its four corners in the demo topology), and the receive antennas are arbitrarily positioned. The $N$ receive elements are sampled from the 2-D dense array, as illustrated in Fig.\ref{fig1_MIMO_array_geometry}. Let $s(k,\bm{r'}_{\rm T},\bm{r'}_{\rm R})$ be the measured scattered field, received at the dense array, from the target scene with $Q$ scatterers. Then, assuming the waves illuminating the target scene are in general spherical waves, the total scattered field under the Born approximation \cite{1953Methods} can be expressed as, 
%Assuming the ultilization of the Born approximation \cite{1953Methods} for
%the scattered field, it can be written as,
%\begin{equation}\label{mimo_scat_wave}
%	s(k,\bm{r'}_{\rm T},\bm{r'}_{\rm R})\!=\! \sum_{q=1}^Q \sigma_q e^{-\mathrm{j}k\left( \left|\bm{r'}_{\rm T}-\bm{r}_q\right|+\left|\bm{r'}_{\rm R}-\bm{r}_q\right|\right)} ,
%\end{equation}
\begin{equation}\label{mimo_scat_wave} 	s(k,\bm{r'}_{\rm T},\bm{r'}_{\rm R})\!=\! \sum_{q=1}^Q \sigma_q\frac{e^{-\mathrm{j}k\left( \left|\bm{r'}_{\rm T}-\bm{r}_q\right|+\left|\bm{r'}_{\rm R}-\bm{r}_q\right|\right)} }{16\pi^2\left|\bm{r'}_{\rm T}-\bm{r}_q\right|\cdot \left|\bm{r'}_{\rm R}-\bm{r}_q\right|}, \end{equation}
where $k=\frac{2\pi f}{c}$ denotes the wavenumber, and $\bm{r'}_{\rm T}$ and $\bm{r'}_{\rm R}$ represent the positions of transmit and receive elements, respectively. $\bm{r}_q$ stands for the position of the $q$th target with $\sigma_q$ being its scattering coefficient. Since the amplitude decay with range has little impact on focusing, it is omitted here. The matrix form of  \eqref{mimo_scat_wave} can be denoted as:
\begin{equation}\label{scat_wave_mat}
	\bm{s}_k\!= \bm{A}_{{\rm R}_k} \bm{A}_{{\rm T}_k}\bm{\sigma},
\end{equation}
where
\begin{equation}\label{s_vec}
\!\!\bm{s}_k\!=\! \left[s\left(k,\bm{r'}_{\rm T},\bm{r'}_{\rm R_1}\right)\!,s\left(k,\bm{r'}_{\rm T},\bm{r'}_{\rm R_2}\right)\!,\dots,s\left(k,\bm{r'}_{\rm T},\bm{r'}_{{\rm R}_N}\right)\right]^T
\end{equation}
%\begin{equation}\label{AR_forward}
%\bm{A}_{{\rm R}_k}\!=\! \begin{bmatrix}		
%e^{-\mathrm{j}k\left|\bm{r'}_{{\rm R}_1}-\bm{r}_{1}\right|}&  \!\!\dots  &e^{-\mathrm{j}k\left|\bm{r'}_{{\rm R}_1}-\bm{r}_{Q}\right|} \\		
%\vdots& \!\!\ddots  &  \vdots \\
%e^{-\mathrm{j}k\left|\bm{r'}_{{\rm R}_N}-\bm{r}_{1}\right|}  & \!\!\dots &   e^{-\mathrm{j}k\left|\bm{r'}_{{\rm R}_N}-\bm{r}_{Q}\right|}
%\end{bmatrix}_{N \times N_Q}
%\end{equation}
%
%
%
%\begin{equation}\label{AT_forward}
%\!\bm{A}_{{\rm T}_k}\!=\!\begin{bmatrix}		
%		e^{-\mathrm{j}k\left|\bm{r'}_{{\rm T}}-\bm{r}_{1}\right|}&   &  & \\
%		&\!\!\!e^{-\mathrm{j}k\left|\bm{r'}_{{\rm T}}-\bm{r}_{2}\right|} &  & \\			
%		& & \!\!\ddots  \!&  \\
%		&  &  &	\! e^{-\mathrm{j}k\left|\bm{r'}_{{\rm T}}-\bm{r}_{Q}\right|}
%	\end{bmatrix}_,
%%_{N_Q \times N_Q,}
%\end{equation}
\begin{equation}\label{AR_forward} \bm{A}_{{\rm R}_k}\!=\! \begin{bmatrix}		 \frac{e^{-\mathrm{j}k\left|\bm{r'}_{{\rm R}_1}-\bm{r}_{1}\right|}}{4\pi\left|\bm{r'}_{{\rm R}_1}-\bm{r}_{1}\right|}&  \!\!\dots  &\frac{e^{-\mathrm{j}k\left|\bm{r'}_{{\rm R}_1}-\bm{r}_{Q}\right|}}{4\pi\left|\bm{r'}_{{\rm R}_1}-\bm{r}_{Q}\right|}  \\		 \vdots& \!\!\ddots  &  \vdots \\ \frac{e^{-\mathrm{j}k\left|\bm{r'}_{{\rm R}_N}-\bm{r}_{1}\right|}}{4\pi\left|\bm{r'}_{{\rm R}_N}-\bm{r}_{1}\right|}  & \!\!\dots &   \frac{e^{-\mathrm{j}k\left|\bm{r'}_{{\rm R}_N}-\bm{r}_{Q}\right|}}{4\pi\left|\bm{r'}_{{\rm R}_N}-\bm{r}_{Q}\right|} \end{bmatrix}_{N \times N_Q} 
\end{equation}
 \begin{equation}\label{AT_forward} \!\bm{A}_{{\rm T}_k}\!=\!\begin{bmatrix}		 		\frac{e^{-\mathrm{j}k\left|\bm{r'}_{{\rm T}}-\bm{r}_{1}\right|}}{4\pi\left|\bm{r'}_{{\rm T}}-\bm{r}_{1}\right|}&   &  & \\ 		&\!\!\!\frac{e^{-\mathrm{j}k\left|\bm{r'}_{{\rm T}}-\bm{r}_{2}\right|}}{4\pi\left|\bm{r'}_{{\rm T}}-\bm{r}_{2}\right|} &  & \\			 		& & \!\!\ddots  \!&  \\ 		&  &  &	\!\frac{e^{-\mathrm{j}k\left|\bm{r'}_{{\rm T}}-\bm{r}_{Q}\right|}}{4\pi\left|\bm{r'}_{{\rm T}}-\bm{r}_{Q}\right|} 	\end{bmatrix}_, %_{N_Q \times N_Q,} 
 \end{equation}

\begin{equation}\label{sigma_vec}
	\bm{\sigma}\!= \left[\sigma_1,\sigma_2,\dots,\sigma_Q\right]^T.
\end{equation}

Then, a sampling matrix $\bm{\Psi}$ is introduced into the dense array, to reduce the number of receive antennas, resulting in,

\begin{equation}\label{sampling_equ}
	\bm{s'}_k\!= \bm{\Psi} \bm{s}_k,
\end{equation}
where
\begin{equation}\label{s'}
\!\!\bm{s'}_k\!=\! \left[s'\left(k,\bm{r'}_{\rm T},\bm{r'}_{\rm R_1}\right)\!,s'\left(k,\bm{r'}_{\rm T},\bm{r'}_{\rm R_2}\right)\!,\dots,s'\left(k,\bm{r'}_{\rm T},\bm{r'}_{{\rm R}_N}\right)\right]^T
\end{equation}
\begin{equation}\label{Psi}
\bm{\Psi}= \begin{bmatrix}		
w_1&   &  & \\
& w_2 &  & \\			
& & \ddots  &  \\
&  &  &w_N
\end{bmatrix}_{N \times N.}
\end{equation}

Here $\bm{\Psi}$ is a diagonal matrix, and most of its diagonal elements are equal to zero. Note that $M$ sampling points are determined on the imaging plane at some specific ranges, and $\bm{r''}_m$ represents the position of the $m$th imaging pixel. The underlying principle of selecting $M$ is detailed in Section II. B. Using the delay-and-sum method \cite{kay1993fundamentals}, the intermediate imaging result concerning the wavenumber, $k$, and the transmit position, $\bm{r'}_{\rm T}$, can be expressed as:
\begin{equation}\label{MIMO_BP_equ}
	\bm{E}_{{\rm T}_k}\!= \bm{\Phi}_{{\rm T}_k} \bm{\Phi}_{{\rm R}_k}\bm{s'}_k,
\end{equation}
where
\begin{equation}\label{I_vec}
	\bm{E}_{{\rm T}_k}\!= \left[E\left(k,\bm{r'}_{\rm T},\bm{r''}_1\right),E\left(k,\bm{r'}_{\rm T},\bm{r''}_2\right),\dots,E\left(k,\bm{r'}_{\rm T},\bm{r''}_M\right)\right]^T,
\end{equation}

%\begin{equation}\label{Phi_T_forward}
%	\!\bm{\Phi}_{{\rm T}_k}\!=\!\begin{bmatrix}		
%		e^{\mathrm{j}k\left|\bm{r''}_1-\bm{r'}_{\rm T}\right|}&   &  & \\
%		&\!\!\!e^{\mathrm{j}k\left|\bm{r''}_2-\bm{r'}_{\rm T}\right|} &  & \\			
%		& & \!\!\ddots  \!&  \\
%		&  &  &	\!\! e^{\mathrm{j}k\left|\bm{r''}_M-\bm{r'}_{\rm T}\right|}
%	\end{bmatrix}_,
%	%_{M \times M}
%\end{equation}
%
%
%\begin{equation}\label{Phi_Rmat}
%	\bm{\Phi}_{{\rm R}_k}= \begin{bmatrix}		
%		e^{\mathrm{j}k\left|\bm{r''}_1-\bm{r'}_{{\rm R}_1}\right|}&  \dots  &e^{\mathrm{j}k\left|\bm{r''}_1-\bm{r'}_{{\rm R}_N}\right|} \\		
%		\vdots& \ddots  &  \vdots \\
%		e^{\mathrm{j}k\left|\bm{r''}_M-\bm{r'}_{{\rm R}_1}\right|} & \dots &   e^{\mathrm{j}k\left|\bm{r''}_M-\bm{r'}_{{\rm R}_N}\right|}
%	\end{bmatrix}_{M \times N.}
%\end{equation}

%\lipsum[2]
%\begin{strip}	

%\rule{1\textwidth}{.4pt}
\begin{equation}\label{Phi_T_forward}
	\!\!\bm{\Phi}_{{\rm T}_k}\!=\! 4\pi \begin{bmatrix}		
		\left|\bm{r''}_1 \!\! - \!\! \bm{r'}_{\rm T}\right| \! e^{\mathrm{j}k\left|\bm{r''}_1-\bm{r'}_{\rm T}\right|}&    & \\
		
		 & \!\!\ddots  \!&  \\
		 &  &	\!\! \left|\bm{r''}_M \!\! - \!\! \bm{r'}_{\rm T}\right| \! e^{\mathrm{j}k\left|\bm{r''}_M-\bm{r'}_{\rm T}\right|}
	\end{bmatrix}_,
	%_{M \times M}
\end{equation}

%\end{strip}
%
%\begin{strip}	
%	\rule{1\textwidth}{.4pt}

\begin{equation}\label{Phi_Rmat}
	\!\!\bm{\Phi}_{{\rm R}_k}\!\!=\! 4\pi \!\! \begin{bmatrix}		
		\left|\bm{r''}_1  \!\! - \!\! \bm{r'}_{{\rm R}_1}\!\right| \! e^{\mathrm{j}k\left|\bm{r''}_1  \! - \! \bm{r'}_{{\rm R}_1}\!\right|}&  \!\!\!\! \dots \!\! &\left|\bm{r''}_1  \!\! - \!\! \bm{r'}_{{\rm R}_N}\!\right| \! e^{\mathrm{j}k\left|\bm{r''}_1  \! - \! \bm{r'}_{{\rm R}_N}\!\right|} \\		
		\vdots& \!\! \ddots \!\! & \!\! \vdots \!\!\\
		\left|\bm{r''}_M  \!\! - \!\! \bm{r'}_{{\rm R}_1}\right| \! e^{\mathrm{j}k\left|\bm{r''}_M \! - \! \bm{r'}_{{\rm R}_1}\!\right|\!} & \!\! \dots &  \!\! \left|\bm{r''}_M  \!\! - \!\! \bm{r'}_{{\rm R}_N}\right| \!\!\!\! e^{\mathrm{j}k\left|\bm{r''}_M \! - \! \bm{r'}_{{\rm R}_N}\!\right|\!}
	\end{bmatrix}_.
%{M \times N.}
\end{equation}
%\rule{1\textwidth}{.4pt}
%\end{strip}

%We discard the  amplitude factors in \eqref{Phi_T_forward} and \eqref{Phi_Rmat}, because they have little impact on focusing \cite{kay1993fundamentals}.

Equation \eqref{MIMO_BP_equ} shows the integrating procedure of imaging. Combining with \eqref{sampling_equ}, \eqref{MIMO_BP_equ} can be written as:
\begin{equation}\label{MIMO_BP_equ2}
	\bm{E}_{{\rm T}_k}\!= \bm{\Phi}_{{\rm T}_k} \bm{\Phi}_{{\rm R}_k} \bm{\Psi}\bm{s}_k.
\end{equation}

Because $\bm{\Psi}$ is a diagonal matrix, and $\bm{s}_k$ is a vector, it is easy to show that,
\begin{equation}\label{MIMO_BP_equ3}
	\bm{E}_{{\rm T}_k}\!= \bm{\Phi}_{{\rm T}_k} \bm{\Phi}_{{\rm R}_k} \bm{S}_k\bm{w},
\end{equation}
where
\begin{equation}\label{S}
\!\!\! \bm{S}_k \!\!=\!\! \begin{bmatrix}		
		s\left(k,\bm{r'}_{\rm T},\bm{r'}_{\rm R_1}\right)\!\!\!&   &  & \\
		&\!\!\! s\left(k,\bm{r'}_{\rm T},\bm{r'}_{\rm R_2}\right) \!\!&  & \\			
		& & \!\!\!\ddots \!\!  &  \\
		&  &  &\!\! s\left(k,\bm{r'}_{\rm T},\bm{r'}_{{\rm R}_N}\right)
	\end{bmatrix}
\end{equation}
\begin{equation}\label{w_vec}
	\bm{w}\!= \left[w_1,w_2,\dots,w_N\right]^T.
\end{equation}

The intermediate imaging results for all the wavenumbers and transmit positions should be integrated. By doing so, the imaging reconstruction procedure can be finally expressed as follows:

\begin{equation}\label{MIMO_BP_equ4}
	\bm{E}\!= \bm{B} \bm{w},
\end{equation}
where
\begin{equation}\label{MIMO_E_sum}
	\bm{E}\!= \sum_k\sum_{\bm{r'}_{\rm T}} \bm{E}_{{\rm T}_k}
\end{equation}
\begin{equation}\label{B_sum}
	\bm{B}\!= \sum_k \sum_{\bm{r'}_{\rm T}} \bm{\Phi}_{{\rm T}_k} \bm{\Phi}_{{\rm R}_k} \bm{S}_k.
\end{equation}

As such, the SAS for the wideband near-field imaging problem can then be expressed as:
\begin{equation}\label{MIMO_cvx_problem1}
	\bm{w}\!= \arg\left(\min\limits_{\bm{w}}\Vert \bm{w} \Vert_0\right) \quad {\rm s.t.} \; \Vert \bm{E}_{\rm ref}-\bm{B}\bm{w}\Vert^2_2\leq \varepsilon,
\end{equation}
where $\bm{w}$ is the nonzero entry of excitation values, and $\bm{E}_{\rm ref}=\left[E_{\rm ref1},E_{\rm ref2},\dots,E_{{\rm ref}M} \right]^T$ is the sampled vector of the reference wideband near-field imaging reconstruction results (referred to as the `reference pattern'). $\varepsilon$ is the tolerance to narrow the gap between the imaging results of the synthesized array and the desired reference patterns. 

However, the inherent nonconvexity behavior of \eqref{MIMO_cvx_problem1} makes it difficult to solve using standard techniques as it is an NP-hard problem. Fortunately, the iterative weighted $l_1$ norm proposed in \cite{candes2008enhancing} can be employed to relax the $l_0$ norm regularization. The $l_1$ norm relaxation for wideband MIMO SAS can be expressed as:

\begin{equation}\label{cvx_problem2}
	\bm{w}\!= \arg\left(\min\limits_{\bm{w}}\Vert \bm{w} \Vert_1\right) \quad {\rm s.t.} \; \Vert \bm{E}_{\rm ref}-\bm{B}\bm{w}\Vert^2_2\leq \varepsilon.
\end{equation}

Here, we employ the online available MATLAB-based CVX toolbox \cite{grant2014cvx} to solve \eqref{cvx_problem2}. Then, the desired receive array layout with the corresponding sparse weights is obtained. If both the positions of transmit and receive antennas need to be synthesized, we can fix the transmit positions, optimize the receive positions, and vice versa. The experimental setup will show the benefit of this MIMO SAS principle.

\subsection{Near-Field Reference Pattern for MIMO Arrays}

According to the aforementioned optimization method, an appropriate  reference pattern is in demand for MIMO wideband SAS. For example, a fixed reference pattern was utilized in \cite{huang2018near} for synthesizing a sparse array for beamforming, which cannot properly cover the aformentioned `PSF challenge' for near-field imaging. Unlike the work in \cite{huang2018near}, we present the principle of setting the reference pattern for near-field imaging within the proposed SAS model, through which the sidelobe levels (SLL) and grating lobe levels can be under control. We also verify the satisfying and superior performance of the sparse MIMO array synthesized with the reference pattern.

Let us consider the design of receive element positions with the fixed transmit array elements. The proposed reference pattern synthesis method can be summarized as follows.

First, we determine the referenced receive array topology (marked as `referenced array' for brevity). The optimization procedure will narrow the performance gap between the 'referenced array' and the 'synthesized sparse array'. We set the weights of 'referenced array' as $\bm{w}_{\rm ref}$ with aperture weighting or apodization to reduce the SLLs \cite{turnbull1991beam,karimkashi2009focused}. Typically, we mark the 'referenced array' as the 'full array' when its element spacing satisfies the Nyquist sampling criterion.

Second, we get the reference pattern for MIMO wideband near-field imaging by integration of the PSFs of scatterers at different imaging positions. Assuming $Q$ scatterers are set as the reference points, then scattered EM waves received by the MIMO array can be written as $\bm{s}_{\rm ref}$ according to \eqref{mimo_scat_wave}. Particularly in our implementation, to overcome the `PSF challenge', the reference pattern is arranged as the continuous point targets covering the interested imaging region as given by the subsequent lemma:

\textbf{\emph{Lemma 1}}: As for the sparse MIMO array synthesized by \eqref{cvx_problem2}, the difference between the imaging result and that of the referenced array is less than $\varepsilon$, for all the target combinations from the reference pattern. 

\textbf{\emph{Proof}}: Please see the Appendix.

Third, the values of $M$ sampling positions: $\bm{r''}_1,\bm{r''}_2,\dots,\bm{r''}_M$ need to be determined. Assuming the imaging region is confined in a rectangular shaped area, the size along the azimuth and height dimensions are $D_x$ and $D_z$, respectively. The azimuth resolution is determined by the extent of spatial frequencies \cite{li2021efficient}, that is,
\begin{equation}\label{resolution_MIMO}
	\delta_{x} =\frac{\pi}{k_{x_{\max}}},
\end{equation}
where $k_x$ is the wavenumber along the azimuth direction, and $k_{x_{\max}}$ is the maximum value of $k_x$. According to \cite{li2021efficient}, we have the following relations:
\begin{equation}
	k_{x_{\max}} =k_{x_{{\rm T}\max}} + k_{x_{{\rm R}\max}},
\end{equation}
\begin{equation}
	k_{x_{{\rm T}\max}} \approx k_c \sin \frac{\Theta_{x_{\rm T}}}{2},
\end{equation}
\begin{equation}
	k_{x_{{\rm R}\max}} \approx k_c \sin \frac{\Theta_{x_{\rm R}}}{2},
\end{equation}
where $k_c$ denotes the center wavenumber of the working EM waves, and $\Theta_{x_{\rm T}}$ and $\Theta_{x_{\rm R}}$ represent the smaller one between the azimuth angle subtended by the transmit or receive array aperture, and the azimuth antenna beamwidth. Thus, \eqref{resolution_MIMO} is rewritten as:
\begin{equation}
	\delta_{x} =\frac{\lambda_c}{2\left(\sin\frac{\Theta_{x_{\rm T}}}{2} +\sin\frac{\Theta_{x_{\rm R}}}{2} \right)},
\end{equation}
where $\lambda_c$ is the wavelength of the center operating frequency. Correspondingly, the resolution along the vertical direction is given by,
\begin{equation}
	\delta_{z} =\frac{\lambda_c}{2\left(\sin\frac{\Theta_{z_{\rm T}}}{2} +\sin\frac{\Theta_{z_{\rm R}}}{2} \right)},
\end{equation}
where $\Theta_{z_{\rm T}}$ and $\Theta_{z_{\rm R}}$ represent the smaller one between the vertical angle subtended by the transmit or receive array, and the vertical antenna beamwidth.

Although the main lobe widths of PSFs change among different positions, the position variations can be omitted since they are usually small enough compared with the imaging region. Hence, the number of the sampling points can be obtained by,
\begin{equation}
\label{M_sampling}
\left\{
\begin{aligned}
 & M_{x}\geq \lfloor\frac{D_x}{\delta_x}\rfloor + 1\\
 & M_{z}\geq \lfloor\frac{D_z}{\delta_z}\rfloor + 1,
\end{aligned}
\right.
\end{equation}
where $\lfloor\cdot \rfloor$ is the round down symbol, and $M_x$ and $M_z$ are the sampling points along the azimuth and vertical directions. The number of all sampling points can be computed as $M= M_x M_z$.

\begin{table}[!t]
	\centering
	\caption{Simulation Parameters for 1-D Sparse Array}
	\setlength{\tabcolsep}{3pt}
	\begin{threeparttable}
		\begin{tabular}{p{200pt}  p{30pt}}
			%\begin{tabular}{|c|c|}
			\hline\hline
			Parameters& Values \\[0.5ex]
			\hline
			Imaging distance $(R_0)$&
			1.0 m\\[0.5ex]		
			Start frequency& 
			30 GHz \\[0.5ex]
			Stop frequency&
			35 GHz \\[0.5ex]
			Number of frequency steps&
			101 \\[0.5ex]
			
			Number of transmit antennas &
			2 \\[0.5ex]		
			Spacing of transmit antennas of the full array&52 cm \\[0.5ex]
			Number of receive antennas of the full array&
			26 \\[0.5ex]	
			Spacing of receive antennas of the full array&
			2.0 cm \\[0.5ex]
			
			Number of receive antennas of the synthesized/ equally spaced sparse array&
			17 \\[0.5ex]
			
			Spacing of receive antennas of the equally spaced sparse array & 3.13 cm
			\\[0.5ex]
			
			\hline
			
		\end{tabular}
	\end{threeparttable}
	\label{tab_linear_arr}
\end{table}

\begin{figure}[t]
	\centering
	\includegraphics[width=2.0in]{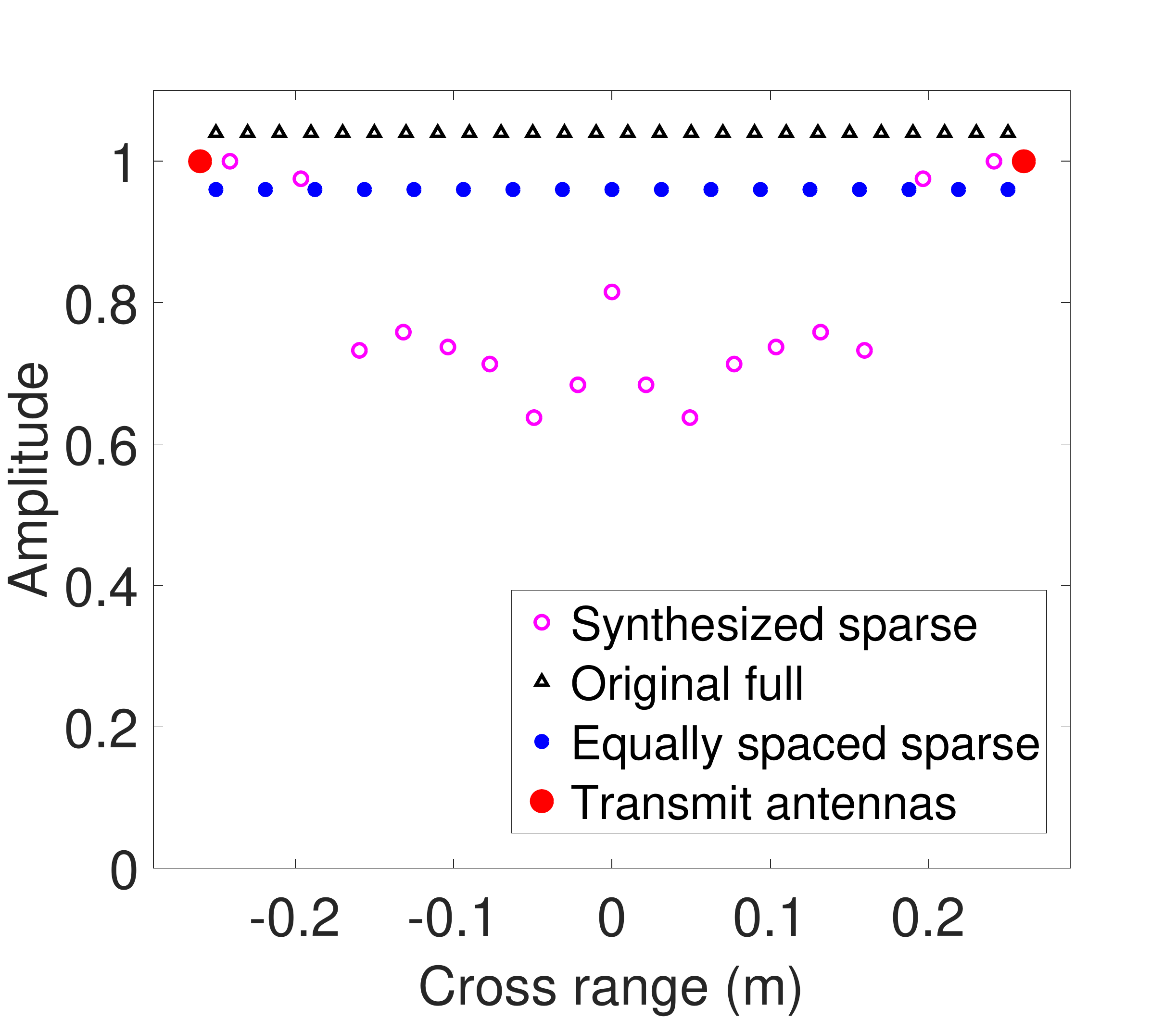}
	\hfill	
	\caption{Topologies for the 1-D synthesized sparse MIMO array, the full MIMO array, and the equally spaced sparse MIMO array.}
	\label{full_1D_array_topologies}
\end{figure}

\section{Numerical Results}

In this section, we compare the imaging results of the arrays generated by the proposed method and those of the arrays with equally spaced antennas.

\subsection{Simulation Results of 1-D synthesized MIMO Array}

We set a scenario of short-range imaging using a 1-D MIMO linear sparse array, with two transmit antennas fixed on the sides. The receive positions and corresponding weights are generated based on the proposed SAS method. The simulation parameters are listed in Table \ref{tab_linear_arr}.

\begin{figure}[!t]
	\centering
	\subfloat[]{\label{a}
		\includegraphics[width=2.0in]{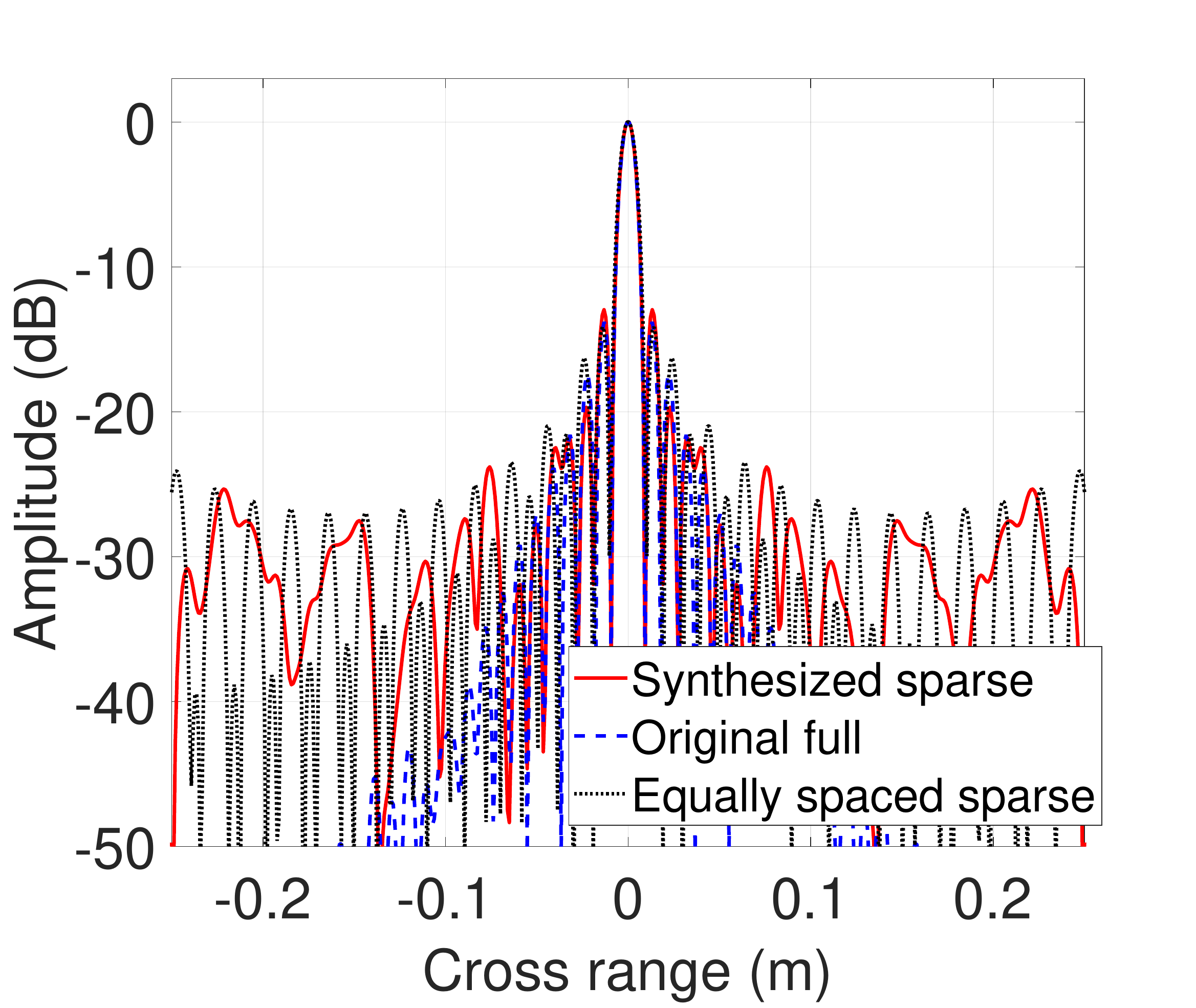}}
	\hfill
	
		\vspace{-0.5mm} 
		
	\subfloat[]{\label{b}
	\includegraphics[width=2.0in]{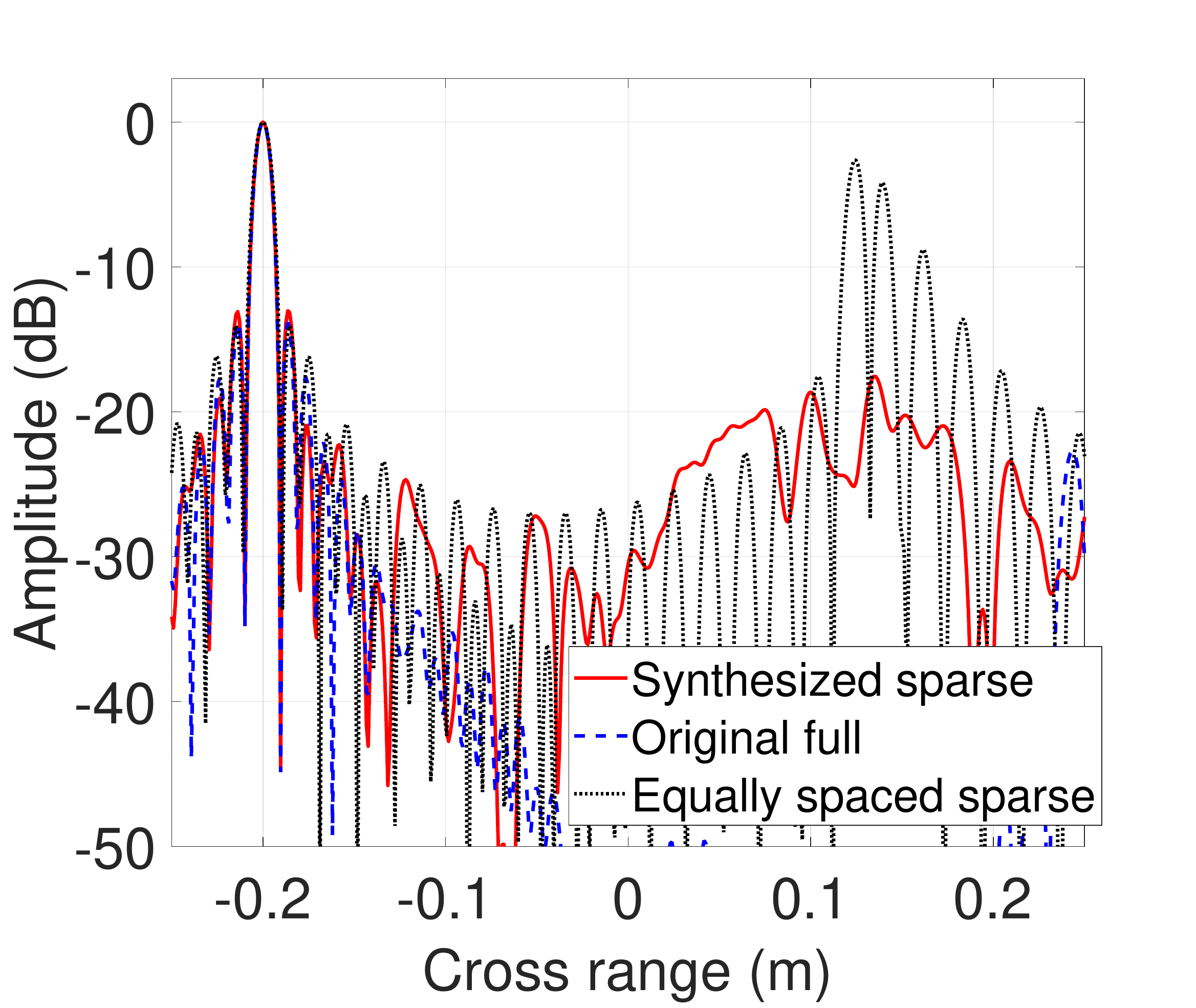}}
	\hfill
	\\	
	\caption{1-D imaging results of the synthesized sparse MIMO array, the full MIMO array, and the equally spaced sparse MIMO array, for the scatterer at (a) the center position, and (b) the edge position.}
	\label{1D_near-field imaging_MIMO_results}
\end{figure}

% with receive antennas satisfying Nyquist sampling principle 

The synthesized linear MIMO array topology is compared with the referenced MIMO array (referred to as the `full MIMO array' for conciseness), and the MIMO array with equally spaced receive antennas with the same number of elements (referred to as the `equally spaced sparse MIMO array'). These array topologies are illustrated in Fig \ref{full_1D_array_topologies}. The imaging results of the full MIMO array are regarded as the standards. 

The cross-range PSFs for different scatterers at the center and edge positions are shown in Fig. \ref{1D_near-field imaging_MIMO_results}. All the MIMO arrays perform well in the center-position case. In the edge-position case, the grating lobes of the synthesized MIMO array are significantly suppressed. In contrast, the equally spaced sparse MIMO array exhibits higher grating lobes close to the main beam, which cannot be efficiently eliminated.
The results verify the efficacy of the proposed method for the 1-D MIMO SAS case.

\begin{table}[!t]
	\centering
	\caption{Simulation Parameters for 2-D Sparse Array}
	\setlength{\tabcolsep}{3pt}
	\begin{threeparttable}
		\begin{tabular}{p{200pt}  p{40pt}}
			%\begin{tabular}{|c|c|}
			\hline\hline
			Parameters& Values \\[0.5ex]
			\hline
			Imaging distance $(R_0)$&
			2 m\\[0.5ex]		
			Start frequency& 
			30 GHz \\[0.5ex]
			Stop frequency&
			35 GHz \\[0.5ex]	
			Number of frequency steps&
			101 \\[0.5ex]			
			Number of transmit antennas &4 \\[0.5ex]		
			Spacing of transmit antennas of the full array& 60 cm \\[0.5ex]
			Number of receive antennas of the full array&
			20$\times$20 \\[0.5ex]	
			Spacing of receive antennas of the full array&
			3.0 cm \\[0.5ex]			
			Number of receive antennas of the synthesized sparse array&
			120 \\[0.5ex]
			Number of receive antenna of the equally spaced sparse array&
			11$\times$11 \\[0.5ex]	
			Spacing of receive antennas of the equally spaced sparse array &
			5.7 cm\\[0.5ex]	
			\hline
		\end{tabular}
	\end{threeparttable}
	\label{tab_planar_arr}
\end{table}

\subsection{Simulation Results of 2-D synthesized MIMO Array}

\begin{figure}[!t]
	\centering
	\vspace{-2.5mm} 
	\includegraphics[width=2.0in]{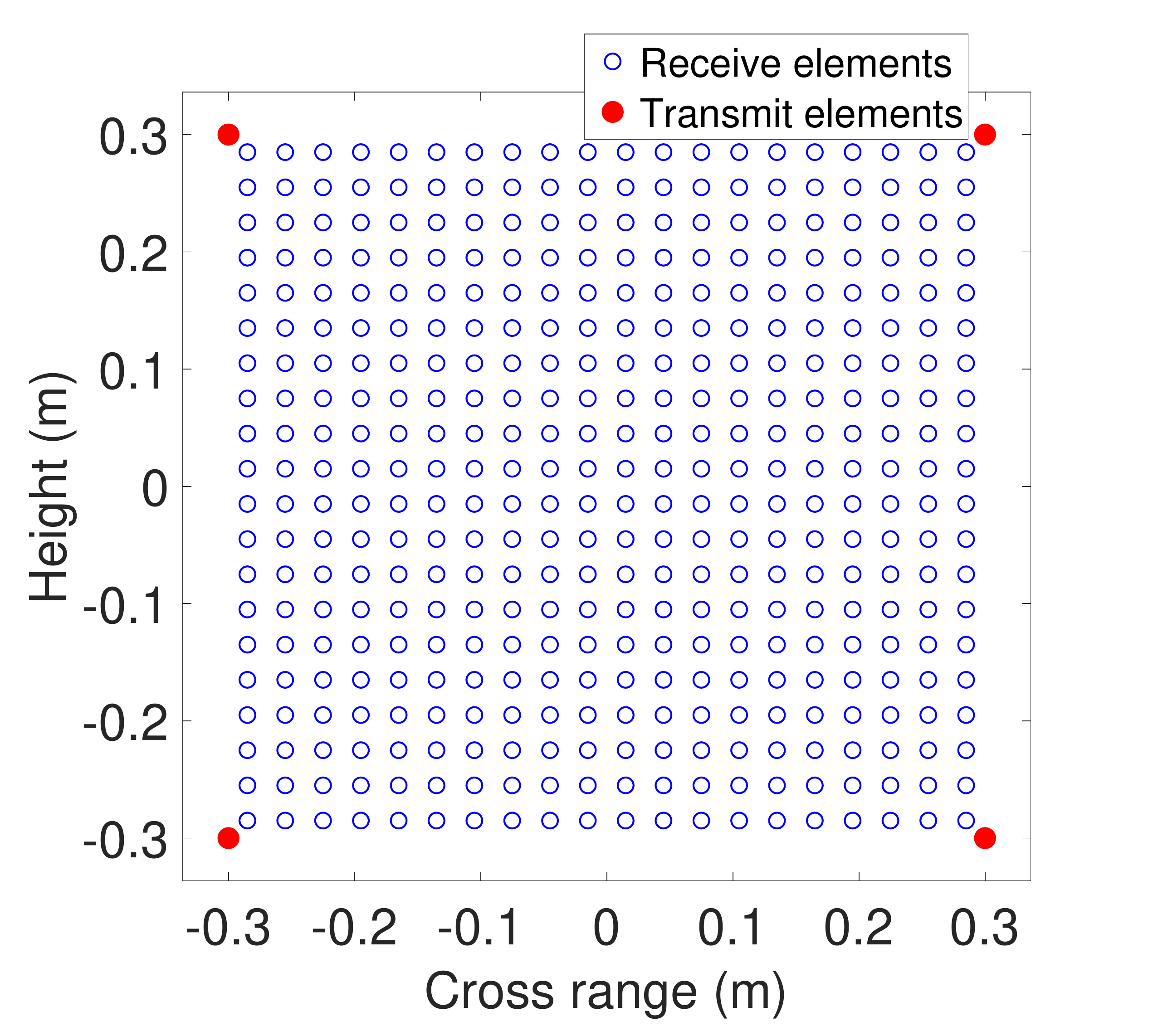}
	\hfill	
	\caption{The topology for the 2-D full MIMO array.}
	\label{full_2D_mimo_array_simu}
\end{figure}

\begin{figure}[!t]
	\centering
	\vspace{-2.5mm} 
	\includegraphics[width=2.0in]{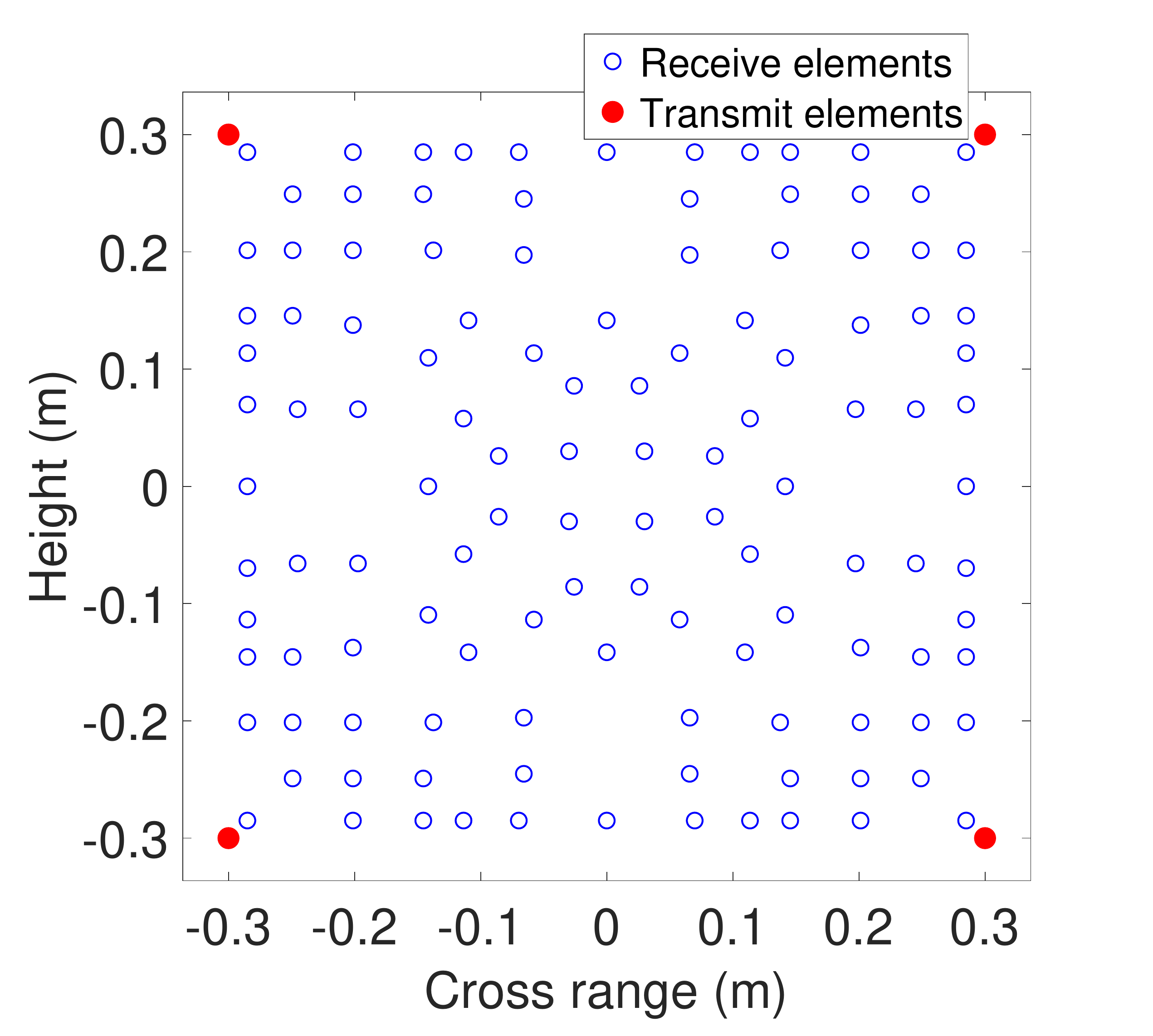}
	\hfill	
	\caption{The topology for the synthesized 2-D sparse MIMO array.}
	\label{sparse_2D_mimo_array_simu}
	\vspace{-1.5mm} 
\end{figure}

\begin{figure}[!t]
	\centering
	\vspace{-2.5mm} 
	\includegraphics[width=2.0in]{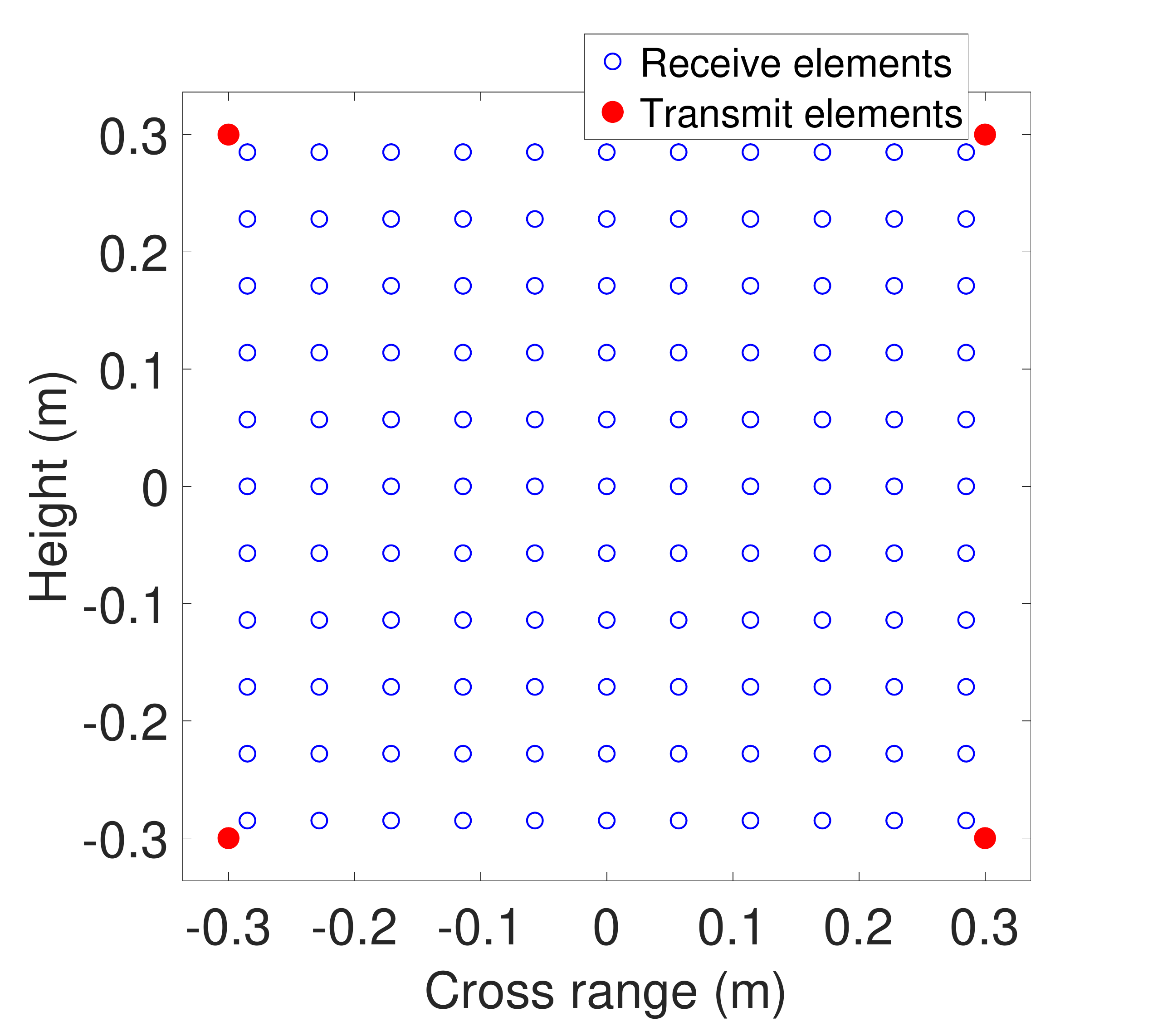}
	\hfill	
	\caption{The topology for the 2-D equally spaced sparse MIMO array.}
	\label{less_2D_mimo_array_simu}
\end{figure}

%\begin{figure}[!t]
%	\centering
%	\vspace{-2.5mm} 
%	\includegraphics[width=3.5in]{raw_vs_merged_pic}
%	% 	\hfill	
%	\vspace{-7.5mm} 
%	\caption{Topologies for the synthesized sparse raw array, and the synthesized sparse merged array.}
%	\label{raw_vs_merged}
%\end{figure}
%
%
\begin{figure}[!t]

	\centering
	\subfloat[]{\label{c}
	\includegraphics[width=1.69in]{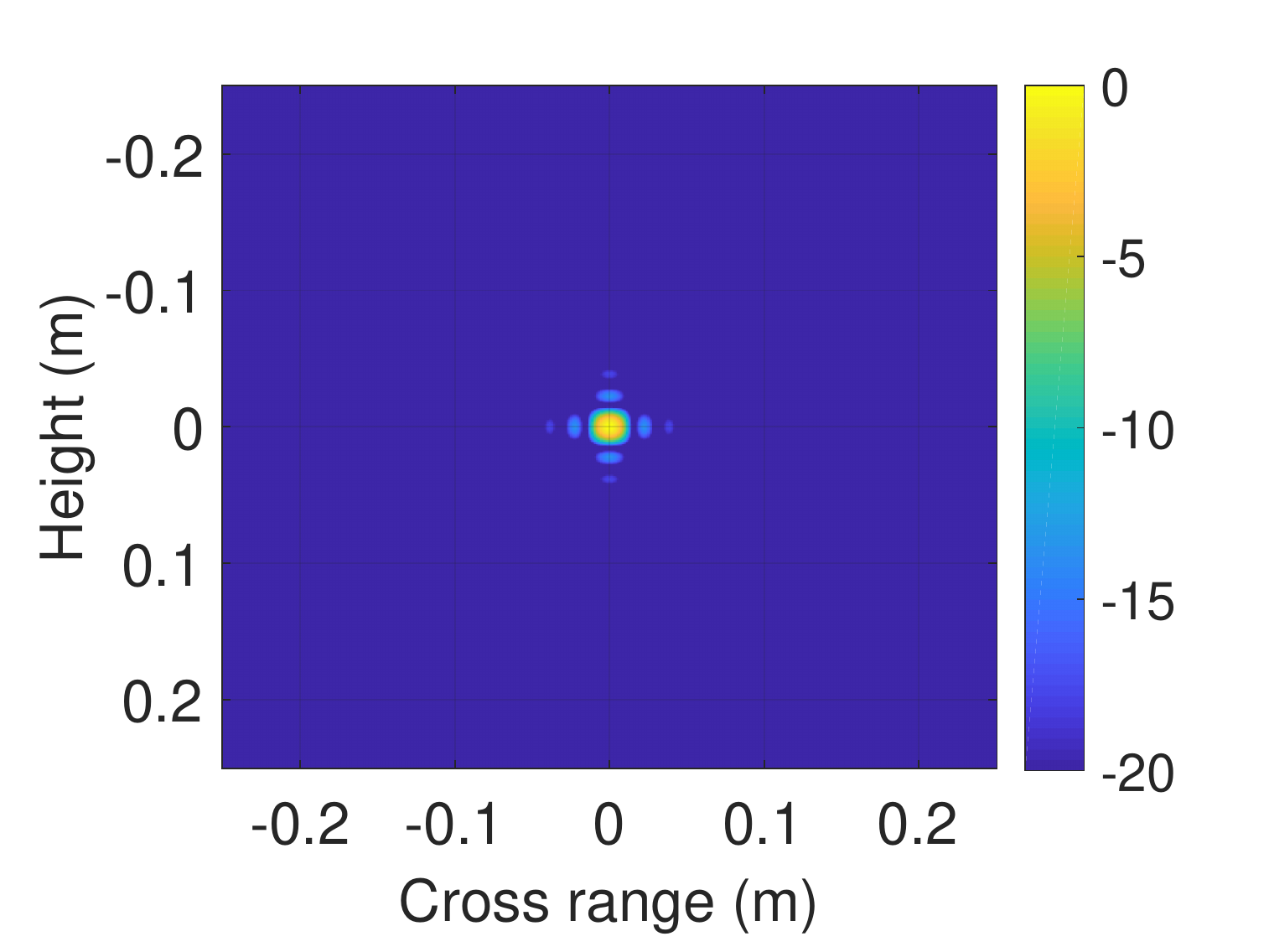}}
	\hfill
	\subfloat[]{\label{d}
	\includegraphics[width=1.69in]{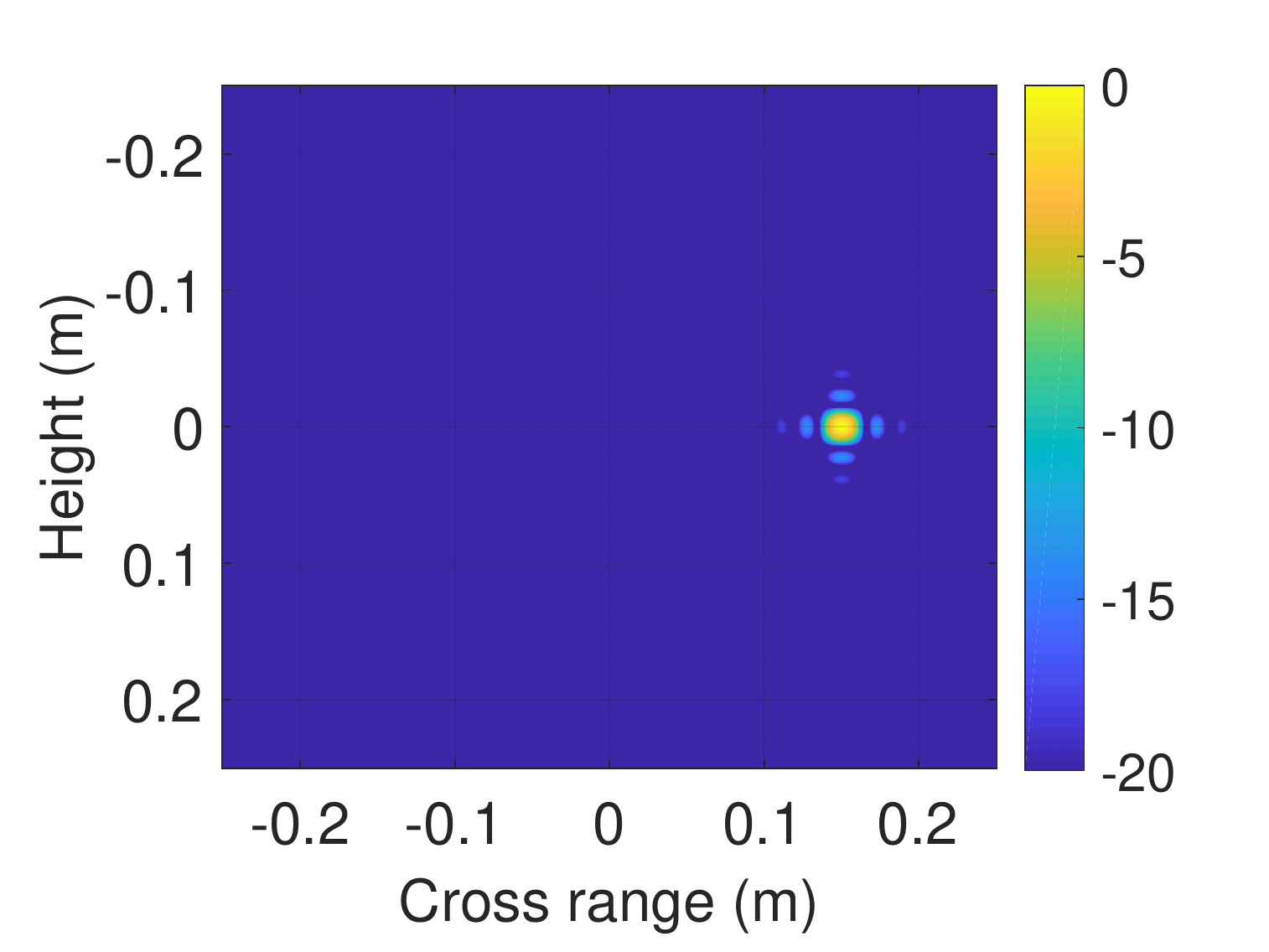}}
	\hfill
	\vspace{-2.5mm} 
		
	\centering
	\vspace{-2.5mm} 
	\subfloat[]{\label{a}
	\includegraphics[width=1.69in]{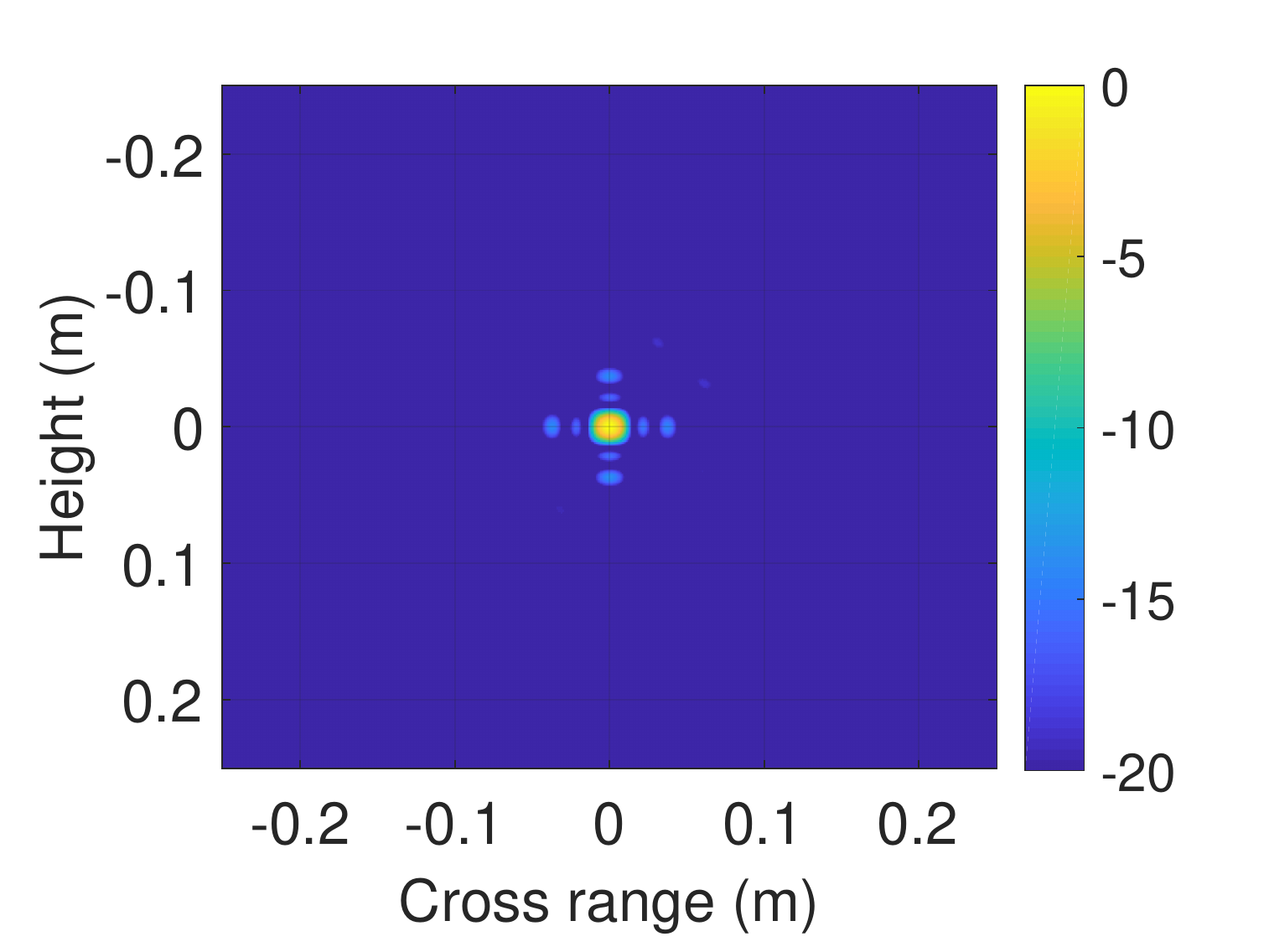}}
	\hfill
	\subfloat[]{\label{b}
	\includegraphics[width=1.69in]{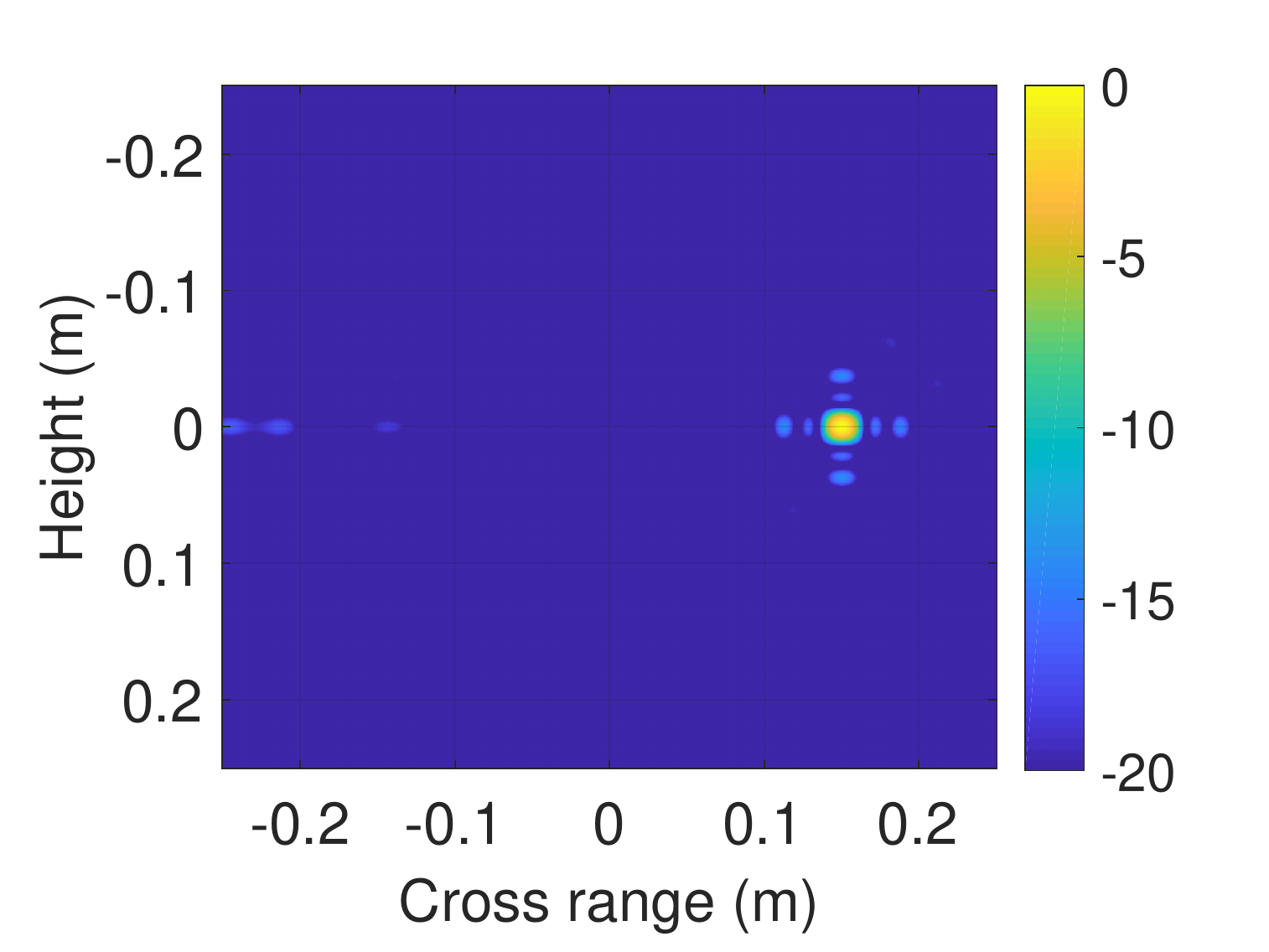}}
	\hfill
	\vspace{-2.5mm} 
		
	\centering
	\subfloat[]{\label{e}
		\includegraphics[width=1.69in]{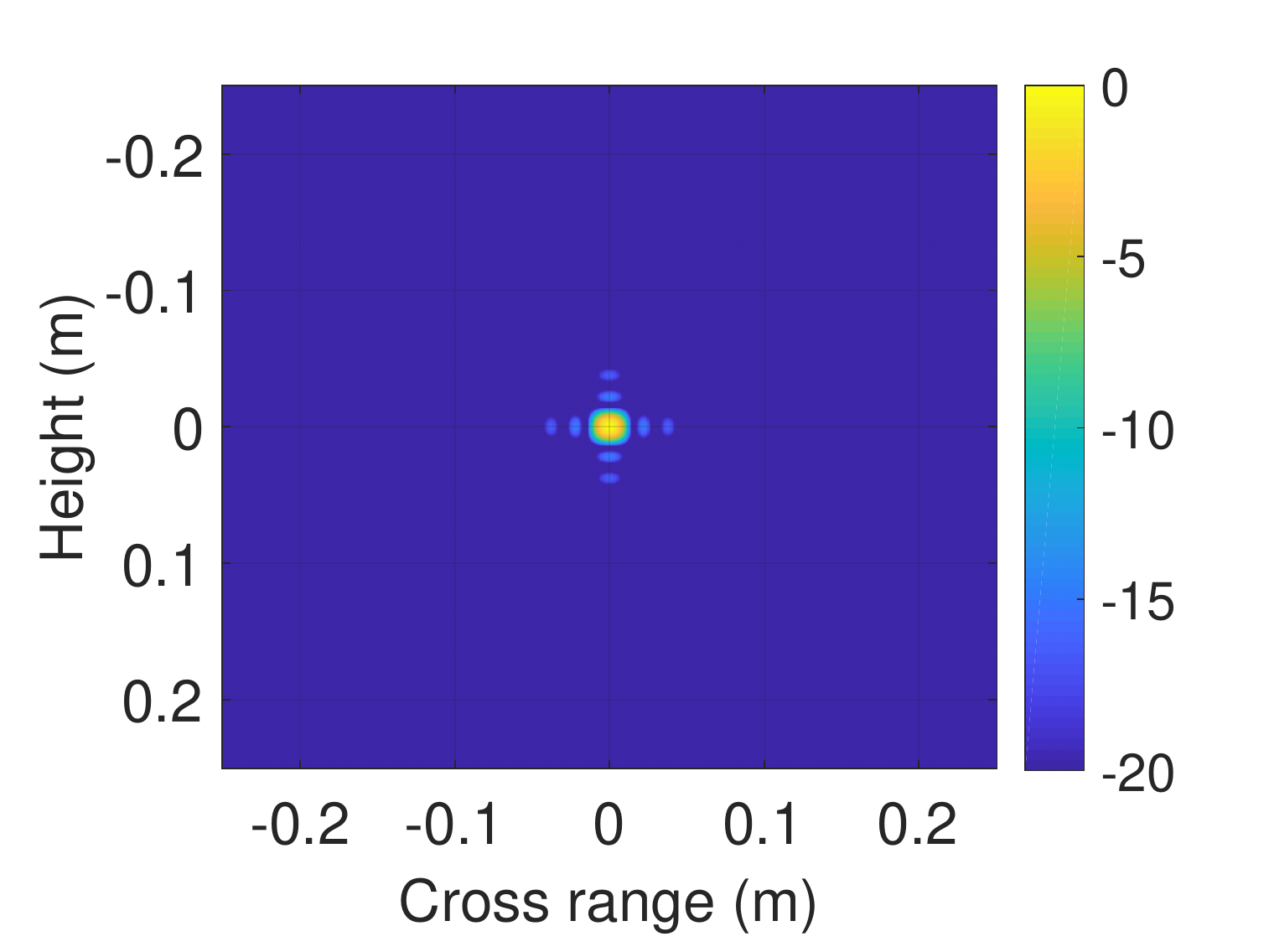}}
	\hfill
	\subfloat[]{\label{f}
		\includegraphics[width=1.69in]{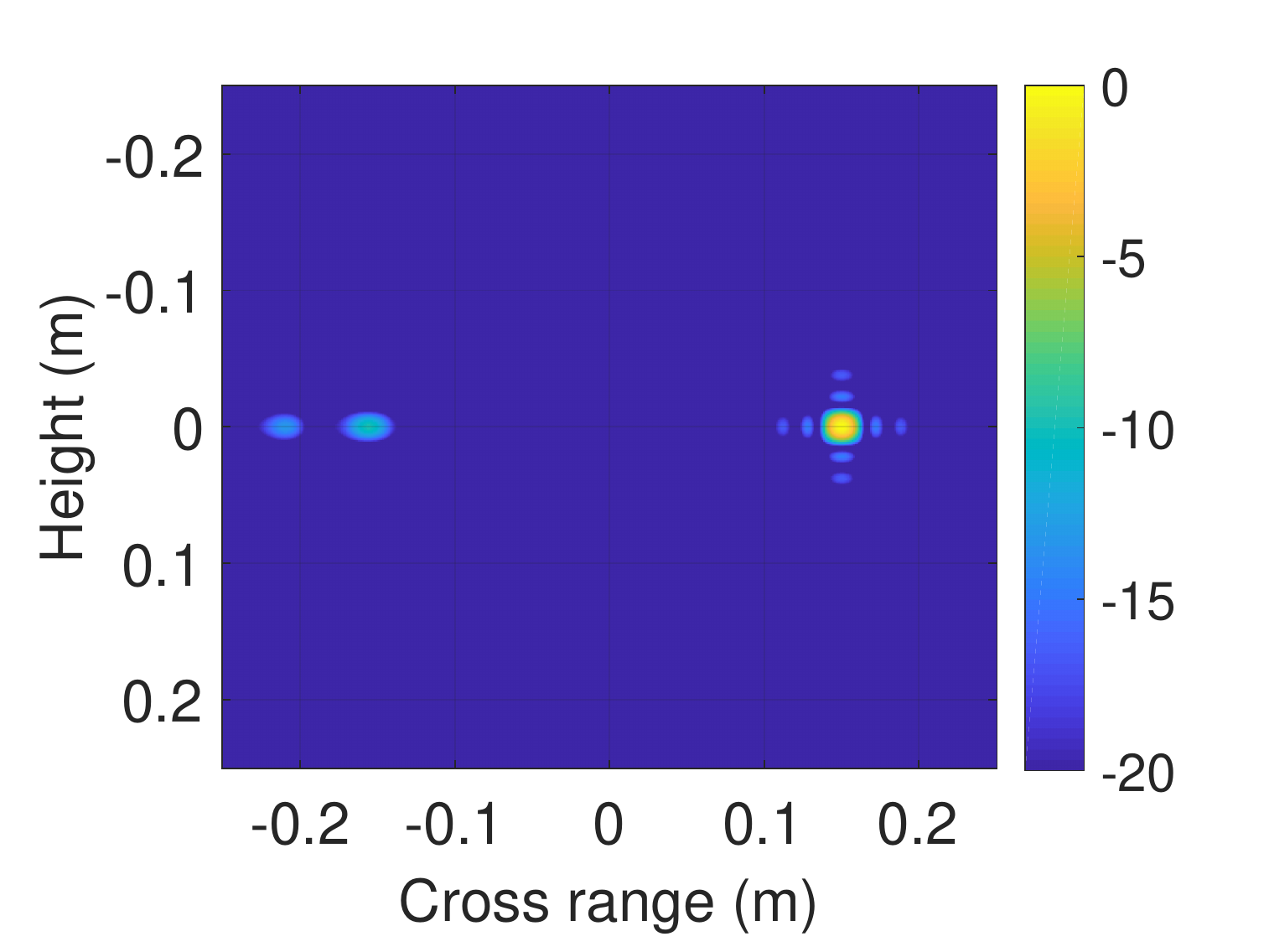}}
	\hfill
	
	\caption{2-D PSFs of the synthesized sparse MIMO array: (a) and (b); of the full MIMO array: (c) and (d), and the equally spaced sparse MIMO array: (e) and (f), for the scatterer at the center position (left column) and the edge position (right column).}
	\label{2d_planar_mimo_comp_simu}
\end{figure}
\begin{figure}[!t]
	\centering
	\subfloat[]{\label{a}
		\includegraphics[width=2.0in]{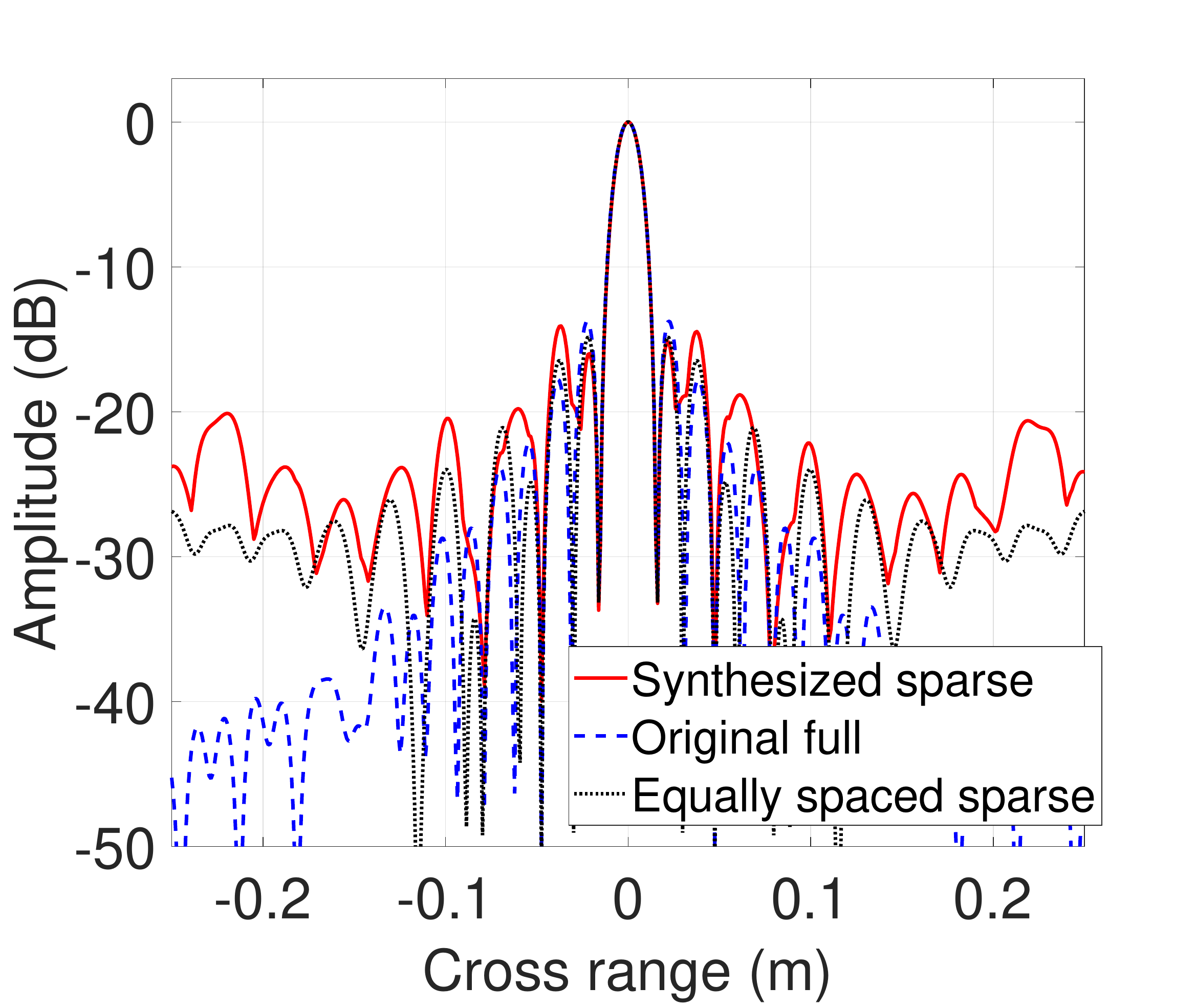}}
	\hfill
	\subfloat[]{\label{b}
		\includegraphics[width=2.0in]{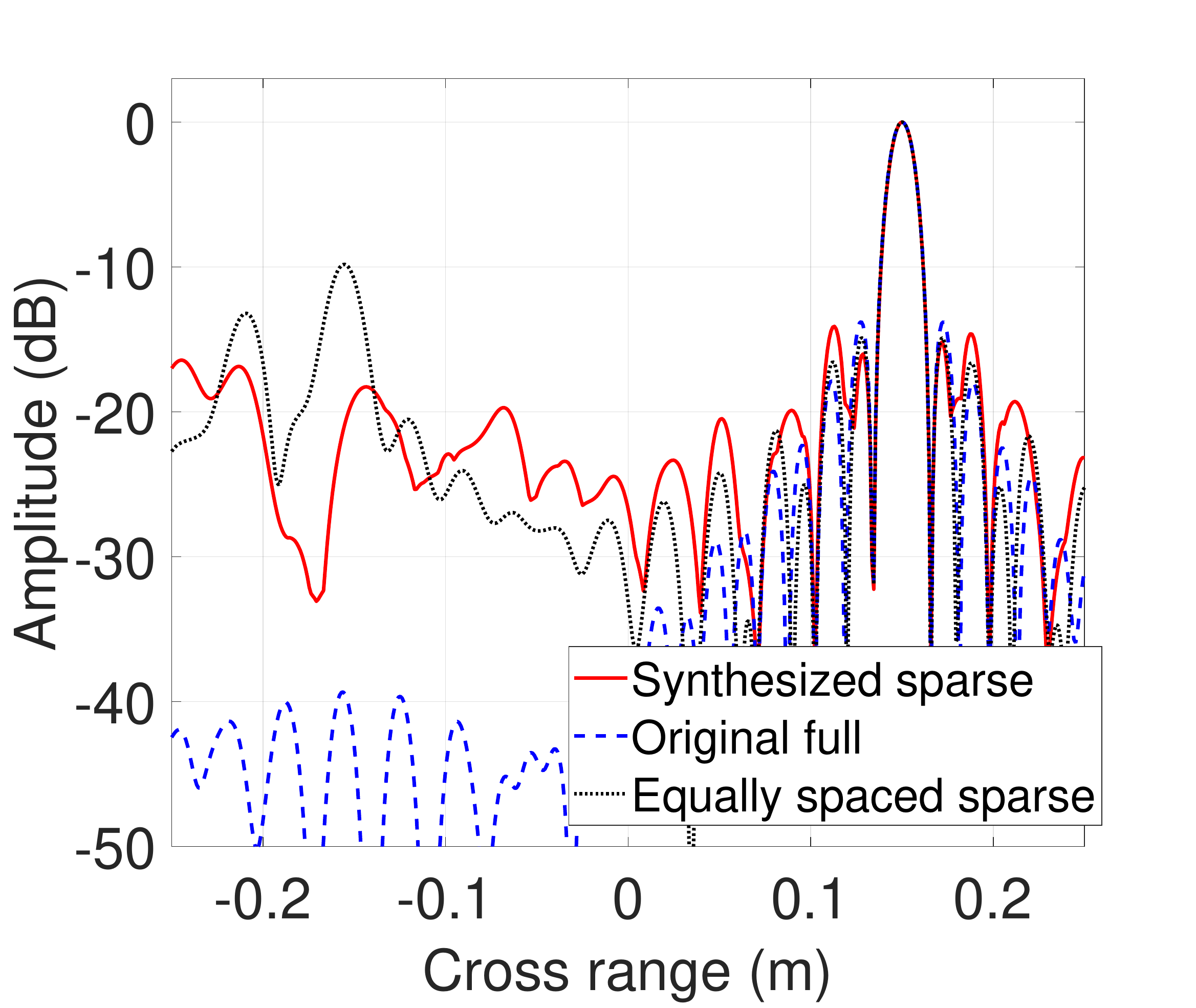}}
	\hfill
	\\	
	\caption{1-D cross-range imaging projection results of the synthesized sparse MIMO array, the full MIMO array, and the equally spaced sparse MIMO array, for the scatterer at (a) the center position, and (b) the edge position.}
	\label{1d_proj_mimo_PSFs}
\end{figure}

To further demonstrate the effectiveness and flexibility of
the proposed method, the performance of the synthesized sparse 2-D planar array is also presented. The number and positions of the transmit elements can be flexible. The aperture of the transmit antennas is usually set to cover the receive antennas in order to get as high resolution as possible. In this simulation, as an example, we fix the transmit elements on the four corners of the topology, and synthesize the 2-D receive array with the proposed SAS method, as is shown in Fig. \ref{sparse_2D_mimo_array_simu}. The 2-D full MIMO and the 2-D equally spaced sparse MIMO array are adopted for comparison, whose topologies are illustrated in Figs. \ref{full_2D_mimo_array_simu} and  \ref{less_2D_mimo_array_simu}, respectively. The related simulation parameters are listed in Table \ref{tab_planar_arr}.

The performance of the aforementioned 2-D MIMO arrays is also evaluated via the PSFs for scatterers at the center and edge positions, as illustrated in Fig. \ref{2d_planar_mimo_comp_simu}. To further show the details, their cross-range projections (selecting the maximum along the height dimension for each cross-range value) are shown in Fig. \ref{1d_proj_mimo_PSFs}, with the dynamic range of 50 dB. The resolutions of the three arrays are approximately the same, due to the same array apertures. Slight sidelobe degradation for the synthesized sparse MIMO array can be observed in both cases compared to the full array and the equally spaced sparse array. Nevertheless, the peak sidelobe level of the synthesized sparse MIMO array is almost the same as those of the other two arrays in the center case, and is less than -14 dB in the edge case. Additionally, compared with the grating lobe (-10 dB) of the equally spaced sparse MIMO array, the optimized array shows suppressed grating lobe (less than -16 dB). Overall, the performance of the sparse MIMO array further indicates the effectiveness of the proposed MIMO SAS method for near-field imaging.

\begin{figure}[!t]
	\centering
	\vspace{-2.5mm} 
	\includegraphics[width=2.0 in]{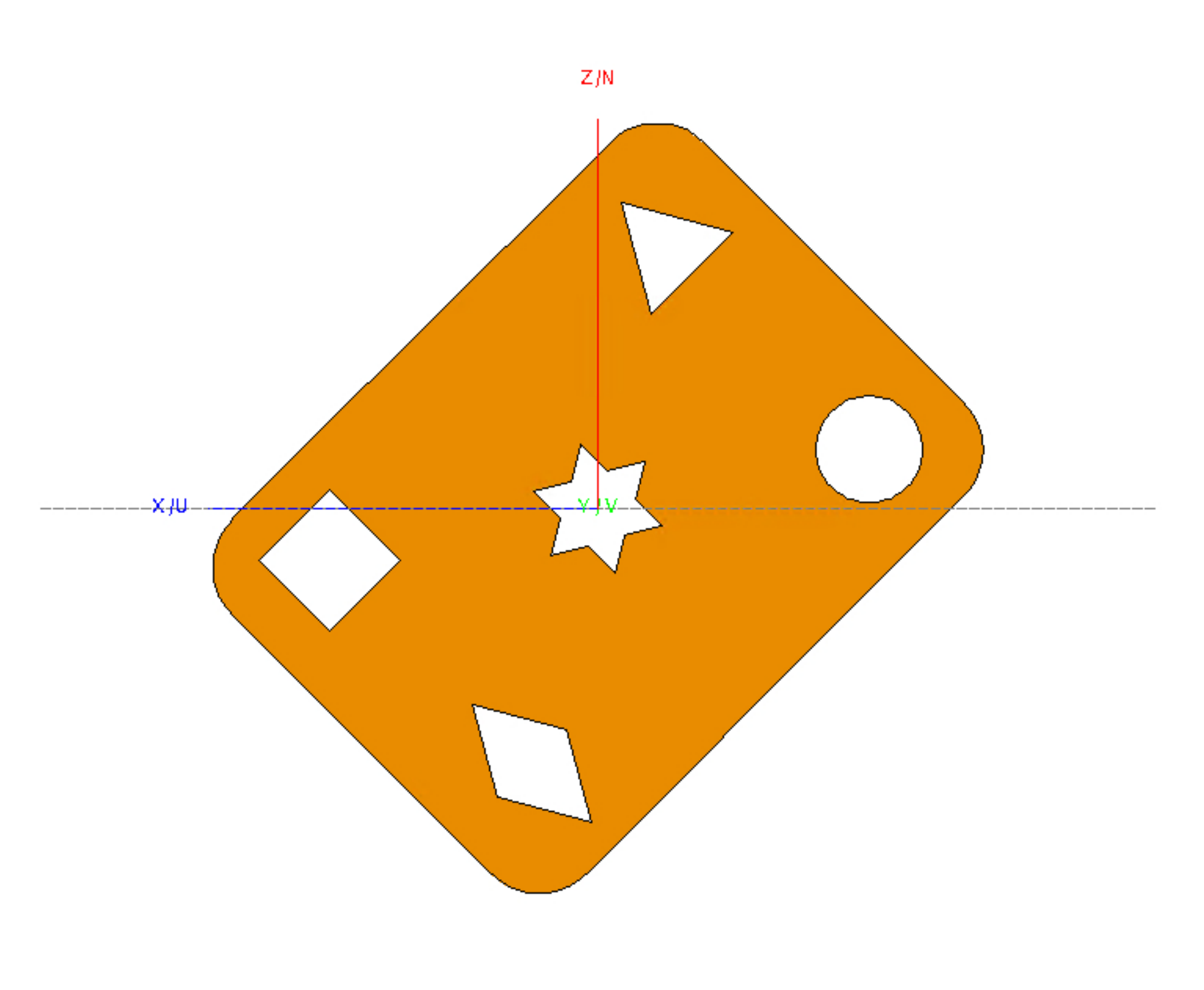}
	\hfill	
	\caption{Simulation target in Feko.}
	\label{feko_tar_sparse_mimo}
\end{figure}

\begin{figure*}[t]	
	%	\begin{adjustwidth}{-\extralength}{0cm}
		%\centering %% If there is a figure in wide page, please release command \centering
		\hspace{0.23\columnwidth}\textbf{Full}
		\hspace{0.30\columnwidth}\textbf{Synthesized sparse} 
		\hspace{0.15\columnwidth}\textbf{Equally spaced sparse} 
		\hspace{0.17\columnwidth}\textbf{Random sparse}
		\\
		%		\centering
		\setlength{\abovecaptionskip}{0pt}
		\vspace{-7mm}
		
		\hfill    
		\subfloat[]{\label{a}\includegraphics[width=1.75in]{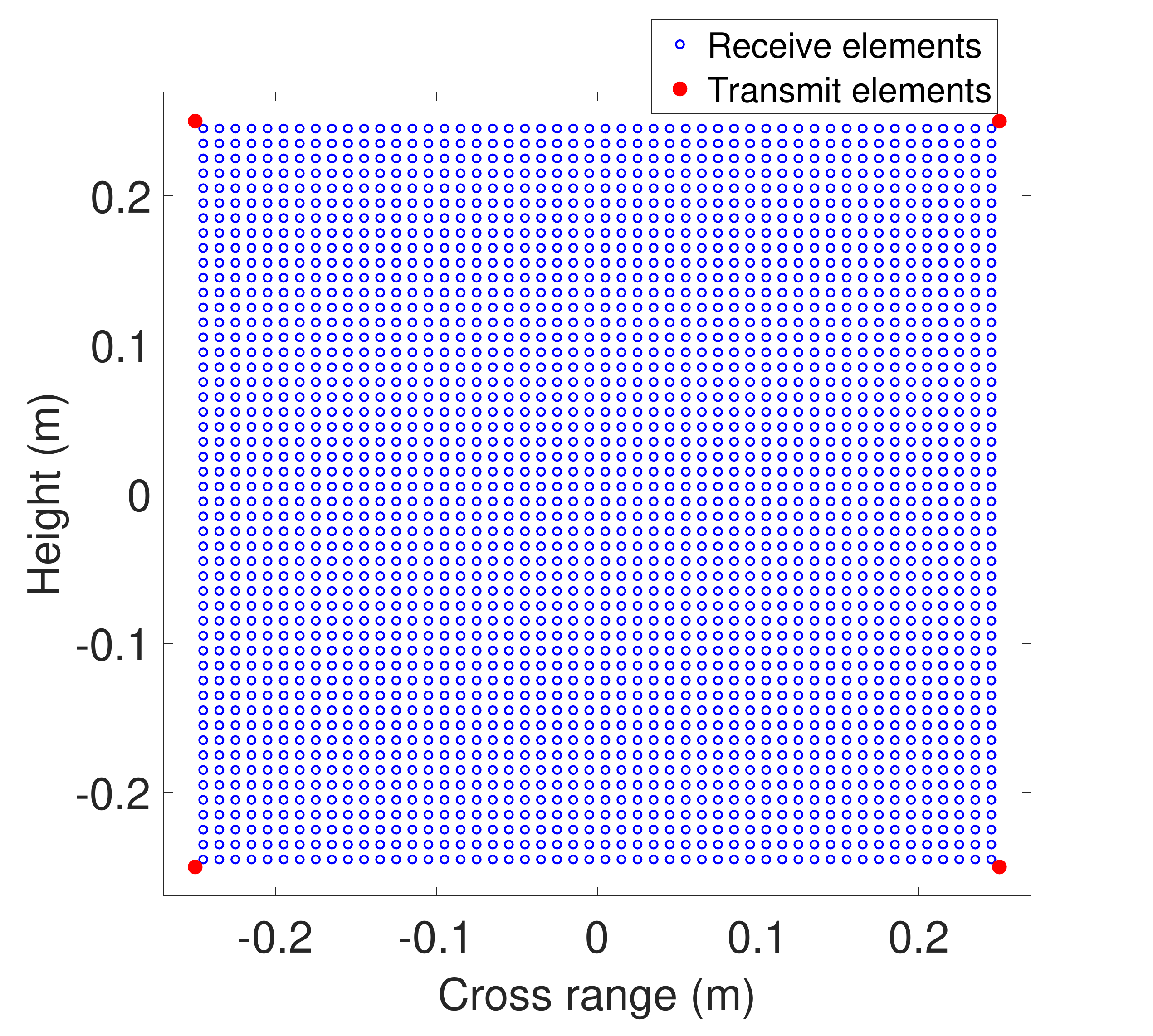}}
		\hfill
		\subfloat[]{\label{b}\includegraphics[width=1.75in]{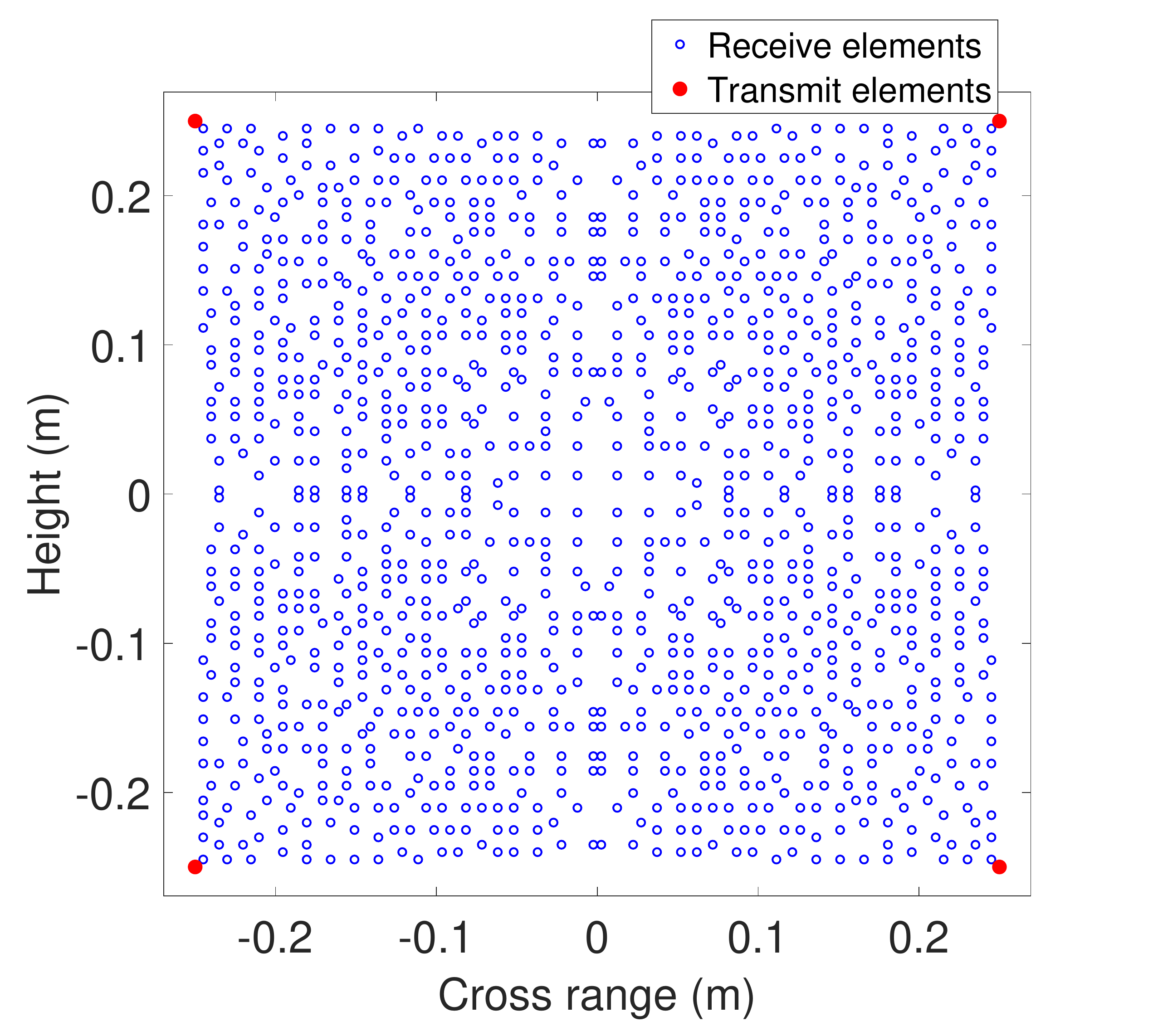}}
		\hfill
		\subfloat[]{\label{c}\includegraphics[width=1.75in]{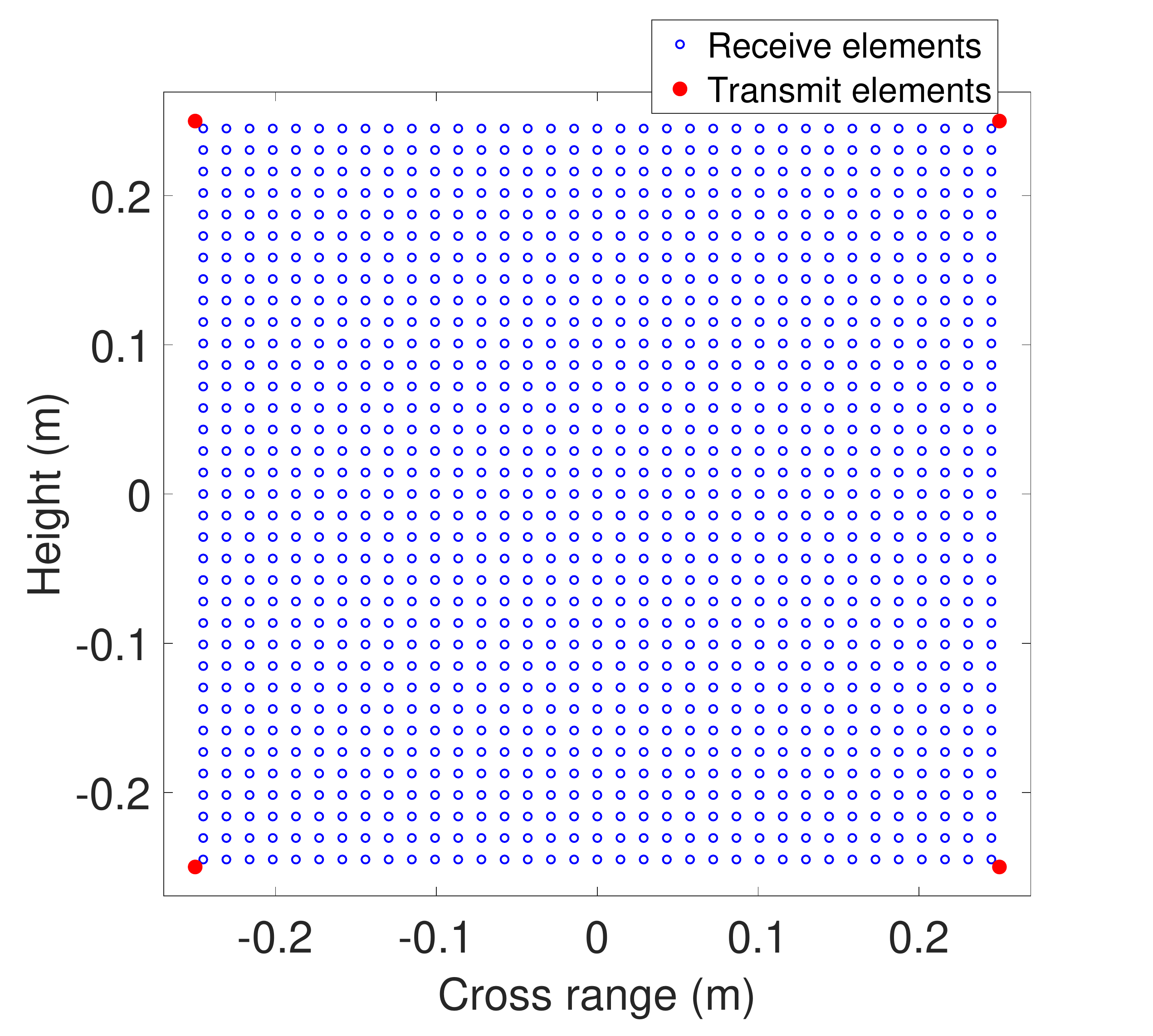}}
		\hfill
		\subfloat[]{\label{d}\includegraphics[width=1.75in]{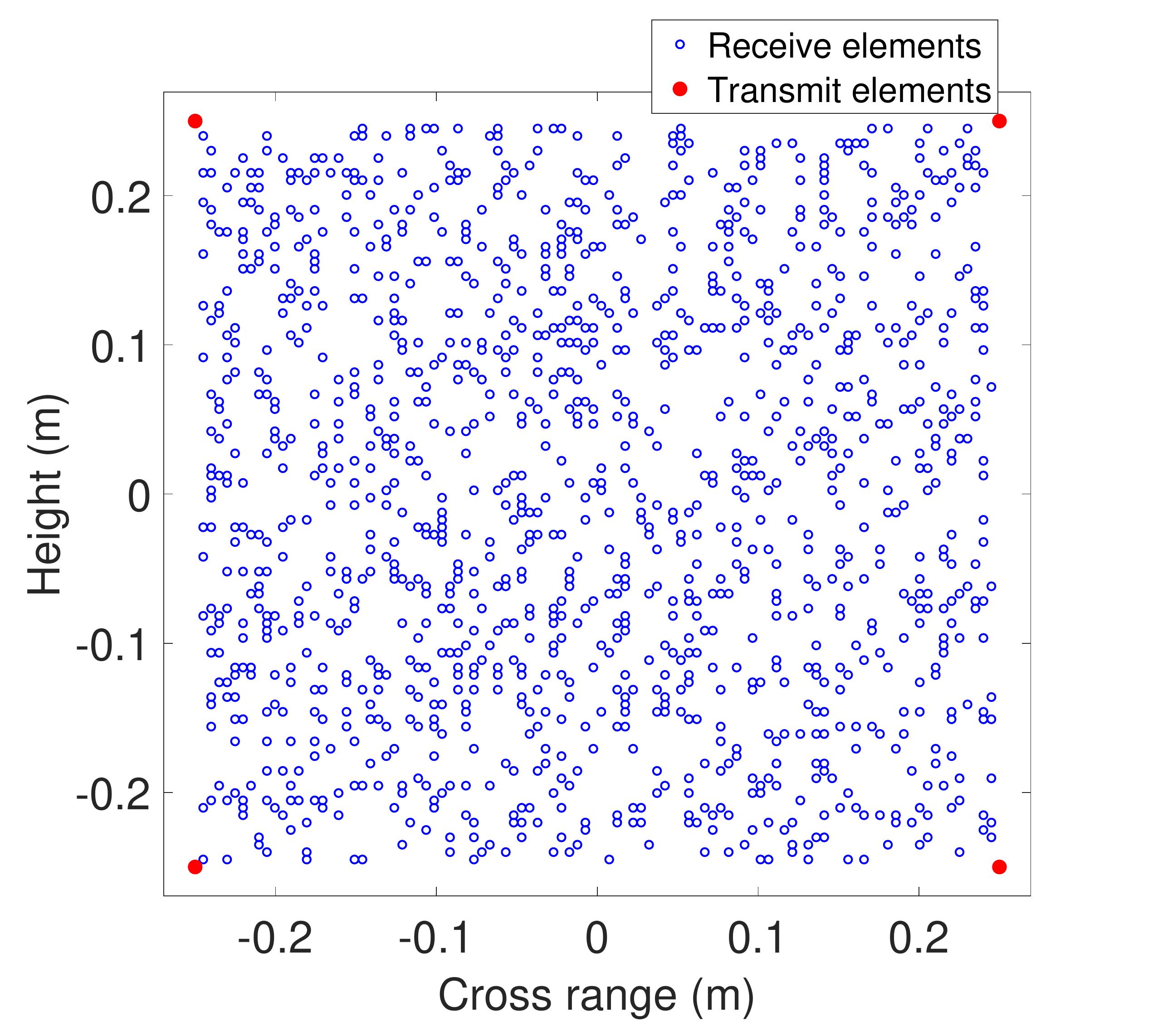}}
		\hfill
		\\
		
		\hfill    
		\subfloat[]{\label{e}\includegraphics[width=1.75in]{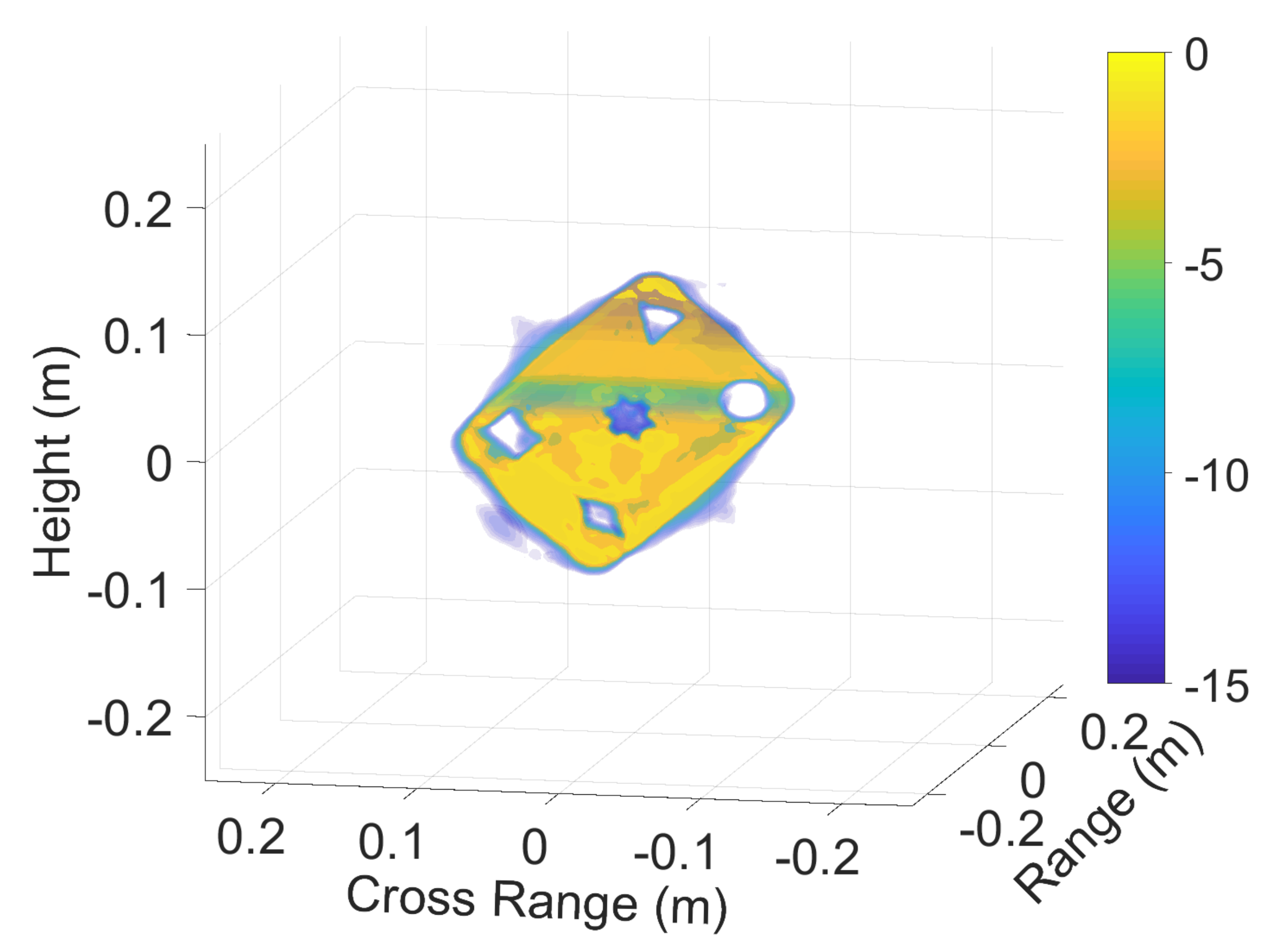}}
		\hfill
		\subfloat[]{\label{f}\includegraphics[width=1.75in]{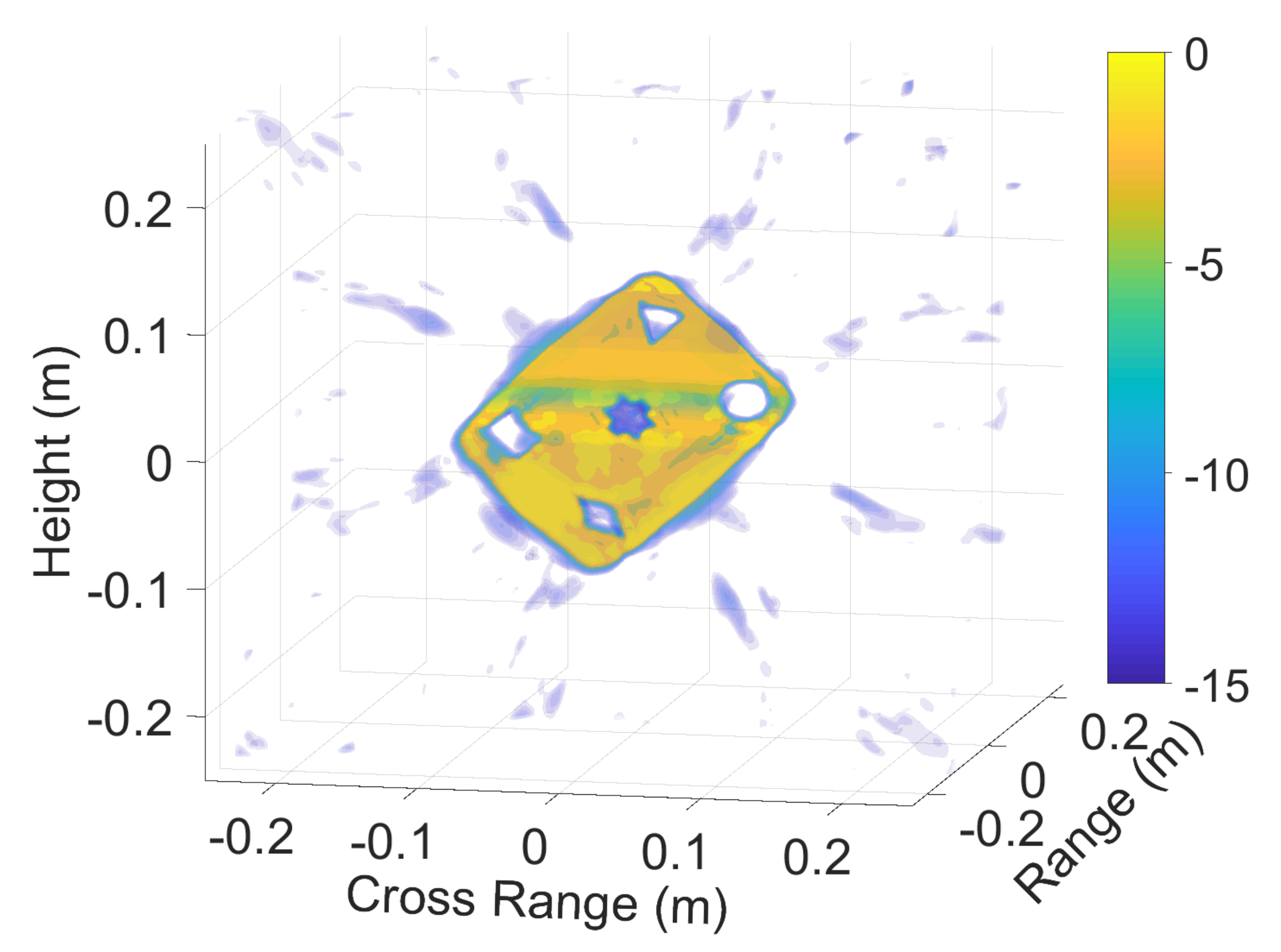}}	
		\hfill
		\subfloat[]{\label{g}\includegraphics[width=1.75in]{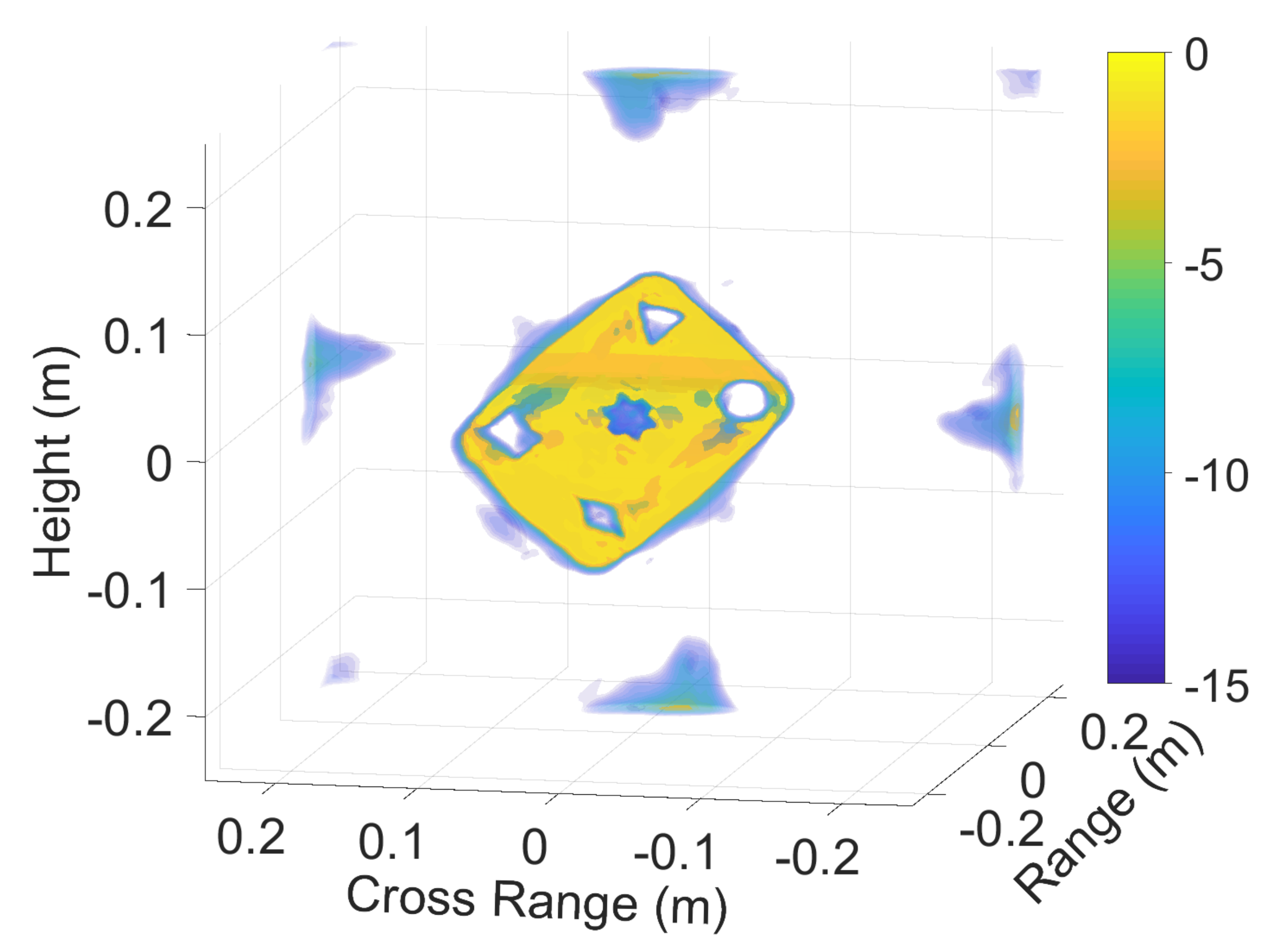}}
		\hfill
		\subfloat[]{\label{h}\includegraphics[width=1.75in]{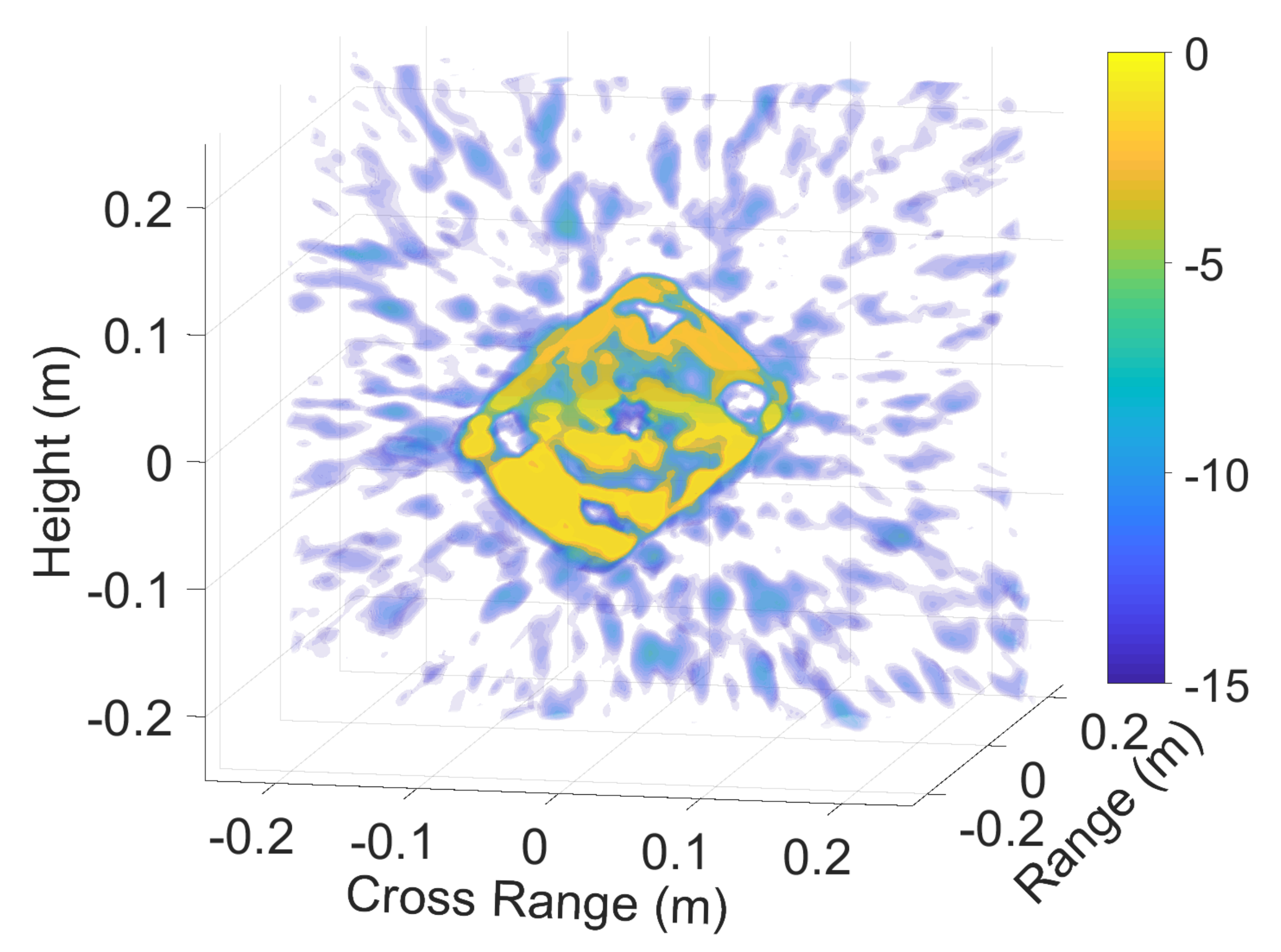}}
		\hfill
		\\
		
		\hfill    
		\subfloat[]{\label{e}\includegraphics[width=1.75in]{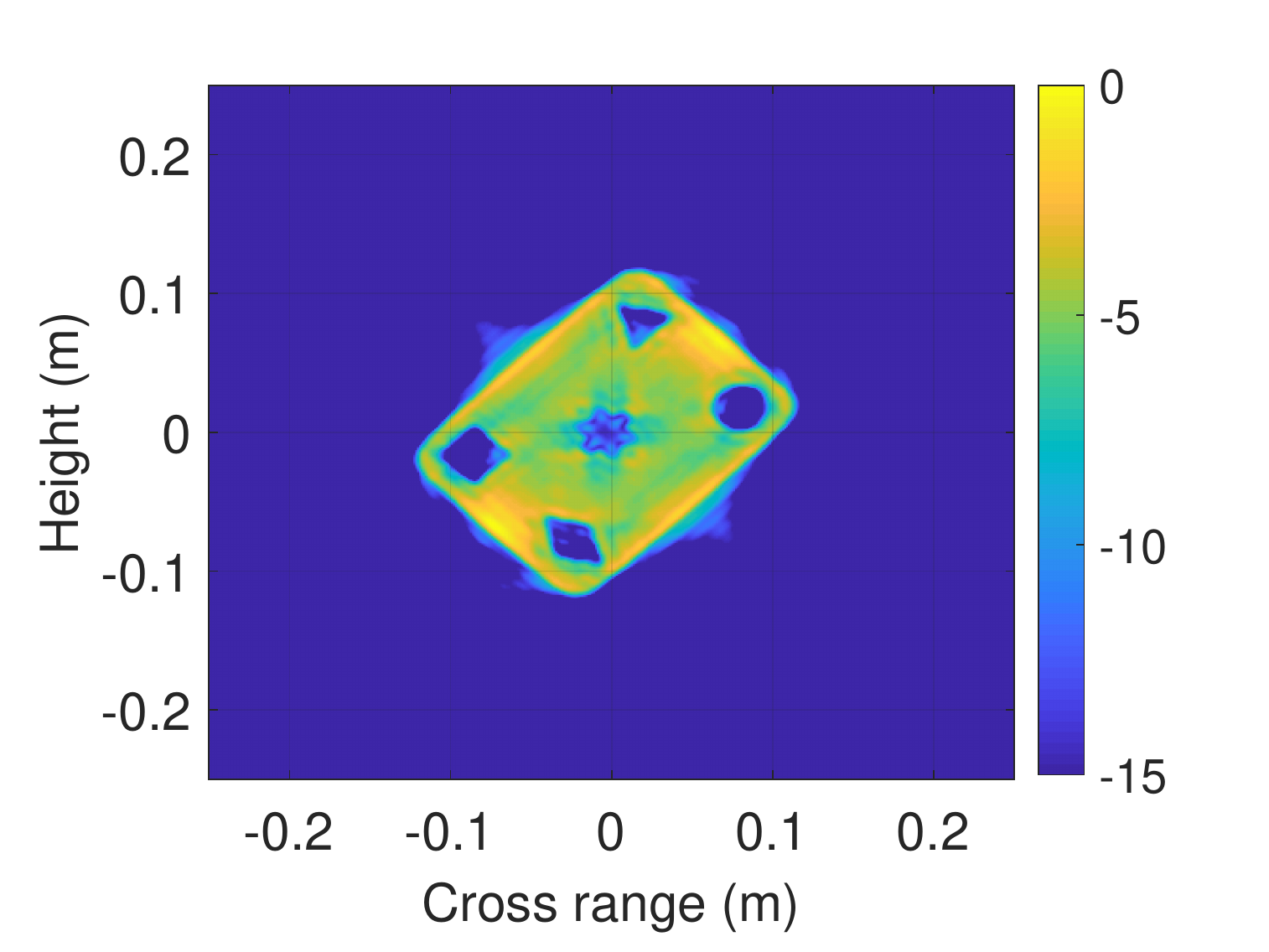}}
		\hfill
		\subfloat[]{\label{f}\includegraphics[width=1.75in]{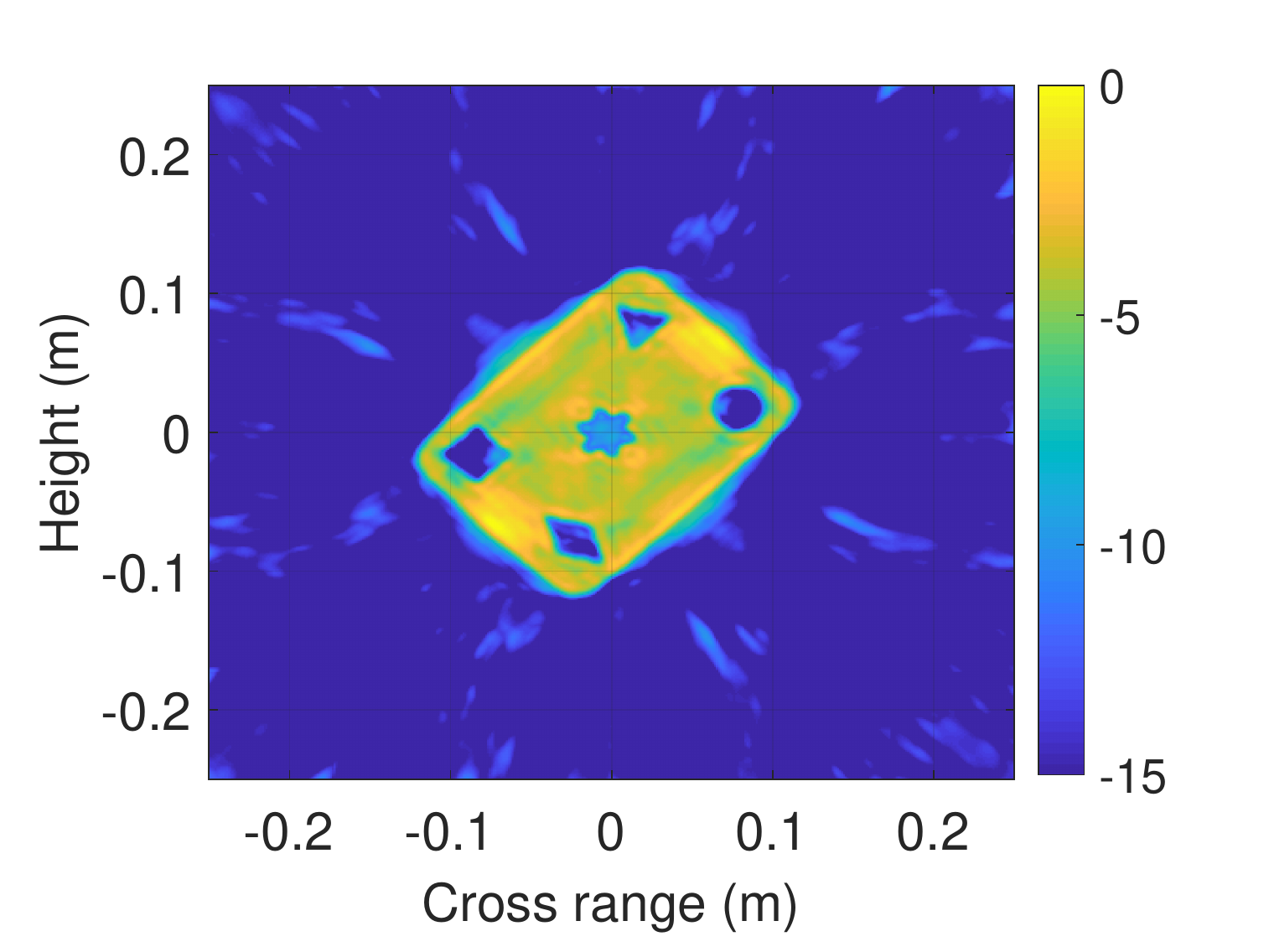}}	
		\hfill
		\subfloat[]{\label{g}\includegraphics[width=1.75in]{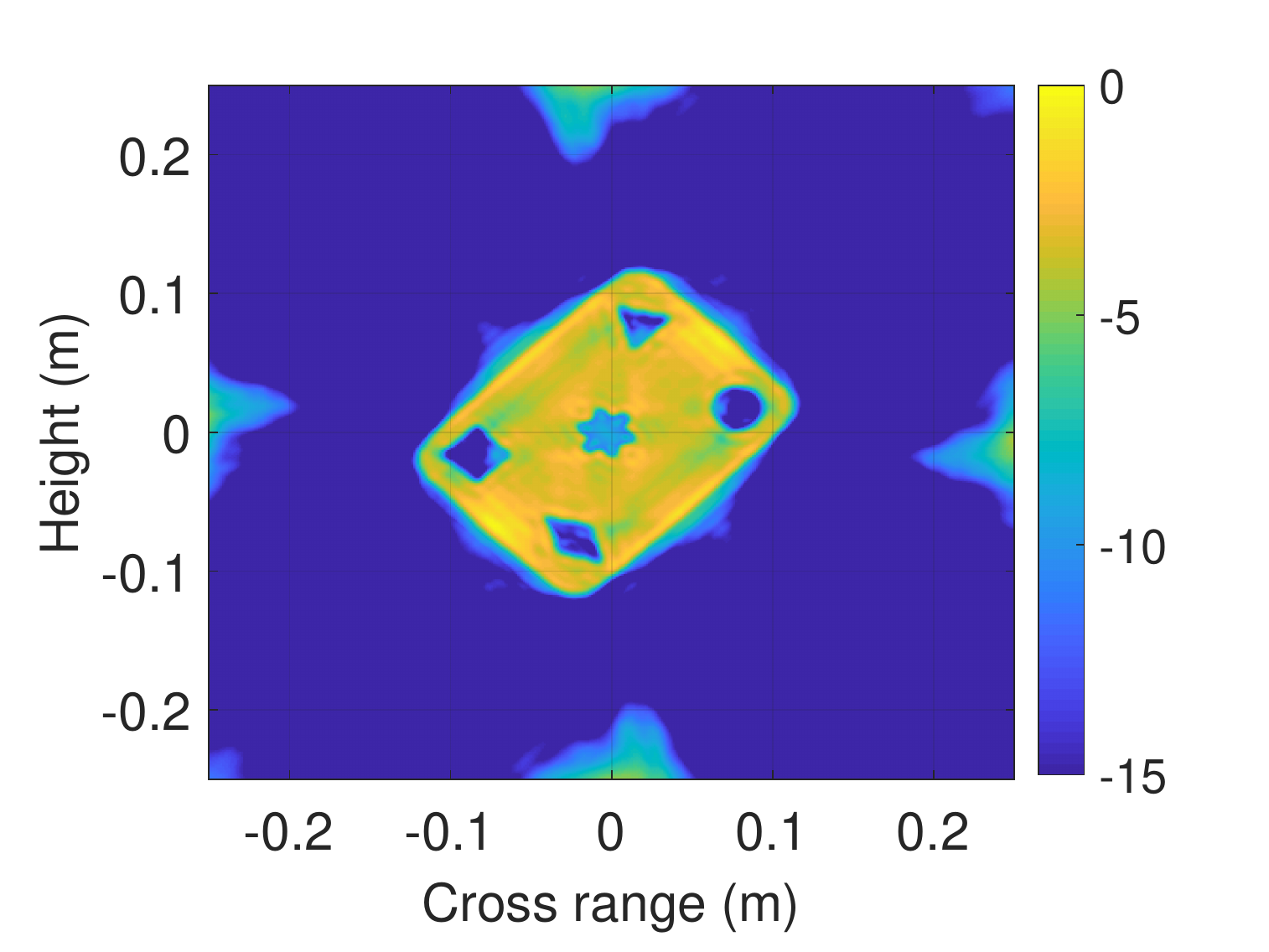}}
		\hfill
		\subfloat[]{\label{h}\includegraphics[width=1.75in]{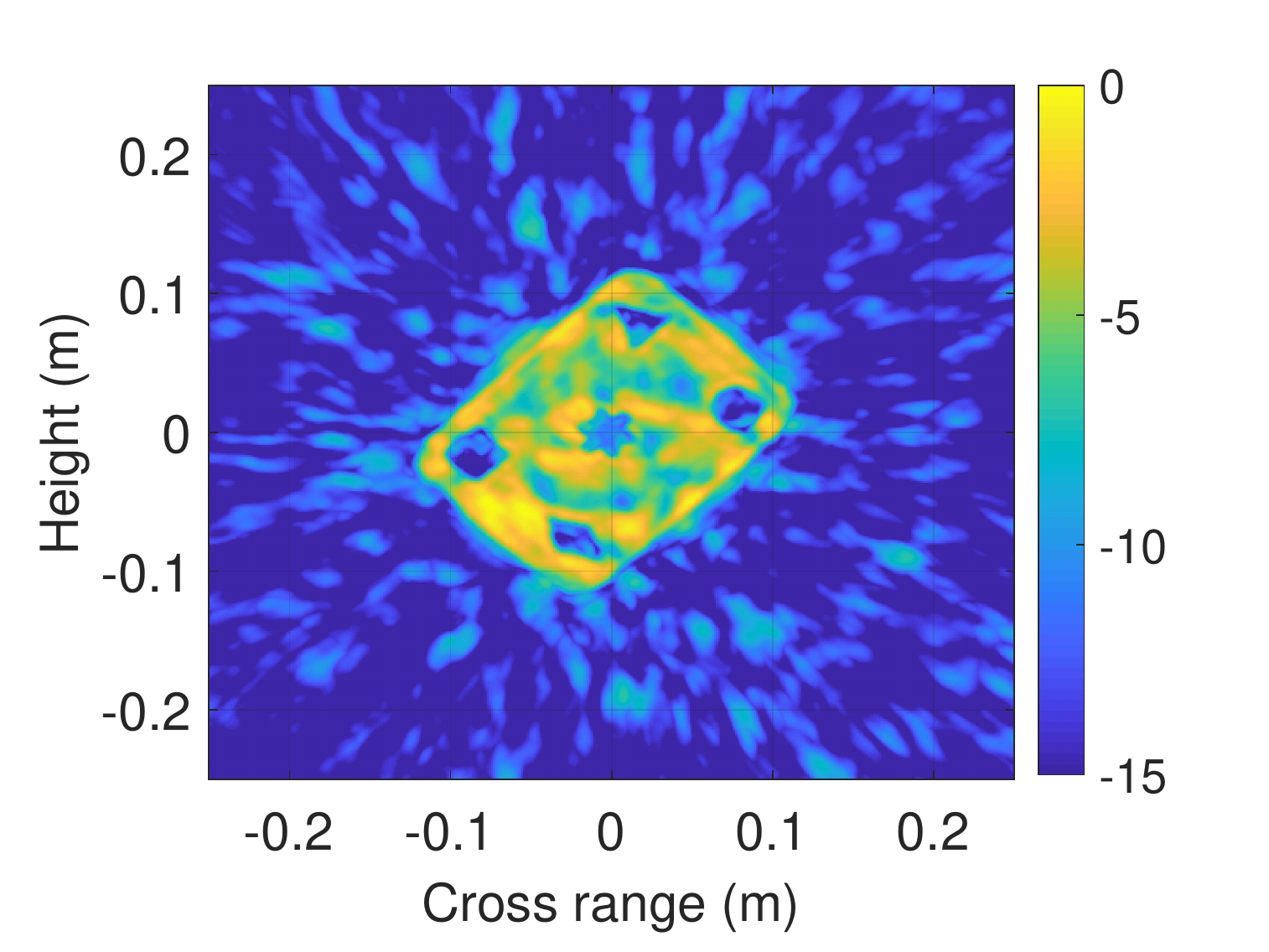}}
		\hfill
		\\
		
		\setlength{\belowcaptionskip}{-0.2cm} 
		%	\end{adjustwidth}
	\caption{Comparison of topologies for Feko simulation and imaging reconstruction results of different arrays, ({a}--{d}) MIMO topologies, ({e}--{h}) 3-D imaging reconstruction results, ({i}--{l}) 2-D imaging results of maximum projection along the range direction. The titles on the top indicate different topologies corresponding to the images column by column.}
	\label{feko_MIMO_results} 
\end{figure*}

\begin{table}[!t]
	\centering
	\caption{Feko Simulation Parameters for 2-D Sparse Array}
	\setlength{\tabcolsep}{3pt}
	\begin{threeparttable}
		\begin{tabular}{p{200pt}  p{40pt}}
			%\begin{tabular}{|c|c|}
			\hline\hline
			Parameters& Values \\[0.5ex]
			\hline
			Imaging distance $(R_0)$&
			0.5 m\\[0.5ex]		
			Start frequency& 
			30 GHz \\[0.5ex]
			Stop frequency&
			35 GHz \\[0.5ex]	
			Number of frequency steps&
			51 \\[0.5ex]			
			Number of transmit antennas &4 \\[0.5ex]		
			Spacing of transmit antennas of the full array& 50 cm \\[0.5ex]
			Number of receive antennas of the full array&
			50$\times$50 \\[0.5ex]	
			Spacing of receive antennas of the full array&
			1.0 cm \\[0.5ex]			
			Number of receive antennas of the synthesized/ equally spaced/ random sparse array&
			1225 \\[0.5ex]

			Spacing of receive antennas of the equally spaced sparse array &
			1.4 cm\\[0.5ex]	
			% 1.44 cm
			\hline
		\end{tabular}
	\end{threeparttable}
	\label{tab_feko_planar_arr}
\end{table}

Furthermore, the computational electromagnetics software package FEKO \cite{feko} is utilized to simulate the scattered EM waves. An openwork patterned metal plate is set as the simulated target, as illustrated in Fig. \ref{feko_tar_sparse_mimo}. The synthesized sparse array topologies are generated for this scenario as shown in Fig. \ref{feko_MIMO_results} \subref{a}, whose parameters are listed in Table. \ref{tab_feko_planar_arr}. The 2-D full MIMO array, the 2-D equally spaced sparse MIMO array, and the sparse MIMO array with randomly distributed receive elements (stated as the `random sparse MIMO array’ for conciseness) are adopted for comparison, whose topologies are illustrated in Figs. \ref{feko_MIMO_results} \subref{b},   \subref{c}, and \subref{d}, respectively. The number of the receive elements in each of the latter two arrays is the same as that of the synthesized sparse MIMO array.

The imaging results of the aforementioned arrays by the back-projection (BP) algorithm \cite{cbp} are demonstrated in Fig. \ref{feko_MIMO_results}. The 3-D volumetric results are shown in the second row. Besides, the maximum value projections of these images along the range direction are shown in the third row. All these reconstruction results are displayed under the dynamic range of 15 dB. Undoubtedly, the reconstruction results of the full array are the clearest. The clutters in the results of the synthesized sparse array are subtly visible. For the equally spaced sparse array, the false targets appear at the edges of the imaging area, which are caused by the grating lobe effects and result in an untruthful image. It is evident that there exists severe noise-like artifacts for the random sparse array. As seen, severe artifacts emerge around the plate and blur its details.

Moreover, to give the quantitative comparison of the aforementioned three sparse MIMO arrays, some metrics, such as Root Mean Squared Error (RMSE), Peak Signal to Noise Ratio (PSNR), Structural Similarity Index (SSIM) \cite{wang2004image} and entropy, are calculated with the imaging result of the full MIMO array as the baseline, as shown in Table \ref{MIMO_feko_indicators}. Here, the  entropy of the image $\textbf{G}$ is defined as:
\begin{equation}\label{image_entropy}
 E(\textbf{G})=-\sum_{m=1}^{M}p_m\left(\textbf{X}_m \right) \log_2 p_m\left(\textbf{X}_m \right),
\end{equation}
where $p_m$ represents the probability distribution, which contains the normalized histogram counts within a fixed $\textbf{G}_m$. $E(\textbf{G})$ is a function of pixel intensity in the image, so that it can depict the degree of image focus\cite{li2019radar}. Apparently, the RMSE and entropy of the synthesized sparse array are the lowest among the three sparse arrays, while its PSNR and SSIM are the highest. We can conclude from these four indicators that the synthesized sparse MIMO array is superior to the other two arrays in near-field imaging.

\begin{table}[!t]
	\centering
	\caption{Quantitative Comparison of Imaging of Different MIMO Topologies for Feko Simulation}
	\setlength{\tabcolsep}{3pt}
	\begin{threeparttable}
		\begin{tabular}{p{80pt}  p{30pt}  p{30pt}  p{30pt}  p{30pt}}
			%\begin{tabular}{|c|c|}
			\hline\hline
			Array Topologies & RMSE & PSNR  & SSIM & entropy \\[0.5ex]
			\hline
			synthesized sparse &12.2 & 26.4 dB & 0.964 & 0.702\\[0.5ex]		
			equally spaced sparse &15.6& 24.3 dB & 0.959 & 0.710\\[0.5ex]	
			random spaced sparse & 23.8 & 20.6 dB & 0.818 & 0.784\\[0.5ex]		
			\hline
			
		\end{tabular}
	\end{threeparttable}
	\label{MIMO_feko_indicators}
	\vspace{-2.5mm} 
\end{table}

\section{EXPERIMENTAL RESULTS}

\begin{figure*}[!t]
	\centering
	\vspace{-0.5mm} 
	\centering
	\includegraphics[width=5.0in]{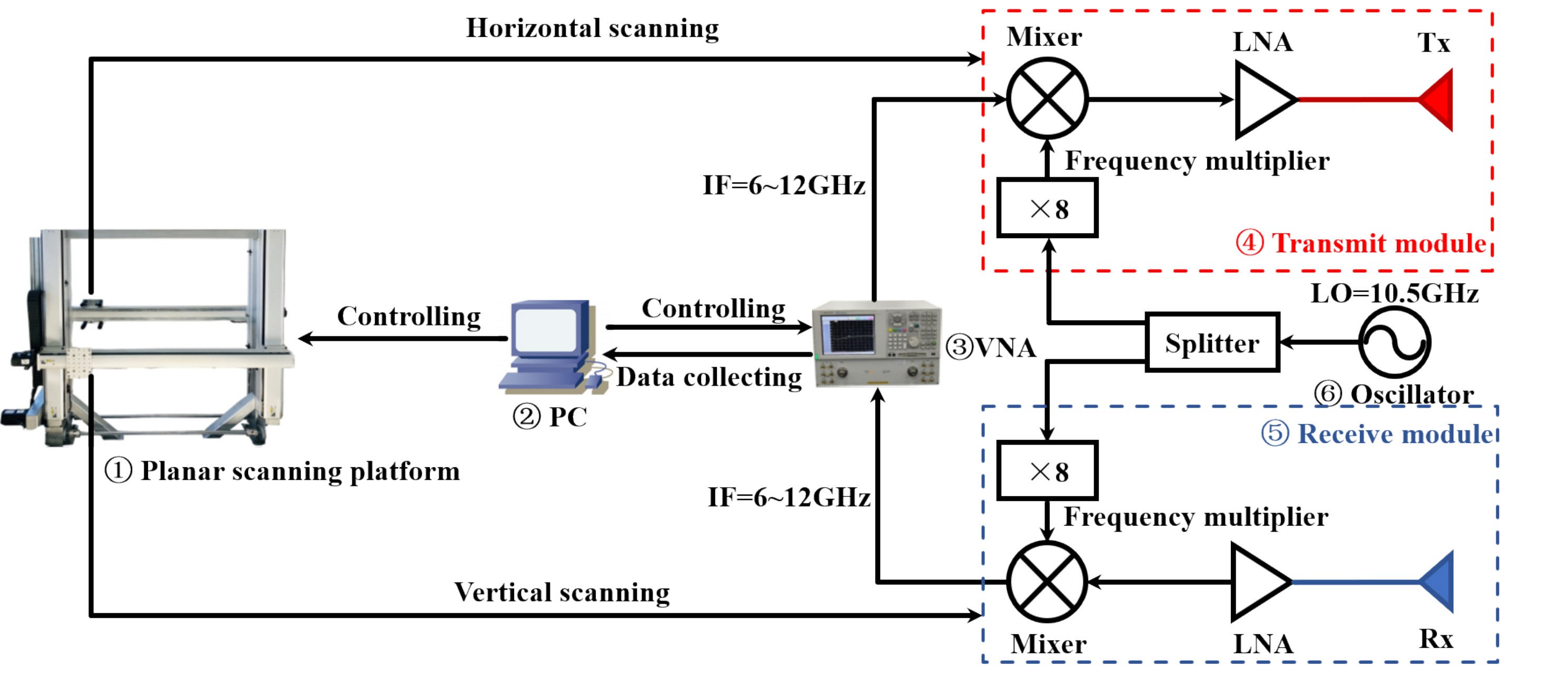}
	
	\vspace{-2.5mm} 
	\caption{Block diagram of the T-shaped MIMO scanning imaging system.}
	\label{T_system_block_diagram}
\end{figure*}

\begin{figure}[!t]
	\centering
	\vspace{-0.5mm} 
	\includegraphics[width=2.6in]{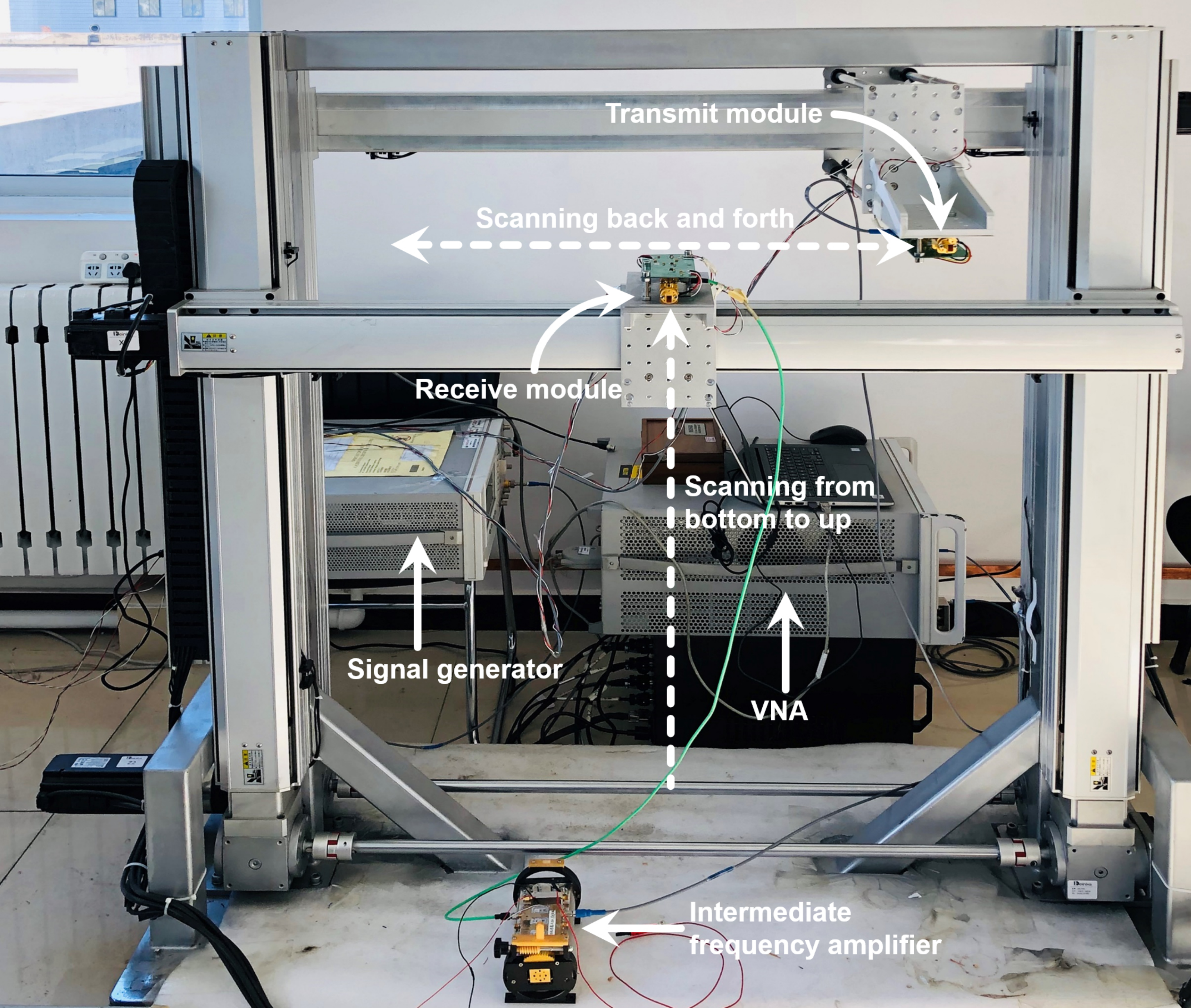}
	\hfill	
	\caption{Prototype of the T-shaped MIMO scanning imaging system.}
	\label{T_shaped_MIMO_system}
\end{figure}

\begin{figure}[!t]
	\centering
	\subfloat[]{\label{a}
		\includegraphics[width=1.69in]{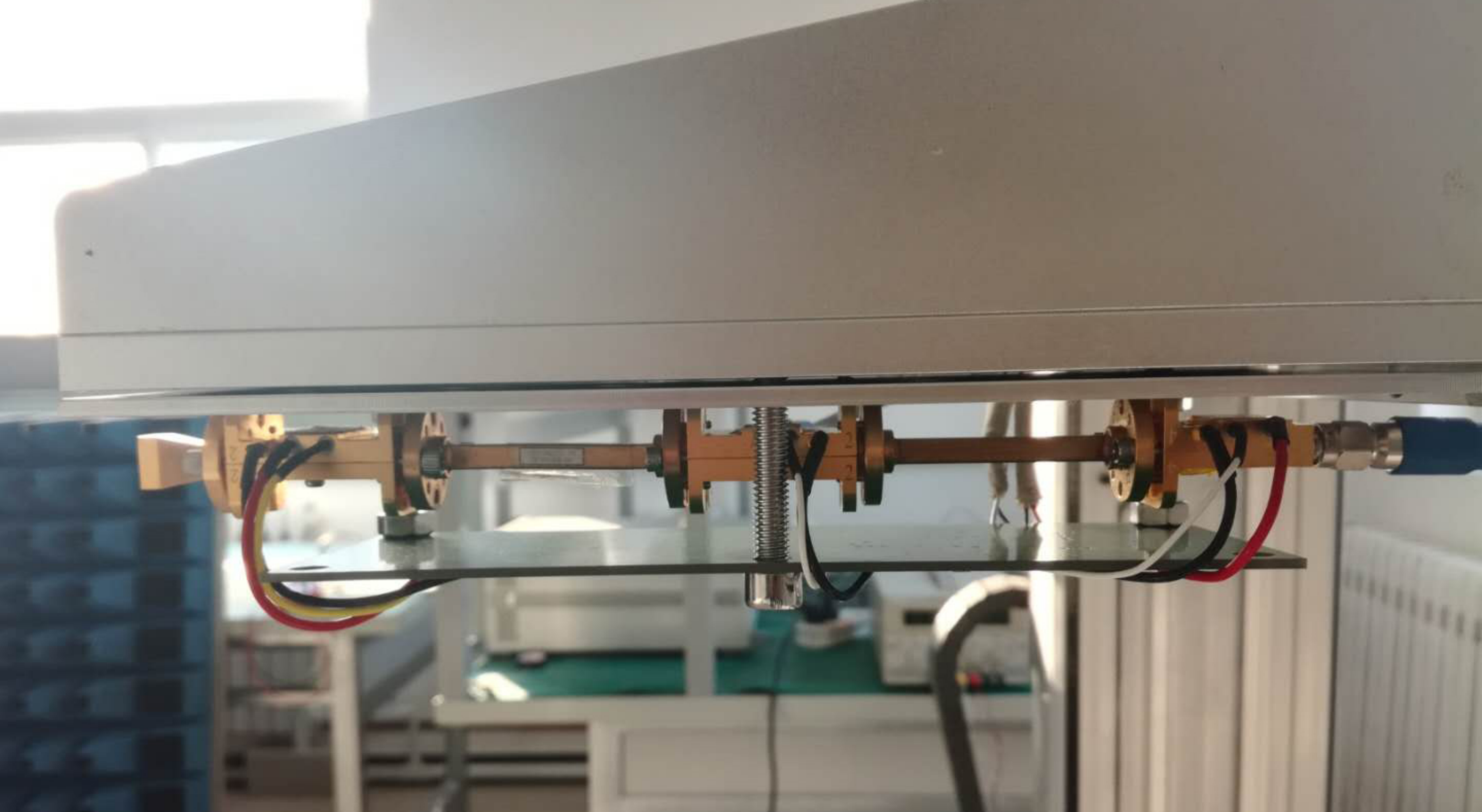}}
	\hfill
	\subfloat[]{\label{b}
		\includegraphics[width=1.69in]{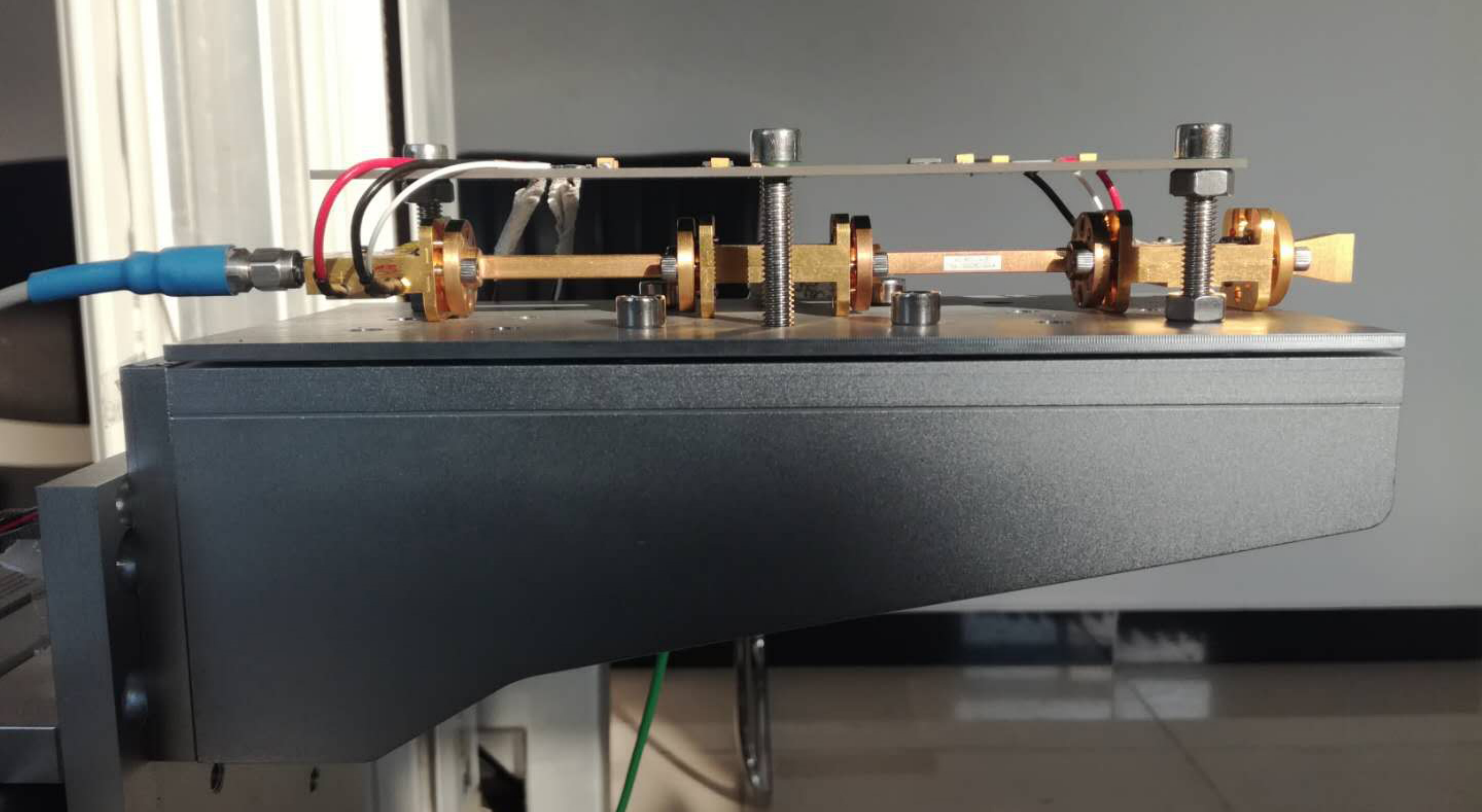}}
	\hfill
	\\	
	\caption{Photographs of (a) the transmit module, and (b) the receive module.}
	\label{Wband_Tx_Rx_module}
\end{figure}

\begin{figure}[!t]
	\centering
	\vspace{-2.5mm} 
	\includegraphics[width=2.0in]{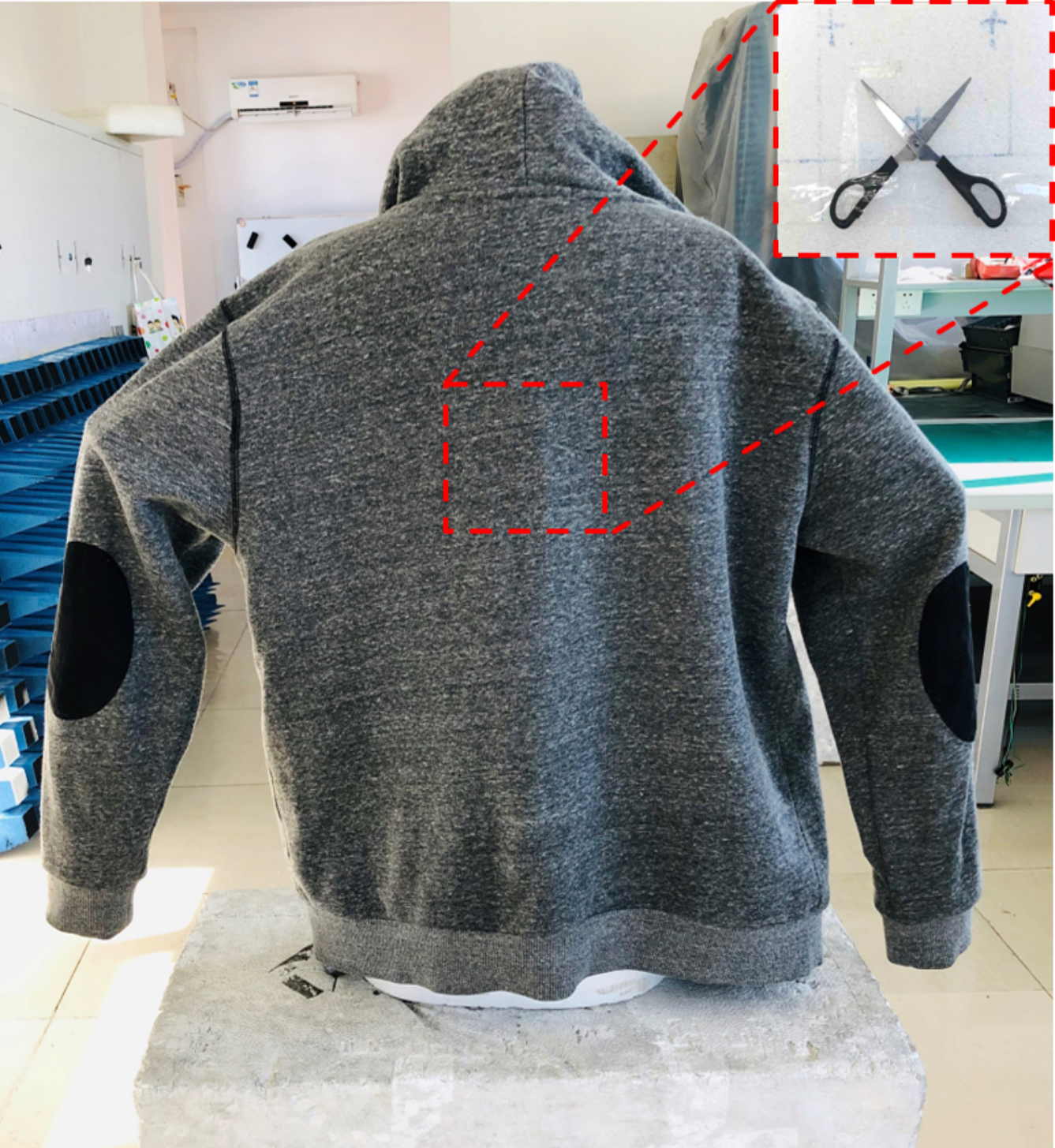}
	\hfill	
	
	\caption{Photographs of the imaging scenario. The scissor is regarded as the target, which is concealed under the cloth.}
	\label{scissor_test_scenario}
	\vspace{-2.5mm} 
\end{figure}

\begin{table}[!t]
	\centering
	\caption{Experimental Parameters}
	\setlength{\tabcolsep}{3pt}
	\begin{threeparttable}
		\begin{tabular}{p{200pt}  p{40pt}}
			%\begin{tabular}{|c|c|}
			\hline\hline
			Parameters& Values \\[0.5ex]
			\hline
			Imaging distance $(R_0)$&
			1.2 m\\[0.5ex]		
			Start frequency& 
			90 GHz \\[0.5ex]
			Stop frequency&
			96 GHz \\[0.5ex]
			Number of frequency steps&
			88 \\[0.5ex]			
			Number of transmit antennas of the full array &
			101 \\[0.5ex]
			Number of receive antennas of the full array &
			101 \\[0.5ex]	
			Spacing of full array antennas &
			5 mm \\[0.5ex]
			Number of transmit antennas of the synthesized/ equally spaced/ random sparse array&
			34 \\[0.5ex]
			Number of receive antennas of the synthesized/ equally spaced/ random sparse array&
			34 \\[0.5ex]
			Spacing of equally spaced sparse array antennas &
			15 mm \\[0.5ex]			
			\hline
			\vspace{-4.5mm} 
		\end{tabular}
	\end{threeparttable}
	\label{tab_T_shape_arr}
	\vspace{-7.5mm} 
\end{table}

We have constructed a two-scanner-based W-band imaging system to further verify the proposed method in a laboratory enviornment. Figure \ref{T_system_block_diagram} illustrates the block diagram of the envisioned imaging system. The system consists of 6 major parts: 1) a planar scanning platform with two independent scanners, through which a T-shaped MIMO topology can be realized by mechanical scanning; 2) a personal computer, which controls the planar scanning platform and the Vector Network Analyzer (VNA); 3) a VNA, which transmits and measures the Intermediate Frequency (IF) signal; 4) the transmit module as illustrated in Fig. \ref{Wband_Tx_Rx_module} \subref{a}, which upconverts the local oscillator (LO) signal, mixes it with the IF signal and transmits the W-band signal; 5) the receive module as illustrated in Fig. \ref{Wband_Tx_Rx_module} \subref{b}, which downconverts the reflected signal to an IF signal; 6) the LO, which provides the stable sinusoidal signal of 10.5 GHz.

Specifically, the transmit and receive modules with two horn antennas working at W-band are fixed on the upper and lower scanners, respectively. During the measurement, the receive module scans vertically from bottom to up. At each receiving module's scanning position, the transmit module scans horizontally back and forth. The PC controls the VNA to transmit the IF signal and record the $S_{21}$ parameter. Finally, the T-shaped MIMO topology can be achieved. The prototype of the T-shaped MIMO scanning imaging system is illustrated in Fig. \ref{T_shaped_MIMO_system}. The target under test is a pair of scissors fixed to a styrene foam block, which is covered by a coat, as shown in Fig. \ref{scissor_test_scenario}.

\begin{figure*}[t]	
	%	\begin{adjustwidth}{-\extralength}{0cm}
		%\centering %% If there is a figure in wide page, please release command \centering
		\hspace{0.23\columnwidth}\textbf{Full}
\hspace{0.30\columnwidth}\textbf{Synthesized sparse} 
\hspace{0.15\columnwidth}\textbf{Equally spaced sparse} 
\hspace{0.17\columnwidth}\textbf{Random sparse}
		\\
		%		\centering
		\setlength{\abovecaptionskip}{0pt}
		\vspace{-7mm}
		
		\hfill    
		\subfloat[]{\label{a}\includegraphics[width=1.75in]{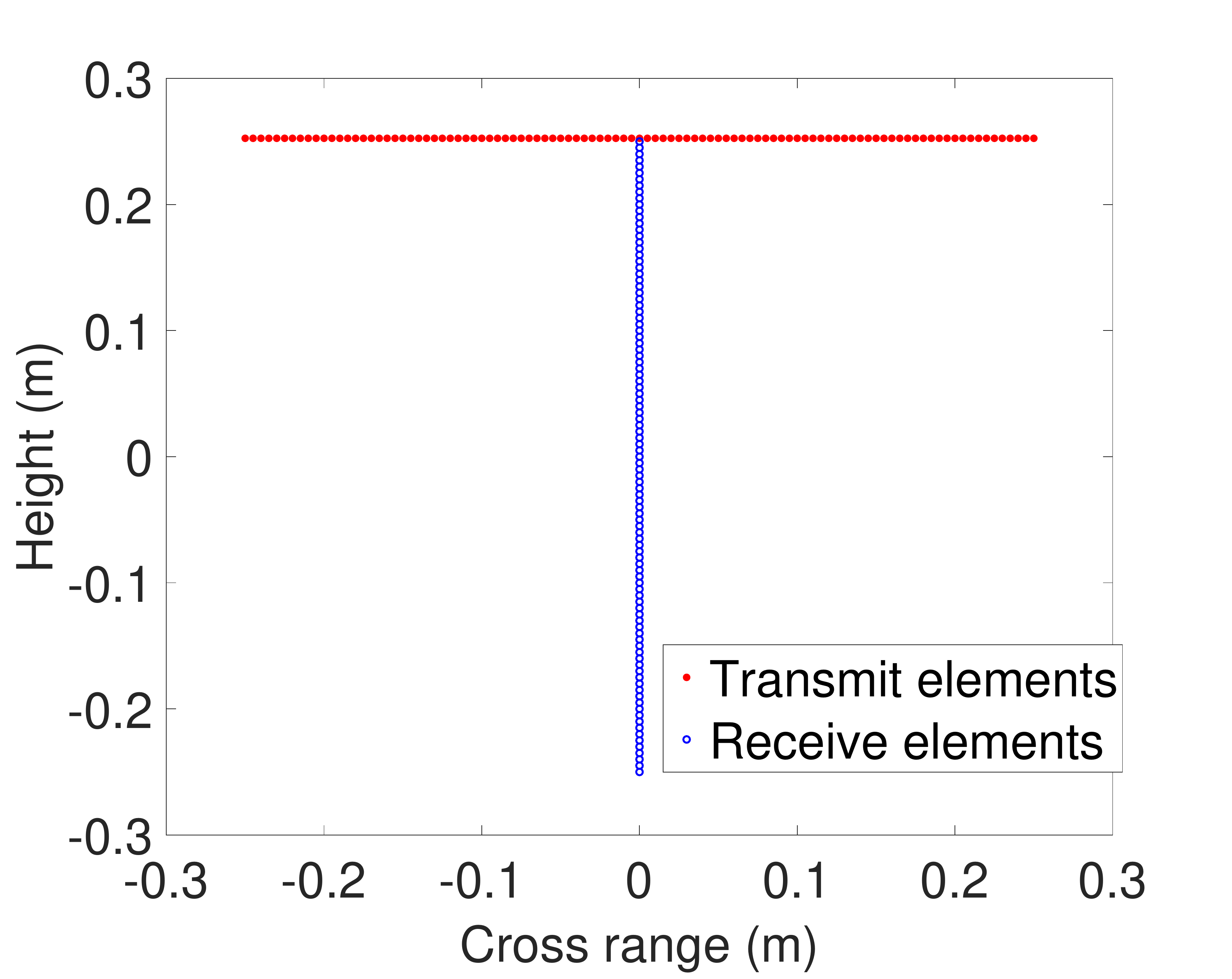}}
		\hfill
		\subfloat[]{\label{b}\includegraphics[width=1.75in]{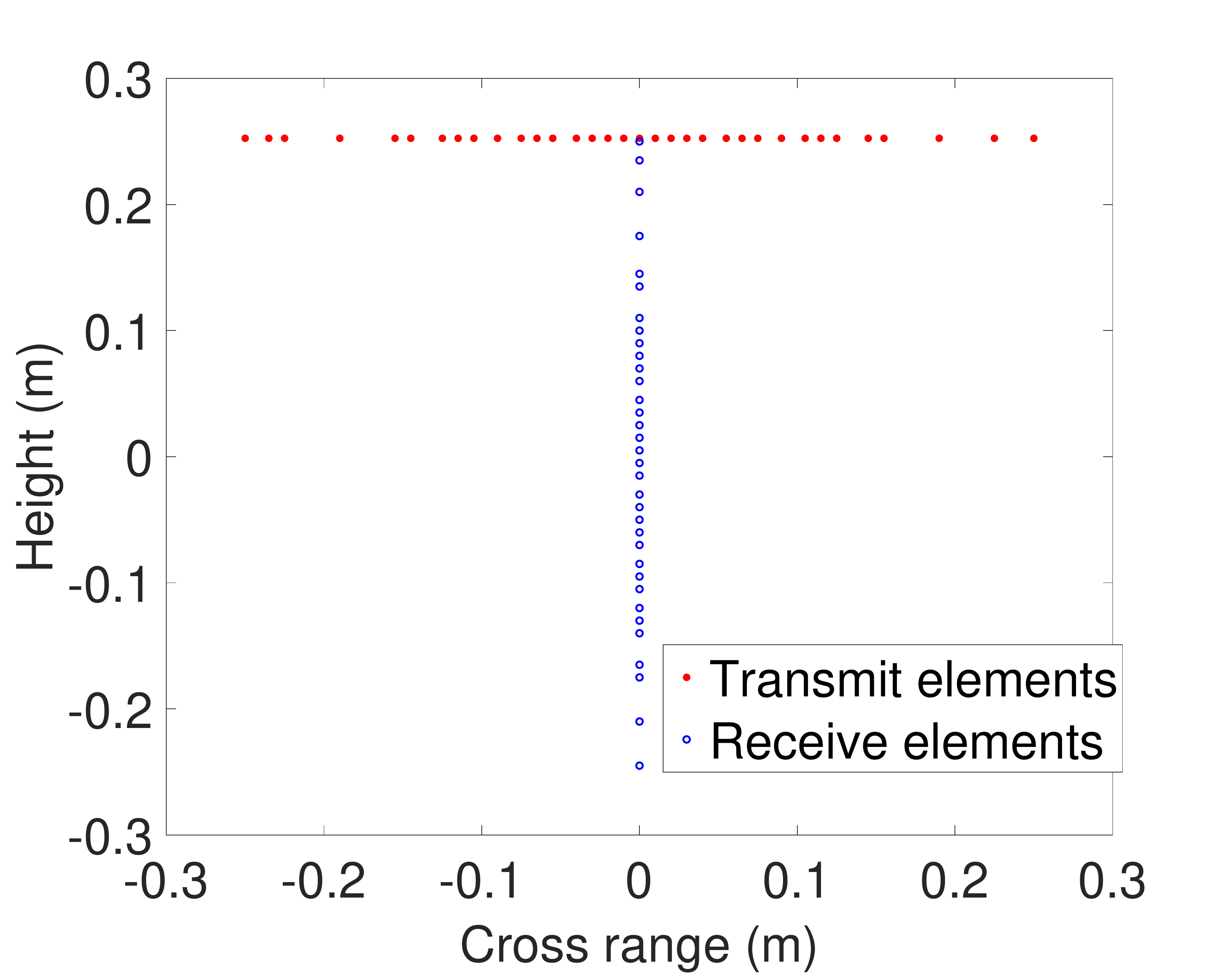}}
		\hfill
		\subfloat[]{\label{c}\includegraphics[width=1.75in]{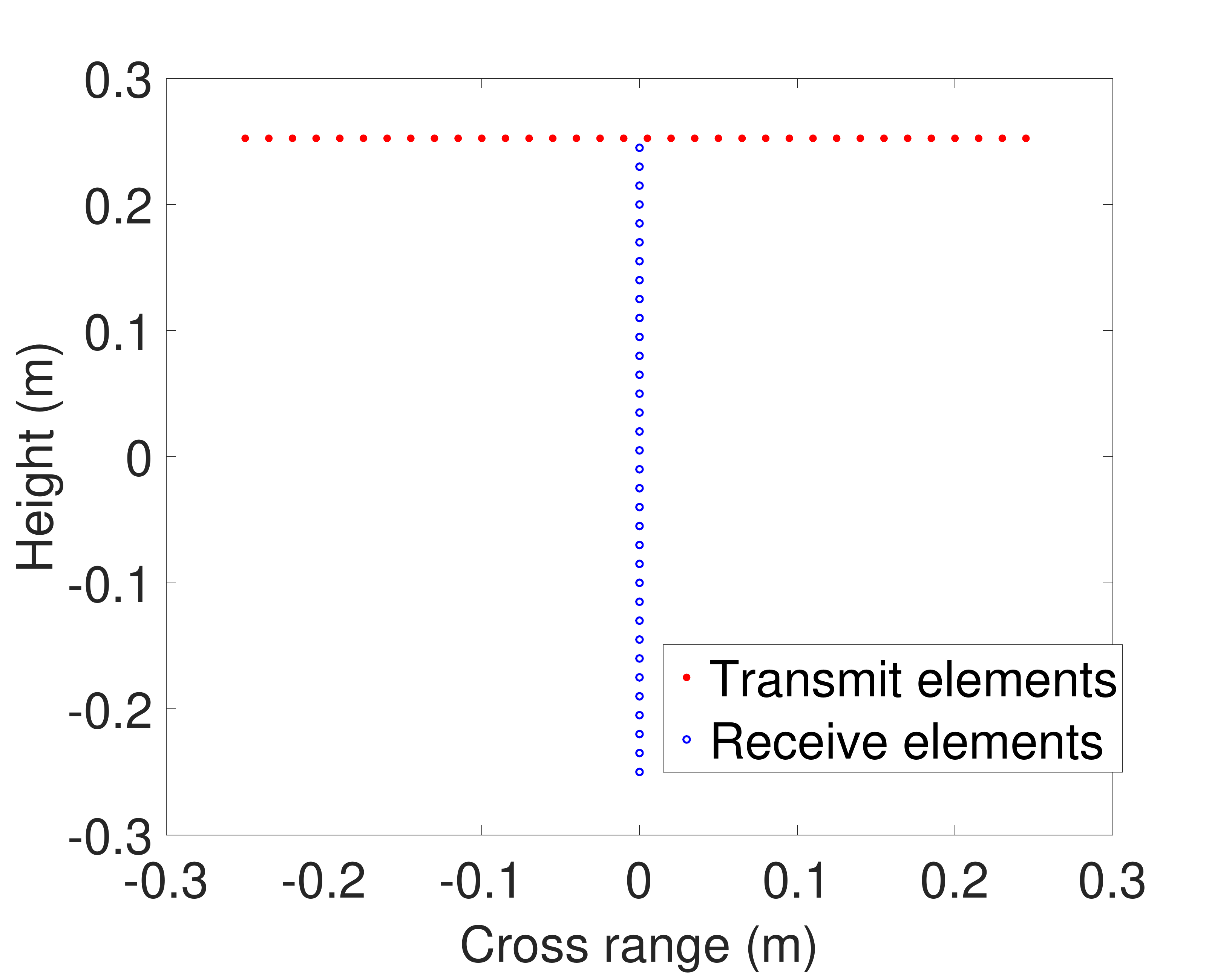}}
		\hfill
		\subfloat[]{\label{d}\includegraphics[width=1.75in]{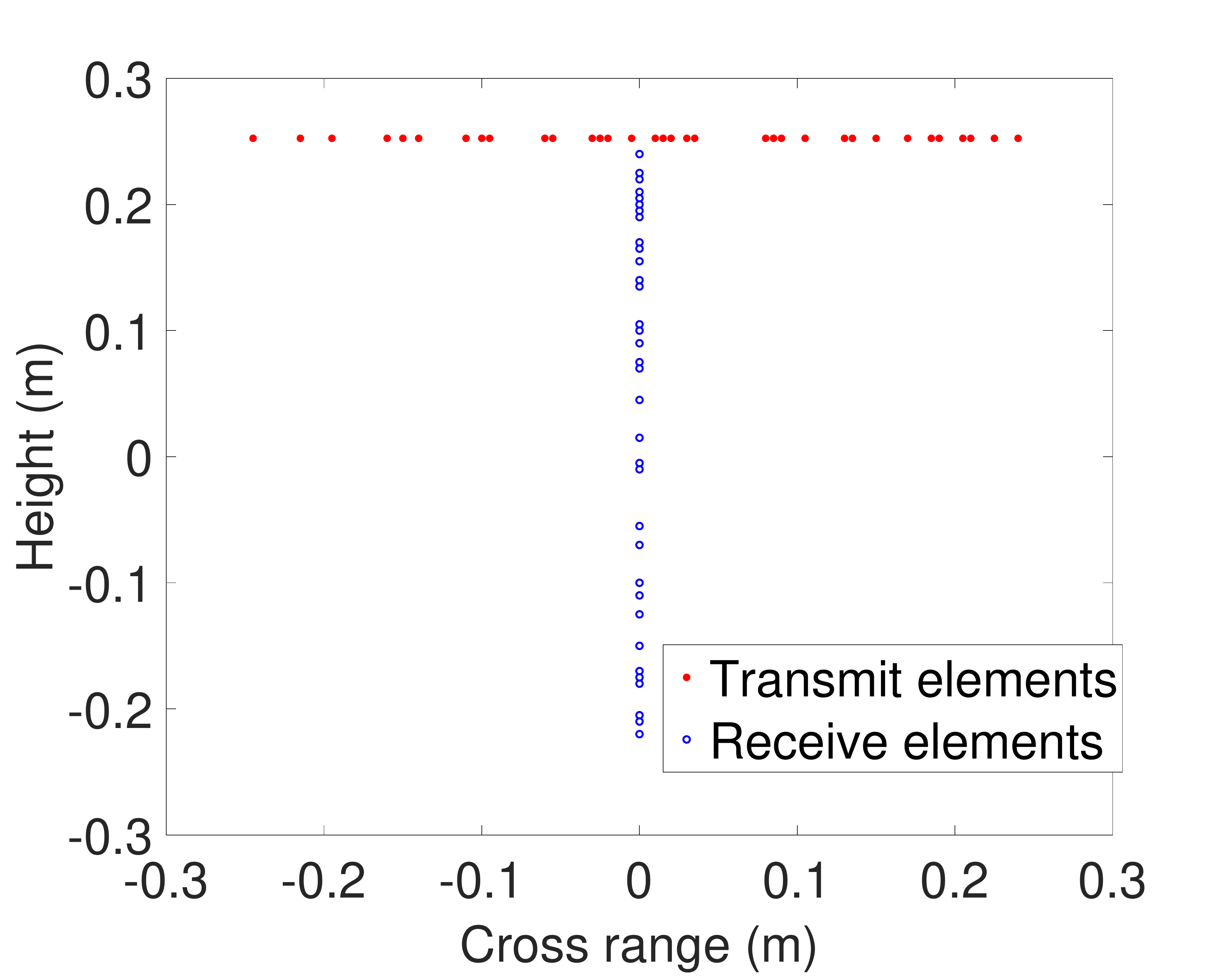}}
		\hfill
		\\
		
		\hfill    
		\subfloat[]{\label{e}\includegraphics[width=1.75in]{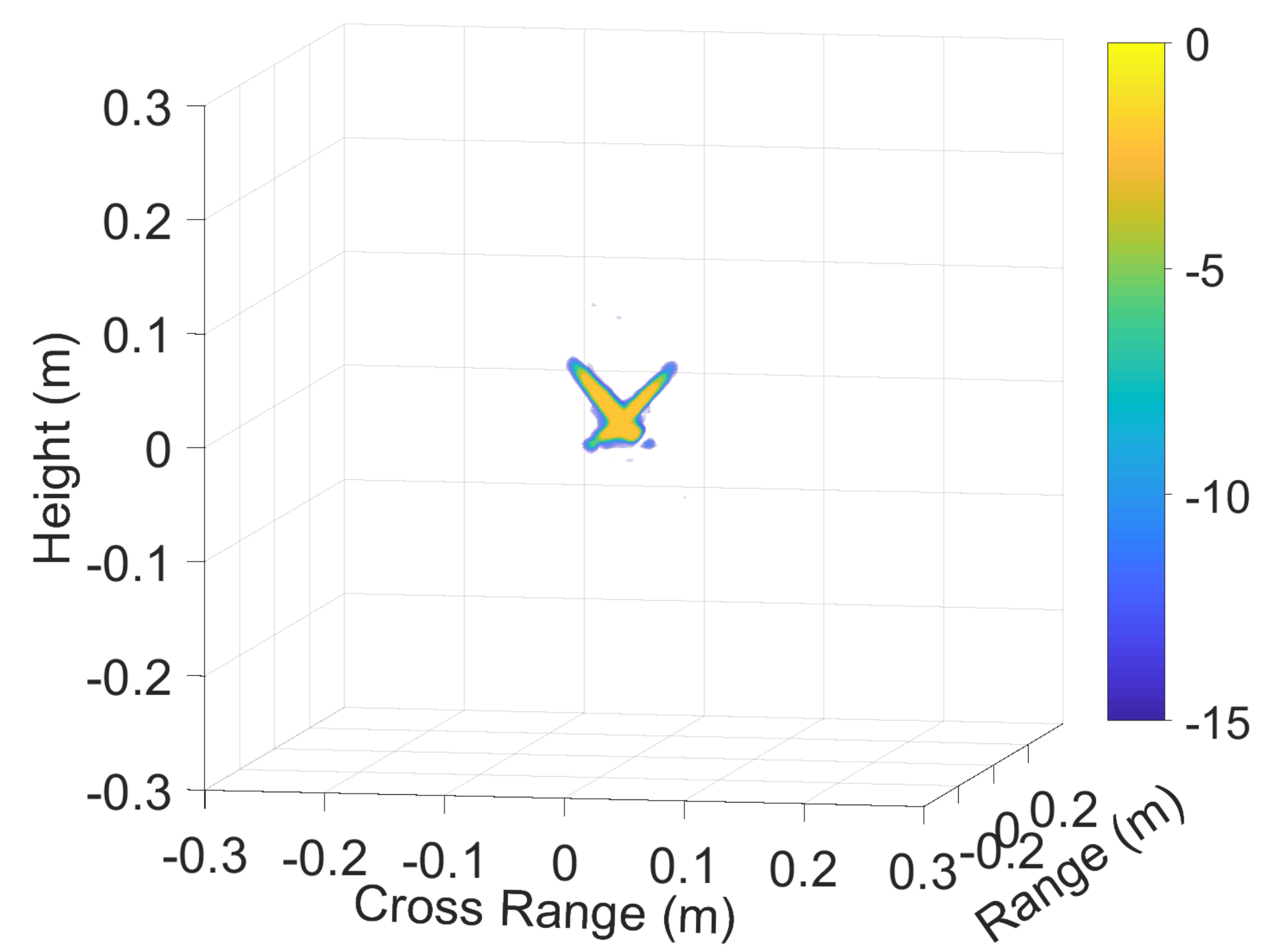}}
		\hfill
		\subfloat[]{\label{f}\includegraphics[width=1.75in]{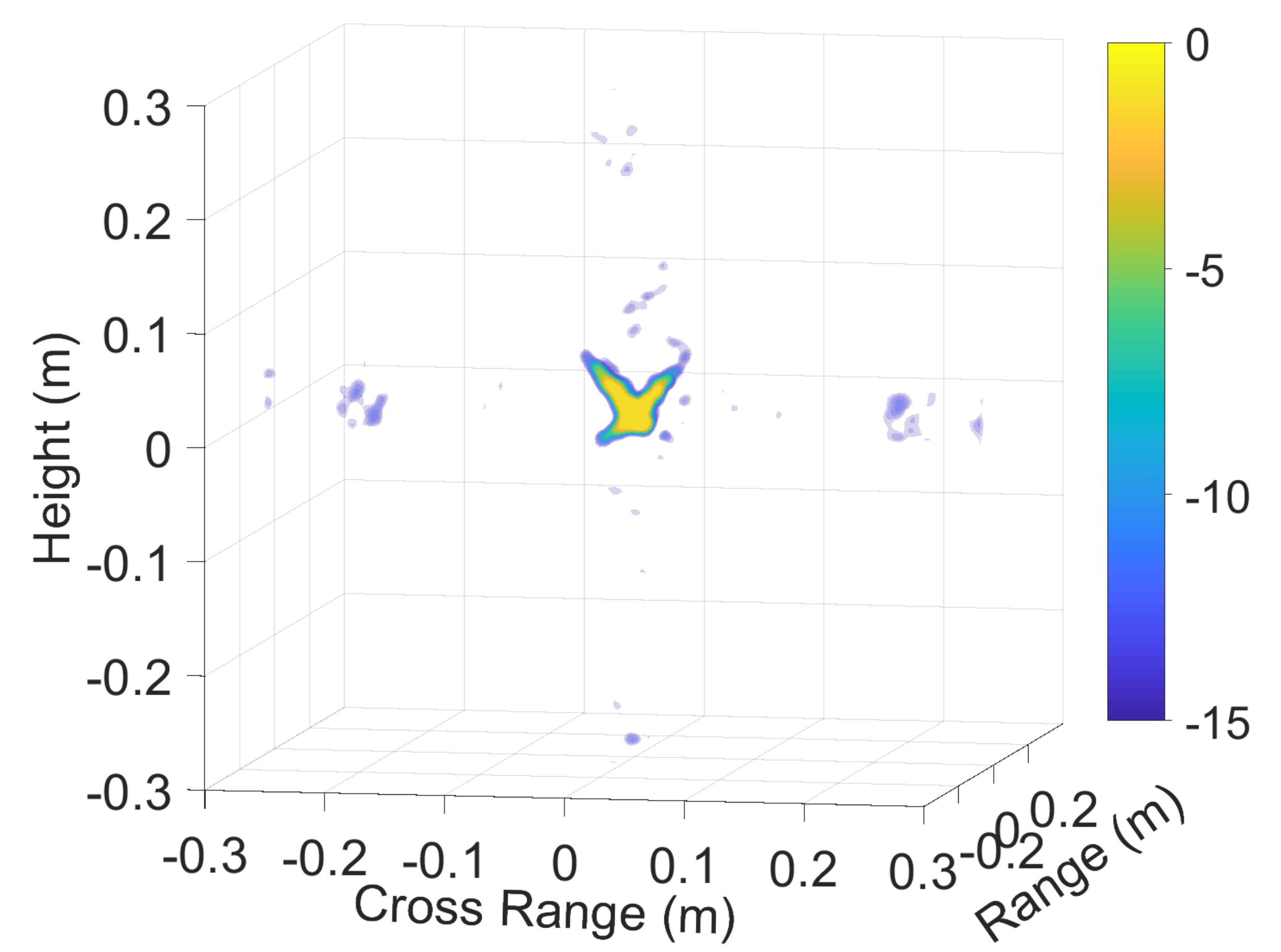}}	
		\hfill
		\subfloat[]{\label{g}\includegraphics[width=1.75in]{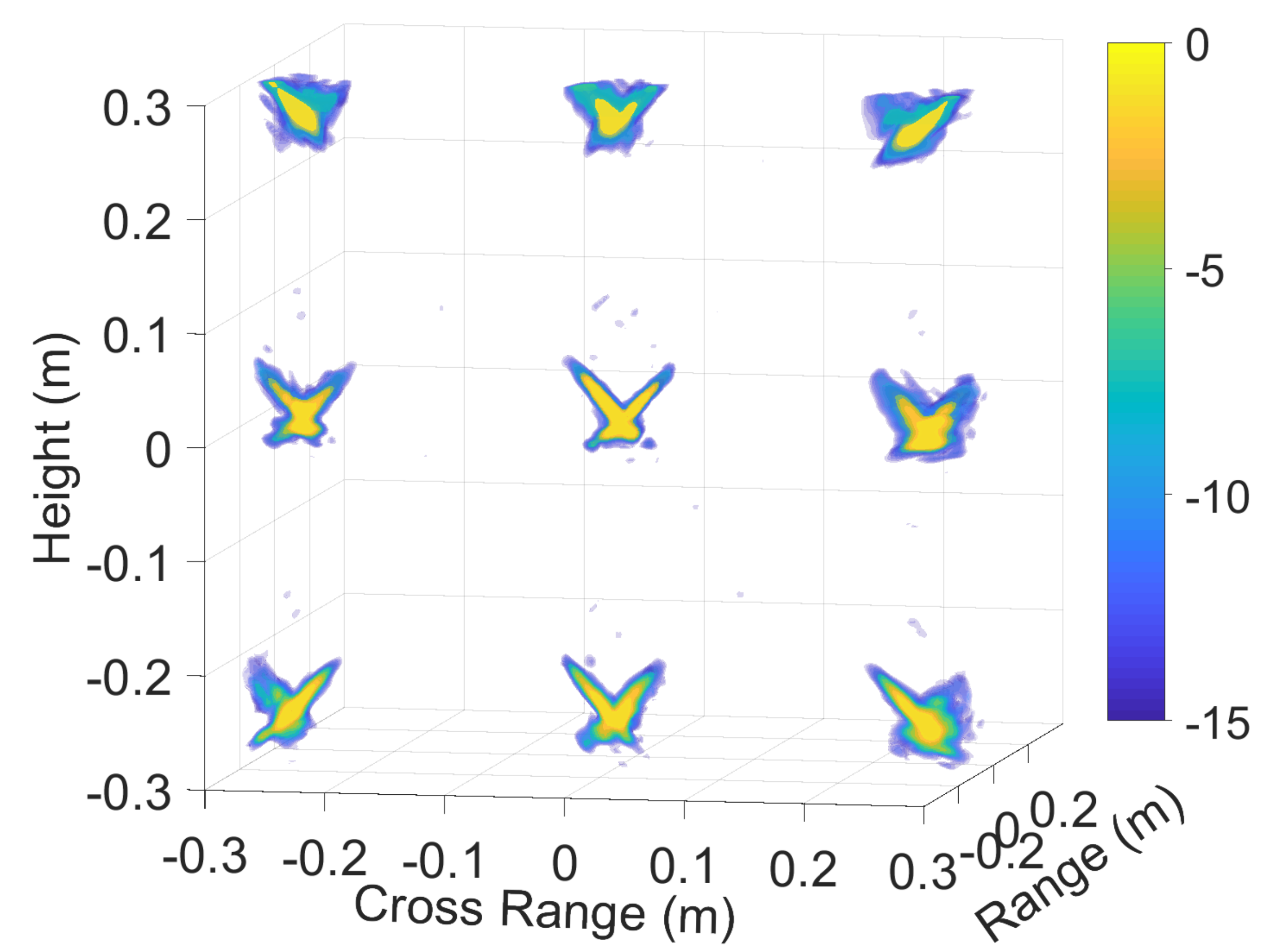}}
		\hfill
		\subfloat[]{\label{h}\includegraphics[width=1.75in]{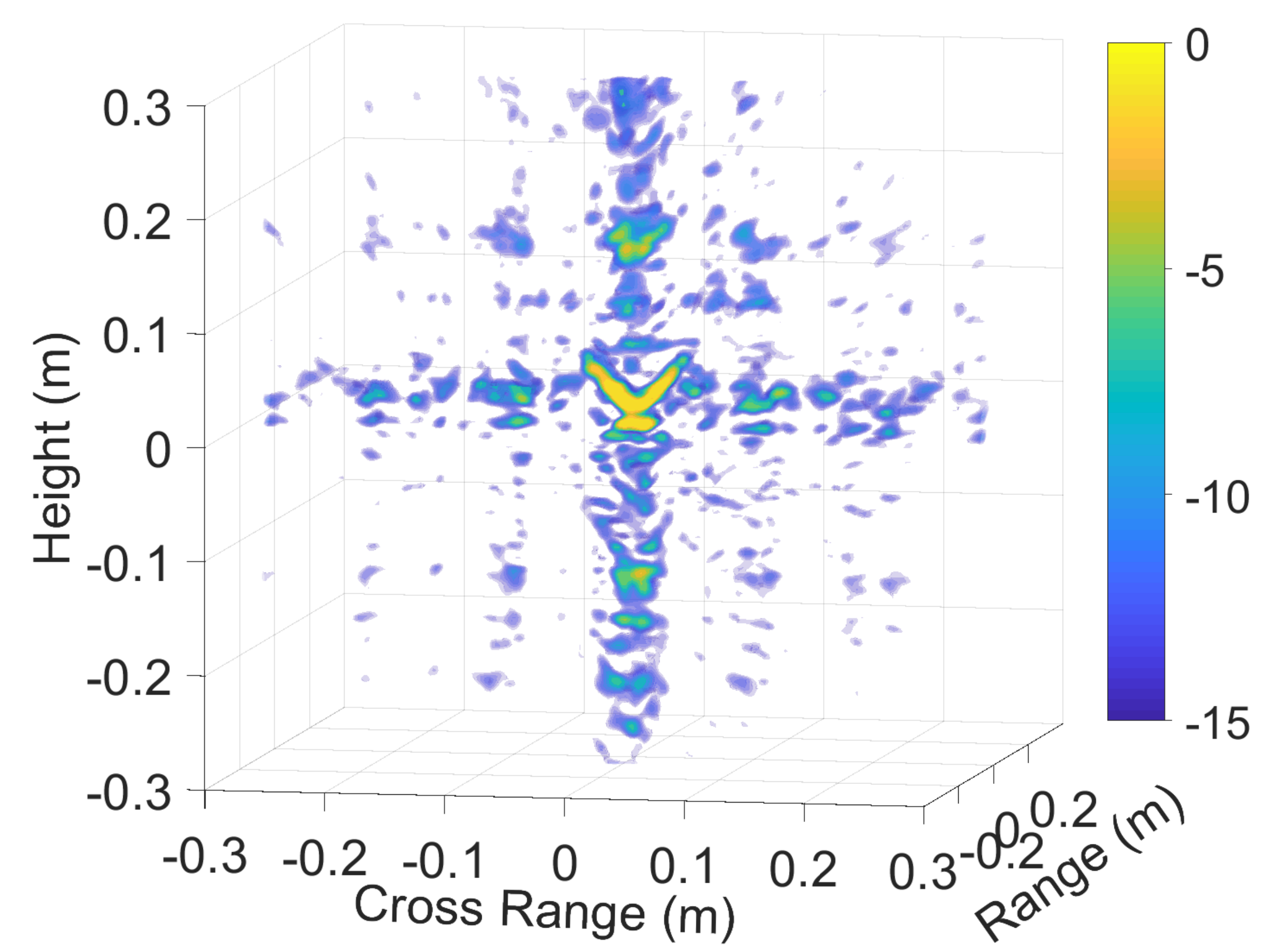}}
		\hfill
		\\
		
		\hfill    
		\subfloat[]{\label{e}\includegraphics[width=1.75in]{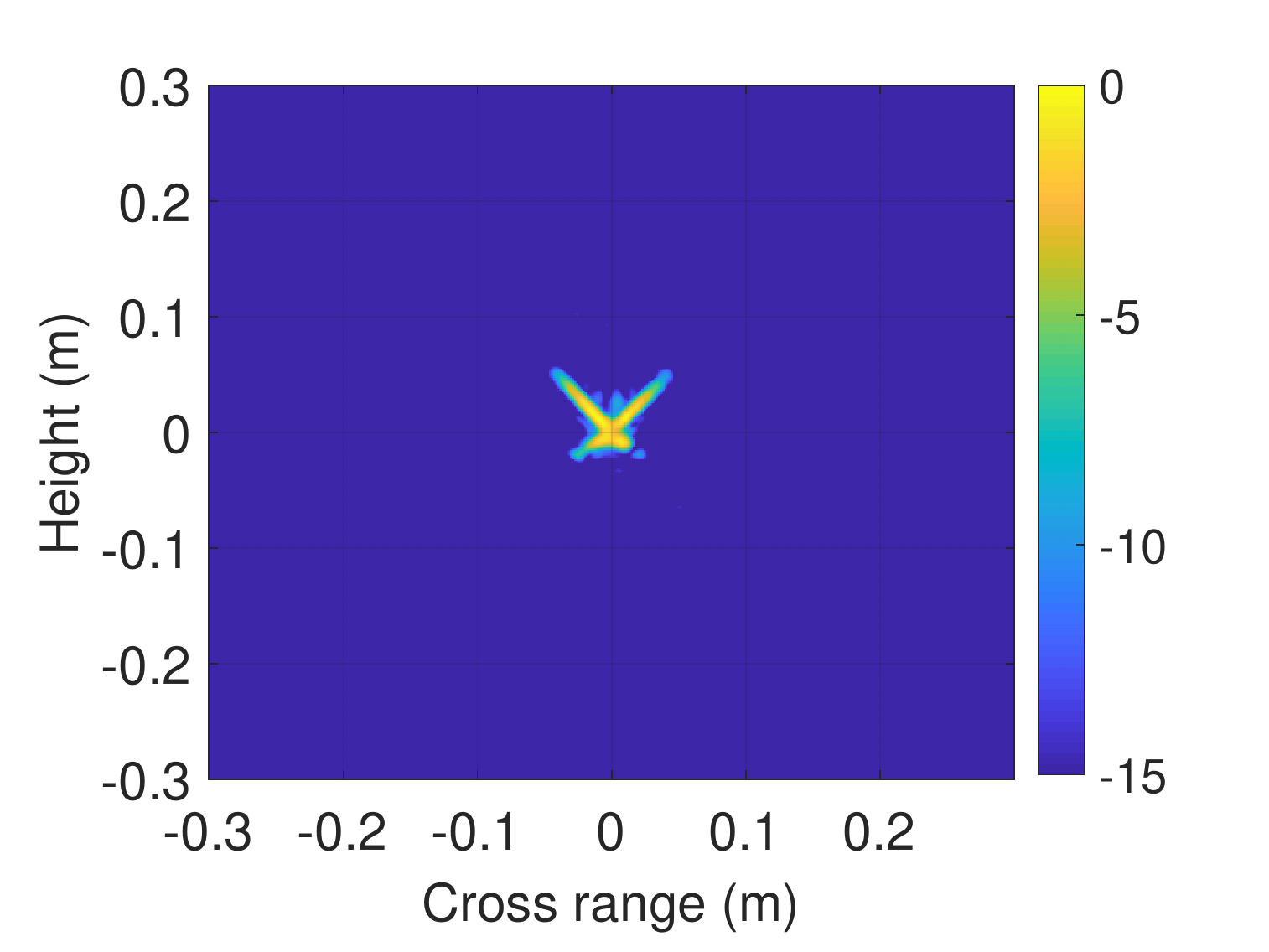}}
		\hfill
		\subfloat[]{\label{f}\includegraphics[width=1.75in]{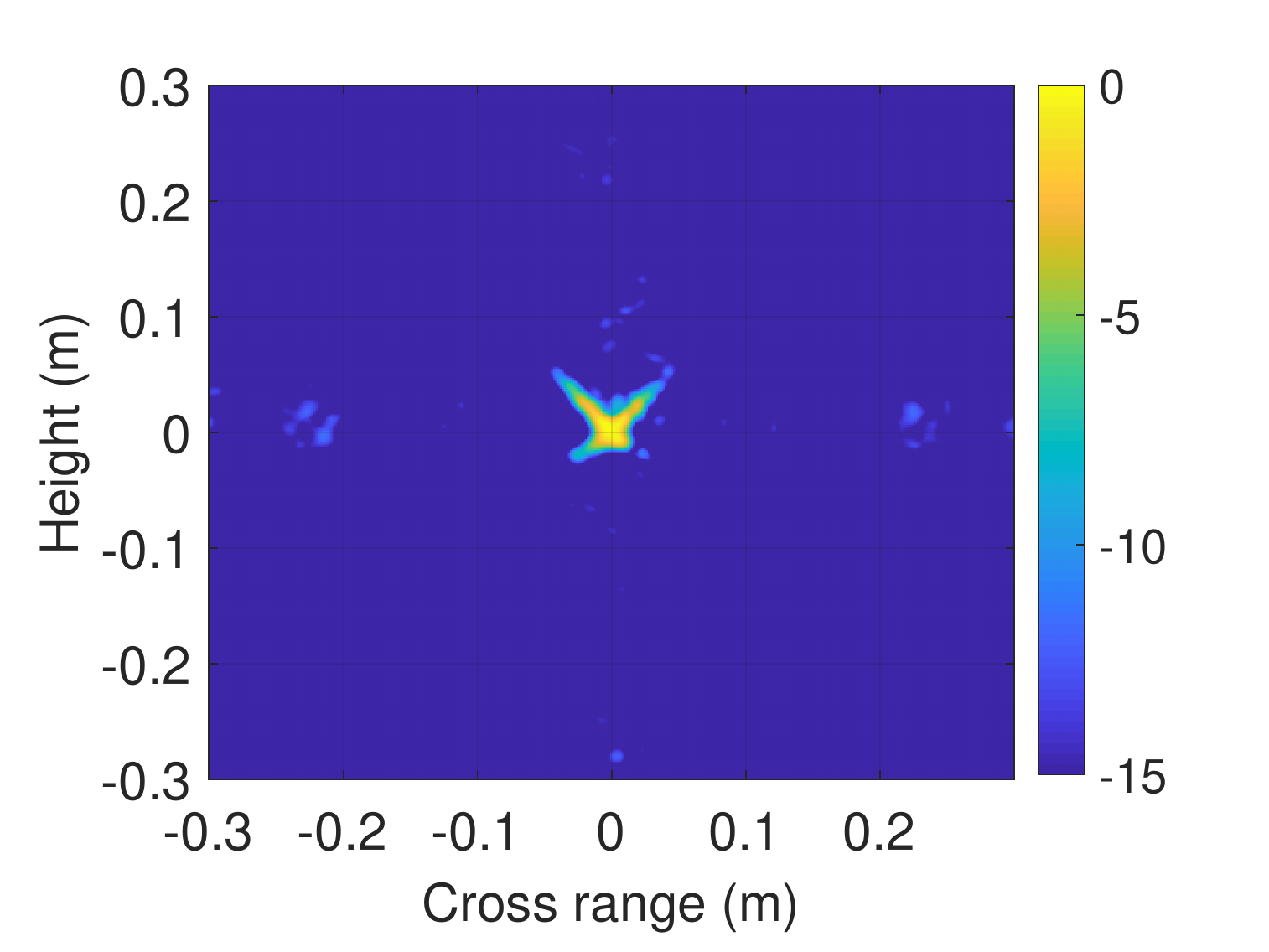}}	
		\hfill
		\subfloat[]{\label{g}\includegraphics[width=1.75in]{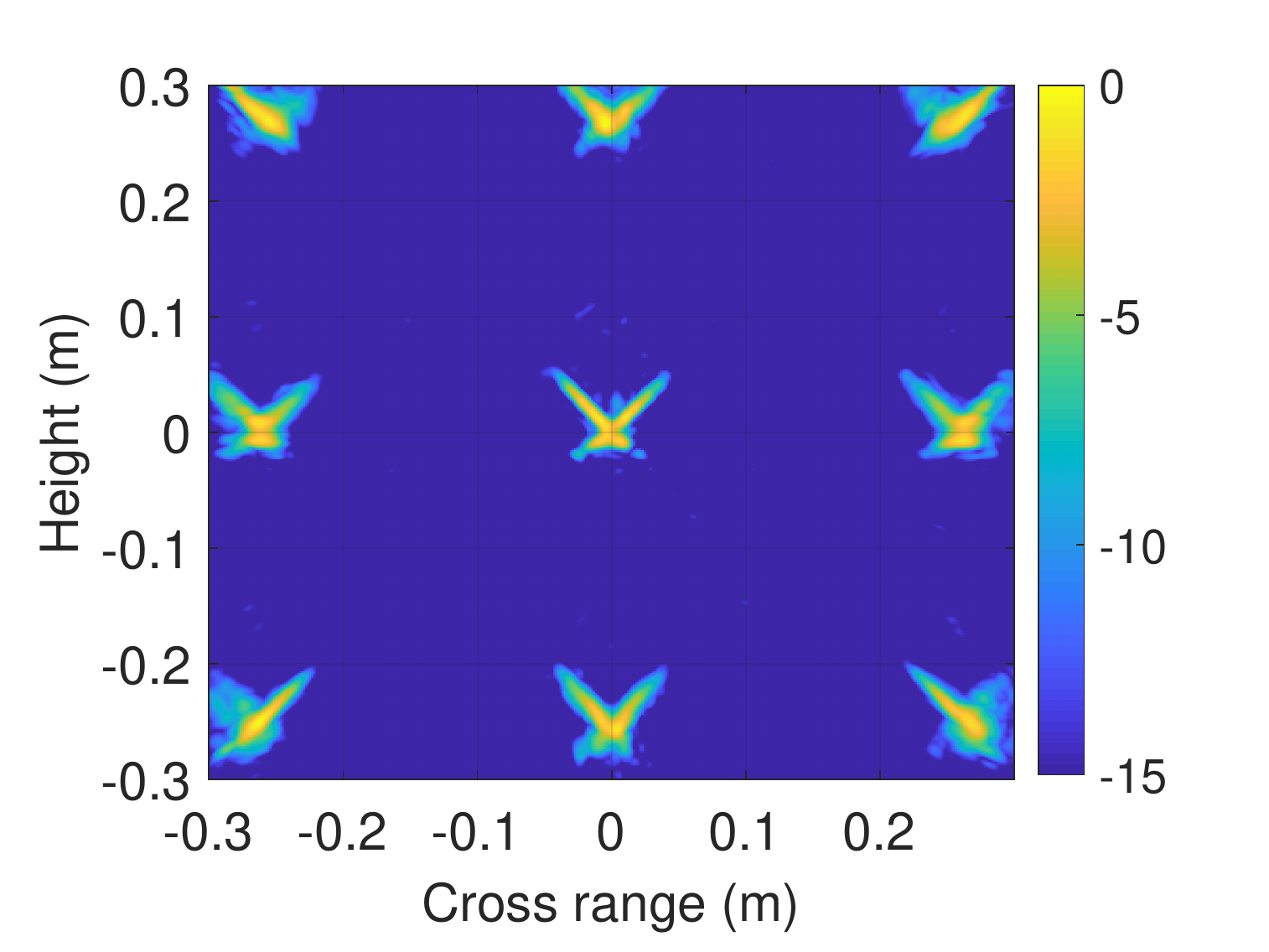}}
		\hfill
		\subfloat[]{\label{h}\includegraphics[width=1.75in]{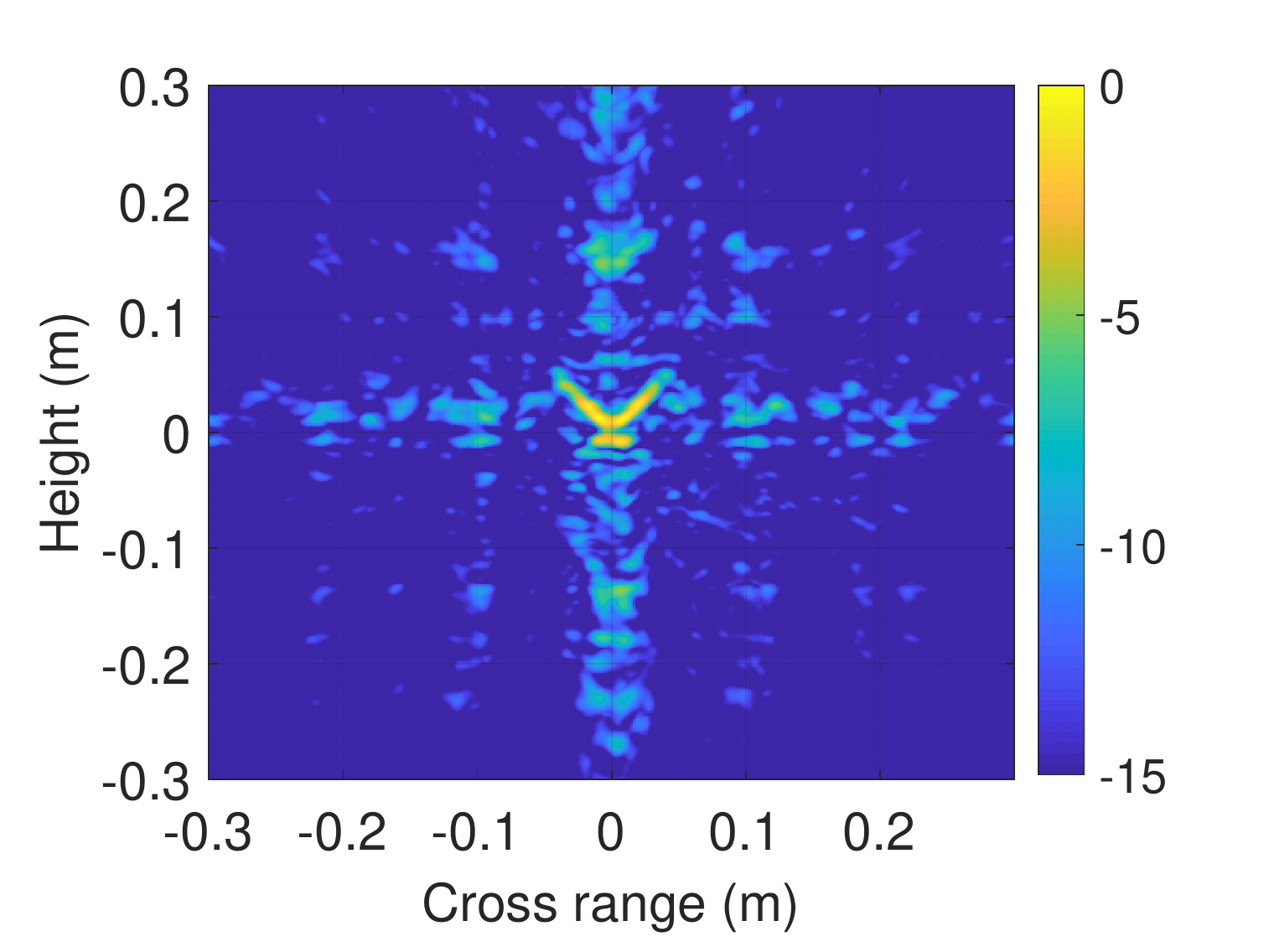}}
		\hfill
		\\
		
		\setlength{\belowcaptionskip}{-0.2cm} 
		%	\end{adjustwidth}
	\caption{Comparison of topologies and imaging reconstruction results of different arrays, ({a}--{d}) T-shaped MIMO topologies, ({e}--{h}) 3-D imaging reconstruction results, ({i}--{l}) 2-D imaging results of maximum projection along the range direction. The titles on the top indicate different topologies corresponding to the images column by column.}
	\label{T_MIMO_results} 
\end{figure*}

In this experiment, the following two steps can achieve the SAS of the T-shaped MIMO array. Firstly, fix the transmit positions and optimize the receive positions. Then optimize the transmit positions with the fixed synthesized  receive positions. The advantages of the synthesized sparse array can be verified with comparisons to the random sparse array, and the equally spaced sparse array (with almost the same number of elements). The result of the full array is employed as a baseline. The corresponding topologies are illustrated in Figs. \ref{T_MIMO_results} \subref{a}-\subref{d}. The experimental parameters are listed in Table \ref{tab_T_shape_arr}.

The second and third rows of Fig. \ref{T_MIMO_results} present the 3-D  reconstruction results and their corresponding 2-D maximum value projections along the range direction by the BP algorithm. The dynamic range of the reconstruction results are set to 15 dB.

Note that the T-shaped full MIMO array shows the best imaging results. The handles of the scissors are invisible, which may be caused by the scotch tape above it. For the T-shaped equally spaced sparse MIMO array, many false scissors' ghosts are reconstructed due to the severe grating lobe effect, making it hard to locate the precise position of the target. In regards to the T-shaped random sparse MIMO array, severe artifacts around the scissors are observed, partially `polluting' its details. Compared with the two designs mentioned above, the synthesized T-shaped sparse MIMO array can mitigate the clutter components and grating lobes and maintain the details of the target.

\begin{table}[!t]
	\centering
	\caption{Quantitative Comparison of Imaging of Different T-shaped MIMO Topologies}
	\setlength{\tabcolsep}{3pt}
	\begin{threeparttable}
		\begin{tabular}{p{80pt}  p{30pt}  p{30pt}  p{30pt}  p{30pt}}
			%\begin{tabular}{|c|c|}
			\hline\hline
			Array Topologies & RMSE & PSNR  & SSIM & entropy \\[0.5ex]
			\hline
			synthesized sparse &4.48 & 35.1 dB & 0.979 & 0.159\\[0.5ex]		
			equally spaced sparse &27.6& 19.3 dB & 0.930 & 0.360\\[0.5ex]	
			random spaced sparse & 26.8 & 19.6 dB & 0.660 & 0.771\\[0.5ex]		
			\hline
			
		\end{tabular}
	\end{threeparttable}
	\label{MIMO_T_actual_indicators}
	\vspace{-2.5mm} 
\end{table}

Similar to the part of the FEKO simulation, the quantitative comparison of the aforementioned T-shaped MIMO array is listed in Table \ref{MIMO_T_actual_indicators}, with the same metrics. The imaging results of the T-shaped full MIMO array are set as the baseline. The results show a consistent superiority of the sparse MIMO arrays generated by our SAS method over the other sparse arrays.

\section{Conclusions}

This paper proposed a CS-based convex optimization method for wideband MIMO array design for near-field imaging. An $l_1$-norm based optimization model was constructed, associated with a new design of the reference pattern. We compared the performance of the synthesized MIMO arrays with the commonly used sparse topologies: the equally spaced and randomly spaced sparse array
configurations. Both the numerical and the experimental imaging results confirm that the synthesized sparse MIMO arrays outperform the other arrays with regards to SLLs and grating lobe suppression. Compared with the fully sampled uniformly spaced array, the developed sparse T-shaped MIMO array generated by the proposed method can reduce the number of elements by more than 64\%. This indicates the potential of the low-cost and simplified wideband sparse MIMO imaging systems for practical applications. Furthermore, the proposed MIMO SAS approach can handle other array schemes, like the cylindrical or the polyline topologies.
  
\appendices
\section{Proof of Lemma 1}

 \renewcommand{\theequation}{A. \arabic{equation}}    
\setcounter{equation}{0} 
We can rewrite \eqref{scat_wave_mat} as the combination of $Q$ different point scatterers:
\begin{equation}
	\label{scat_wave_mat2}
	\bm{s}_k\!= \bm{a}_{k,1} \sigma_1+\bm{a}_{k,2} \sigma_2+\dots+\bm{a}_{k,q} \sigma_q+\dots+\bm{a}_{k,Q} \sigma_Q,
\end{equation}
where
%\begin{align}
%	\begin{split}
%\!\!\!\bm{a}_{k,q}\!=\! e^{-\mathrm{j}k\left|\bm{r'}_{\rm T}\! -\bm{r}_q\right|} \!\! \left[e^{-\mathrm{j}k\left|\bm{r'}_{{\rm R}_1}\!\!-\bm{r}_q\right|},\dots,e^{-\mathrm{j}k\left|\bm{r'}_{{\rm R}_N}\!\!-\bm{r}_q\right|}\right]^T_,\\ 
%	\end{split}\label{a}
%\end{align}
\begin{align} 	\begin{split} \!\!\!\bm{a}_{k,q}\!=\! \frac{e^{-\mathrm{j}k\left|\bm{r'}_{\rm T}\!\!-\bm{r}_q\right|}}{4\pi\left|\bm{r'}_{\rm T}\!\!-\! \bm{r}_q\right|} \!\! \left[\frac{e^{-\mathrm{j}k\left|\bm{r'}_{{\rm R}_1}\!\!-\bm{r}_q\right|}}{4\pi\left|\bm{r'}_{{\rm R}_1}\!\!-\bm{r}_q\right|},\dots,\frac{e^{-\mathrm{j}k\left|\bm{r'}_{{\rm R}_N}\!\!-\bm{r}_q\right|}}{4\pi\left|\bm{r'}_{{\rm R}_N}\!\!-\bm{r}_q\right|}\right]^T_,\\ 	\end{split}\label{a} 
\end{align}
and $k$ denotes the wavenumber, and $q\in \left\lbrace 1,2,\dots,Q\right\rbrace$. Mark the set $\mathbb{Q}$ and $\mathbb{A}_k$ as
\begin{align}\label{set_A}
	\mathbb{Q}\!&=  \left\lbrace  1,2,\dots,Q \right\rbrace \\	
	\mathbb{A}_k\!&=  \left\lbrace  \bm{a}_{k,1},\bm{a}_{k,2},\dots,\bm{a}_{k,Q} \right\rbrace.
\end{align}

Denoting $\bm{s}_{k,q}=\bm{a}_{k,q} \sigma_q$, we can get
\begin{equation}\label{Sq}
	\bm{S}_{k,q}= \! \begin{bmatrix}		
		s_{k,q}\left(\bm{r'}_{{\rm R}_1}\right)\!\!&\!   &  & \\
		&\!\!\! s_{k,q}\left(\bm{r'}_{{\rm R}_2}\right) &  & \\			
		&\! & \!\!\ddots \! &  \\
		&\!  &  &\!\!\! s_{k,q}\left(\bm{r'}_{{\rm R}_N}\right)
	\end{bmatrix}_{N \times N}
\end{equation}

Similarly, we can get the sub-imaging results for the $q \rm th$ scatterer using \eqref{MIMO_BP_equ3}: 
\begin{equation}\label{BP_equ_q}
	\bm{E}_{{\rm T}_{k,q}}\!= \bm{\Phi}_{{\rm T}_k} \bm{\Phi}_{{\rm R}_k} \bm{S}_{k,q}\bm{w}.
\end{equation}

With the fixed transmit array, for any receive array topology $\bm{w}$ synthesized by \eqref{cvx_problem2}, we get the following constraints:
\begin{equation}\label{deduce1}
	\Vert \sum_k \sum_{\bm{r'}_{\rm T}} \sum_{q=1}^Q \left[ \bm{\Phi}_{{\rm T}_k} \bm{\Phi}_{{\rm R}_k} \bm{S}_{k,q} \left(  \bm{w}_{\rm ref}- \bm{w}\right) \right] \Vert^2_2\leq \varepsilon.
\end{equation}

Define $\bm{w}_{\rm diff}=\bm{w}_{\rm ref}-\bm{w}$, and with the same conversion from \eqref{MIMO_BP_equ2} to \eqref{MIMO_BP_equ3}, \eqref{deduce1} can be rewritten as:

\begin{equation}\label{deduce2}
	\Vert \sum_k \sum_{\bm{r'}_{\rm T}} \sum_{q=1}^Q  \sigma_q \bm{\Phi}_{{\rm T}_k} \bm{\Phi}_{{\rm R}_k} \bm{W}_{\rm diff}\bm{a}_{k,q} \Vert^2_2\leq \varepsilon.
\end{equation}
where
\begin{equation}\label{W}
	\bm{W}_{\rm diff}= \begin{bmatrix}		
		{w}_{\rm ref}\!\left(1\right)\!-\!{w}\!\left(1\right)&  & \\		
		&\ddots  &  \\
		& &{w}_{\rm ref}\!\left(N\right)\!-\!{w}\!\left(N\right)
	\end{bmatrix}
\end{equation}

In the near-field, according to the imaging principle of delay-and-sum method\cite{kay1993fundamentals}, the following assumption for orthogonal relationship should be approximately satisfied for any $\bm{a}_{k,i}\in \mathbb{A}_k$ and $\bm{a}_{k,j}\in \mathbb{A}_k\setminus\bm{a}_{k,i}$,
\begin{equation}\label{bot}
	\bm{a}_{k,i} \ \bot \ \bm{a}_{k,j}
\end{equation}

Therefore, for any subsets $\mathbb{A}_{\rm sub}\subseteq\mathbb{A}_k$ or $\mathbb{Q}_{\rm sub}\subseteq\mathbb{Q}$ , we get
\begin{equation}\label{deduce3}
	\begin{aligned}
	&\Vert \sum_k \sum_{\bm{r'}_{\rm T}} \sum_{i\in\mathbb{Q}_{\rm sub}} \sigma_i \bm{\Phi}_{{\rm T}_k} \bm{\Phi}_{{\rm R}_k} \bm{W}_{\rm diff}\bm{a}_{k,i} \Vert^2_2\leq	\\
	&\Vert \sum_k \sum_{\bm{r'}_{\rm T}} \sum_{q=1}^Q  \sigma_q   \bm{\Phi}_{{\rm T}_k} \bm{\Phi}_{{\rm R}_k} \bm{W}_{\rm diff}\bm{a}_{k,q} \Vert^2_2\leq \varepsilon.
\end{aligned}
\end{equation}

Then \eqref{deduce3} can be rewritten as
\begin{equation}\label{deduce4}
	\Vert \sum_k \sum_{\bm{r'}_{\rm T}} \sum_{i\in\mathbb{Q}_{\rm sub}} \left[ \bm{\Phi}_{{\rm T}_k} \bm{\Phi}_{{\rm R}_k} \bm{S}_{k,i} \left(  \bm{w}_{\rm ref}- \bm{w}\right) \right] \Vert^2_2\leq \varepsilon.
\end{equation}

Now, let us define
\begin{equation}\label{E_pie_ref}
\bm{E}'_{\rm ref}\!=\!\!\! \sum\limits_k  \sum_{\bm{r'}_{\rm T}} \sum\limits_{i\in\mathbb{Q}_{\rm sub}}\! \bm{\Phi}_{{\rm T}_k} \bm{\Phi}_{{\rm R}_k} \bm{S}_{k,i}\bm{w}_{\rm ref}
\end{equation}
\begin{equation}\label{B_pie_ref} \bm{B}'\!=\!\!\! \sum\limits_k \sum_{\bm{r'}_{\rm T}} \sum\limits_{i\in\mathbb{Q}_{\rm sub}}\! \bm{\Phi}_{{\rm T}_k} \bm{\Phi}_{{\rm R}_k} \bm{S}_{k,i},
\end{equation}
which then leads one to finally express \eqref{deduce4} as
\begin{equation}\label{cvx_problem_end}
	\Vert \bm{E}'_{\rm ref}-\bm{B}'\bm{w}\Vert^2_2\leq \varepsilon.
\end{equation}

Therefore, one can conclude that for any possible combination in the subset of the reference pattern, the difference between the imaging results of the synthesized array and the referenced array is less than $\varepsilon$.

%\begin{appendices}	
%\end{appendices}

\ifCLASSOPTIONcaptionsoff
\newpage
\fi

\bibliography{IEEEabrv,full}
\bibliographystyle{IEEEtran}

%\bibliography{./bibtex/bib/IEEEabrv,./bibtex/bib/full}
%\bibliographystyle{./bibtex/bib/IEEEtran}

\end{document}